\documentclass{aastex61}

\usepackage{xspace}
\usepackage{color}
\usepackage{hyperref}
\usepackage{afterpage}
\usepackage{makecell}
\hypersetup{
   pdftitle={Survey Design, Pipeline, and Supernova Discoveries from the \textit{HST} See Change Program}, 
   pdfauthor={Brian Hayden},
   colorlinks=true,        
   linkcolor=blue,         
   citecolor=blue,         
   urlcolor=blue           
}

\begin{document}
\shorttitle{See Change Survey}
\shortauthors{Hayden et al.}

\renewcommand{\arraystretch}{1.75}

\newcommand{\myemail}{bhayden@stsci.edu}
\newcommand{\ang}{\AA\xspace}
\newcommand{\note}[1]{\textcolor{red}{#1}}
\newcommand{\rrg}[1]{\textcolor{violet}{#1}}
\newcommand{\scw}[1]{\textcolor{orange}{#1}}
\newcommand{\code}[1]{\texttt{#1}}
\newcommand{\HST}{\textit{HST}\xspace}
\newcommand{\Hubble}{\textit{Hubble Space Telescope}\xspace}
\newcommand{\lambdar}{\textsc{lambdar}\xspace}
\newcommand{\LCDM}{$\Lambda$CDM\xspace}
\newcommand{\SNeIa}{SNe~Ia\xspace}
\newcommand{\SNIa}{SN~Ia\xspace}
\newcommand{\gold}{Ia\xspace}
\newcommand{\silver}{likely Ia\xspace}
\newcommand{\bronze}{possible Ia\xspace}
\newcommand{\pyrite}{non-Ia\xspace}
\newcommand{\ntransients}{57\xspace}
\newcommand{\ngold}{27\xspace}
\newcommand{\nsilver}{0\xspace}
\newcommand{\nIa}{27\xspace} 
\newcommand{\ncosmo}{26\xspace} 
\newcommand{\ncosmospec}{21\xspace}
\newcommand{\ncosmonospec}{6\xspace}
\newcommand{\nbronze}{7\xspace}
\newcommand{\npyrite}{22\xspace}
\newcommand{\ngoldspec}{21\xspace}
\newcommand{\nsilverspec}{0\xspace}
\newcommand{\pkg}[1]{\texttt{#1}\xspace}
\newcommand{\orbitPercent}{45\%\xspace}
\newcommand{\SnePerOrbit}{0.2\xspace}
\newcommand{\PIDList}{11597, 11663, 12203, 12477, 12994, 13306, 13677, 13747, 13823, 13845, and 14327\xspace}

\newcommand{\stsci}{\affiliation{Space Telescope Science Institute, 3700 San Martin Drive Baltimore, MD 21218, USA}}
\newcommand{\lbnl}{\affiliation{E.O. Lawrence Berkeley National Laboratory, 1 Cyclotron Rd., Berkeley, CA, 94720, USA}}
\newcommand{\uhawaii}{\affiliation{Department of Physics and Astronomy, University of Hawai`i at M{\=a}noa, Honolulu, Hawai`i 96822, USA}}
\newcommand{\ucberkeley}{\affiliation{Department of Physics, University of California Berkeley, Berkeley, CA 94720, USA}}
\newcommand{\humboldt}{\affiliation{Institut f\"ur Physik, Newtonstr. 15, 12489 Berlin, Humboldt-Universit\"at zu Berlin, Germany}}
\newcommand{\umkc}{\affiliation{Department of Physics and Astronomy, University of Missouri-Kansas City, Kansas City, MO 64110, USA}}
\newcommand{\jpl}{\affiliation{Jet Propulsion Laboratory, California Institute of Technology, Pasadena, CA, 91109, USA}}
\newcommand{\usf}{\affiliation{Department of Physics and Astronomy, University of San Francisco, San Francisco, CA 94117-108, USA}}
\newcommand{\uf}{\affiliation{Department of Astronomy, University of Florida, Gainesville, FL 32611, USA}}
\newcommand{\lancaster}{\affiliation{Physics Department, Lancaster University, Lancaster LA1 4YB, United Kingdom}}
\newcommand{\finca}{\affiliation{Finnish Centre for Astronomy with ESO (FINCA), Quantum, Vesilinnantie 5, University of Turku, FI-20014 Turku, Finland}}
\newcommand{\turku}{\affiliation{Department of Physics and Astronomy, University of Turku, FI-20014 Turku, Finland}}
\newcommand{\ANUrsaa}{\affiliation{Research School of Astronomy and Astrophysics, The Australian National University, Canberra, ACT 2601, Australia}}
\newcommand{\ANUcga}{\affiliation{Centre for Gravitational Astrophysics, College of Science, The Australian National University, ACT 2601, Australia}}
\newcommand{\oskarklein}{\affiliation{The Oskar Klein Centre, Department of Physics, AlbaNova, Stockholm University, SE-106 91 Stockholm, Sweden}}
\newcommand{\pilarone}{\affiliation{Institute of Cosmos Sciences, University of Barcelona, E-08028 Barcelona, Spain}}
\newcommand{\pilartwo}{\affiliation{Instituto de Fisica Fundamental, CSIC, E-28006 Madrid, Spain}}
\newcommand{\floridastate}{\affiliation{Department of Physics, Florida State University, Tallahassee, FL 32306, USA}}
\newcommand{\ucdavis}{\affiliation{Department of Physics, University of California Davis, One Shields Avenue, Davis, CA 95616, USA}}
\newcommand{\ipmu}{\affiliation{Kavli Institute for the Physics and Mathematics of the Universe, University of Tokyo, Kashiwa, 277-8583, Japan}}
\newcommand{\mcgill}{\affiliation{Department of Physics, McGill University, 3600 rue University, Montr\'eal, Qu\'ebec, H3P 1T3, Canada}}
\newcommand{\uwastronomy}{\affiliation{Department of Astronomy, University of Washington, Seattle, WA 98195, USA}}
\newcommand{\romeobservatory}{\affiliation{INAF-Osservatorio Astronomico di Roma (OAR), Via Frascati 33, 00040, Monteporzio Catone, Italy}}
\newcommand{\ucriverside}{\affiliation{Department of Physics \& Astronomy, University of California Riverside, 900 University Avenue, Riverside, CA 92521, USA}}
\newcommand{\mpi}{\affiliation{Max-Planck-Institut f\"ur extraterrestrische Physik, Giessenbachstrasse, Garching D-85748, Germany}}
\newcommand{\kavlistanford}{\affiliation{Kavli Institute for Particle Astrophysics and Cosmology and Department of Physics, Stanford University, Stanford, CA 94305, USA}}
\newcommand{\slac}{\affiliation{SLAC National Accelerator Laboratory, Menlo Park California 94025}}
\newcommand{\uchicago}{\affiliation{Department of Astronomy \& Astrophysics, The University of Chicago, 5640 South Ellis Avenue, Chicago, IL 60637, USA}}
\newcommand{\kicp}{\affiliation{Kavli Institute for Cosmological Physics at the University of Chicago, 5640 South Ellis Avenue, Chicago, IL 60637, USA}}
\newcommand{\yorkuniversity}{\affiliation{Department of Physics and Astronomy, York University 4700 Keele St., Toronto, Ontario, Canada MJ3 1P3}}
\newcommand{\ruhr}{\affiliation{Ruhr University Bochum, Faculty of Physics and Astronomy, Astronomical Institute (AIRUB), German Centre for Cosmological Lensing, 44780 Bochum, Germany}}
\newcommand{\bonn}{\affiliation{Argelander-Institut f\"ur Astronomie, Auf dem H\"ugel 71, 53121, Bonn, Germany}}
\newcommand{\kwazulunatal}{\affiliation{Astrophysics \& Cosmology Research Unit, School of Mathematics, Statistics \& Computer Science, University of KwaZulu-Natal, Westville Campus, Durban 4041, South Africa}}
\newcommand{\uvictoria}{\affiliation{Department of Physics and Astronomy, University of Victoria, Victoria, BC, V8W 2Y2, Canada}}
\newcommand{\leiden}{\affiliation{Leiden Observatory, Leiden University, Niels Bohrweg 2, 2333 CA Leiden, Netherlands}}
\newcommand{\umich}{\affiliation{Department of Physics, University of Michigan, Ann Arbor, MI 48109, USA}}
\newcommand{\yonsei}{\affiliation{Yonsei University, Department of Astronomy, Seoul, Republic of Korea}}
\newcommand{\stanfordphysics}{\affiliation{Department of Physics, Stanford University, 382 Via Pueblo Mall, Stanford, CA 94305, USA}}
\newcommand{\livermore}{\affiliation{Lawrence Livermore National Laboratory, Livermore, CA 94551}}
\newcommand{\lpnhe}{\affiliation{Laboratoire de Physique Nucleaire et de Hautes Energies, Universite Pierre et Marie Curie, 75252 Paris Cedec 05, France}}
\newcommand{\cral}{\affiliation{Centre de Recherche Astrophysique de Lyon, Universit\'e Lyon 1, 9 Avenue Charles Andr\'e, F-69561 Saint Genis Laval Cedex, France}}
\newcommand{\ferrara}{\affiliation{Dipartimento di Fisica e Scienze della Terra, Universit{\`a} degli Studi di Ferrara, Via Saragat 1, I-44122 Ferrara, Italy}}
\newcommand{\arizona}{\affiliation{Department of Physics, University of Arizona, Tucson, AZ 85721, USA}}
\newcommand{\microsoft}{\affiliation{Microsoft, Lisbon - Portugal}}
\newcommand{\cunhaaffil}{\affiliation{Robert Bosch LLC. 384 Santa Trinita Ave, Sunnyvale, CA 94085, USA}}
\newcommand{\diracwash}{\affiliation{DIRAC Institute, Department of Astronomy, University of Washington, 3910 15th Ave NE, Seattle, WA 98195, USA}}

\title{The HST See Change Program: I. Survey Design, Pipeline, and Supernova Discoveries\footnote{Based on observations with the NASA/ESA \textit{Hubble Space Telescope}, obtained at the Space Telescope Science Institute, which is operated by AURA, Inc., under NASA contract NAS 5-26555, under programs 13677, 14327}}

\author[0000-0001-9200-8699]{Brian Hayden}
\stsci
\lbnl
\ucberkeley
\email{bhayden@stsci.edu}

\author[0000-0001-5402-4647]{David Rubin}
\uhawaii
\stsci
\lbnl

\author[0000-0002-5828-6211]{Kyle Boone}
\lbnl\ucberkeley\diracwash

\author{Greg Aldering}
\lbnl

\author{Jakob Nordin}
\humboldt

\author{Mark Brodwin}
\umkc

\author{Susana Deustua}
\stsci

\author{Sam Dixon}
\lbnl
\ucberkeley

\author{Parker Fagrelius}
\lbnl
\ucberkeley

\author{Andy Fruchter}
\stsci

\author{Peter Eisenhardt}
\jpl

\author{Anthony Gonzalez}
\uf

\author{Ravi Gupta}
\lbnl
\ucberkeley

\author{Isobel Hook}
\lancaster

\author{Chris Lidman}
\ANUcga
\ANUrsaa

\author{Kyle Luther}
\ucberkeley

\author{Adam Muzzin}
\yorkuniversity

\author{Zachary Raha}
\usf

\author{Pilar Ruiz-Lapuente}
\pilarone
\pilartwo

\author{Clare Saunders}
\lbnl
\ucberkeley

\author{Caroline Sofiatti}
\lbnl
\ucberkeley

\author{Adam Stanford}
\ucdavis

\author{Nao Suzuki}
\ipmu

\author{Tracy Webb}
\mcgill

\author{Steven C. Williams}
\lancaster
\finca
\turku

\author{Gillian Wilson}
\ucriverside

\author{Mike Yen}
\lbnl
\ucberkeley

\author{Rahman Amanullah}
\oskarklein

\author{Kyle Barbary}
\lbnl
\ucberkeley

\author{Hans B{\"o}hringer}
\mpi

\author{Greta Chappell}
\floridastate

\author{Carlos Cunha}
\cunhaaffil

\author{Miles Currie}
\stsci
\uwastronomy

\author{Rene Fassbender}
\romeobservatory

\author{Michael Gladders}
\uchicago
\kicp

\author{Ariel Goobar}
\oskarklein

\author{Hendrik Hildebrandt}
\ruhr
\bonn

\author{Henk Hoekstra}
\leiden

\author{Xiaosheng Huang}
\usf
\lbnl

\author{Dragan Huterer}
\umich

\author{M. James Jee}
\yonsei
\ucdavis

\author{Alex Kim}
\lbnl

\author{Marek Kowalski}
\humboldt

\author{Eric Linder}
\lbnl
\ucberkeley

\author{Joshua E. Meyers}
\livermore

\author{Reynald Pain}
\lpnhe

\author{Saul Perlmutter}
\lbnl
\ucberkeley

\author{Johan Richard}
\cral

\author{Piero Rosati}
\ferrara

\author{Eduardo Rozo}
\arizona

\author{Eli Rykoff}
\slac

\author{Joana Santos}
\microsoft

\author{Anthony Spadafora}
\lbnl

\author{Daniel Stern}
\jpl

\author{Risa Wechsler}
\stanfordphysics
\kavlistanford
\slac

\author{The Supernova Cosmology Project}

\begin{abstract}

The See Change survey was designed to make $z>1$ cosmological measurements by efficiently discovering high-redshift Type~Ia supernovae (\SNeIa) and improving cluster mass measurements through weak lensing. This survey observed twelve galaxy clusters with the \Hubble spanning the redshift range $z=1.13$ to $1.75$, discovering \ntransients likely transients and \nIa likely SNe~Ia at $z\sim 0.8-2.3$. As in similar previous surveys \citep{dawson09}, this proved to be a highly efficient use of \HST for SN observations; the See Change survey additionally tested the feasibility of maintaining, or further increasing, the efficiency at yet higher redshifts, where we have less detailed information on the expected cluster masses and star-formation rates. We find that the resulting number of SNe Ia per orbit is a factor of $\sim 8$ higher than for a field search, and \orbitPercent of our orbits contained an active SN Ia within 22 rest-frame days of peak, with one of the clusters by itself yielding 6 of the SNe Ia. We present the survey design, pipeline, and SN discoveries. Novel features include fully blinded SN searches, the first random forest candidate classifier for undersampled IR data (with a 50\% detection threshold within 0.05 magnitudes of human searchers), real-time forward-modeling photometry of candidates, and semi-automated photometric classifications and follow-up forecasts. We also describe the spectroscopic follow-up, instrumental in measuring host-galaxy redshifts. The cosmology analysis of our sample will be presented in a companion paper.
\end{abstract}


\section{Introduction}

Type~Ia supernovae (\SNeIa) serve as standardizable candles, enabling distance measurements over cosmological scales. Several dozen such SNe provided the first strong evidence that the expansion of the universe is accelerating \citep{riess98, perlmutter99}. SN samples are now reaching the thousands, with better control of systematic uncertainties, providing better constraints on the physical nature of dark energy \citep{suzuki12, scolnic18, riess18, descosmoIa, companionpaper}.

Due to the much fainter infrared (IR) sky background in space, the \Hubble (\HST) has superior sensitivity compared to ground-based facilities,  making \HST observations important for SNe at $z\gtrsim 1$ (where most of the SN light redshifts out of the observer-frame optical). However, due to the small fields of view (FoVs) of \HST instruments, untargeted \HST SN searches result in a large majority of observations taken when no SN is active. SN cosmology surveys leveraging the observing power of \HST have historically employed one of three strategies to maximize the science return of \HST orbits:
\begin{enumerate}
    \item Discover the SNe with wide-field, ground-based facilities and use \HST for critical measurements near peak brightness \citep{perlmutter98hst, garnavich98hst}
    \item Perform the SN survey as part of a wide, multi-visit \HST program, (e.g., \citealt{riess04})
    \item Search for SNe in galaxy clusters to greatly increase the SN rate per pointing \citep{dawson09, suzuki12}\footnote{A related strategy of searching in the fields around galaxy clusters has also been used on the ground during the era in which ground-based CCD camera FoVs were small \citep{norgaard89, perlmutter95}.}
\end{enumerate}

The See Change survey (GO 13677 and 14327, PI: Perlmutter) was designed to build further on the latter strategy, seeking to improve our constraints on time-variable dark energy by observing distant, massive galaxy clusters with \HST Wide-Field Camera 3 (WFC3) Near-infrared channel (IR) and UV/Visual channel (UVIS). These observations also improve weak-lensing measurements of cluster masses \citep{jee17, kim19, finner20} and increase the number of SNe~Ia discovered per unit orbit by a factor of several (Figure~\ref{fig:snRates}) for better cosmological constraints \citep{companionpaper}. Beyond the supernova science, the See Change survey also provides a unique resource for galaxy cluster science.  Mass determinations are typically the limiting systematic uncertainty for cosmological constraints derived from evolution in the cluster abundances. See Change provides a galaxy cluster sample with the requisite data for weak lensing mass determinations prior to the era of accelerated expansion. During this epoch clusters are undergoing significant assembly, and it is important to compare different mass estimation techniques to assess the consistency of different methods.  \citet{jee17} use this sample to pioneer the use of WFC3 IR for measuring weak lensing masses. In addition, the epoch $z\sim 1$--2 is a period of active galaxy assembly and evolution of stellar populations in cluster galaxies. The See Change data set has been used for a number of studies of the cluster galaxy population at this epoch, including evolution of brightest cluster galaxies and intracluster light \citep{webb15a,webb15b,webb17,ko18,demaio19}; the evolution of the more general galaxy population \citep{noble17, delahaye17,foltz18,noble19,sazonova19,strazzullo19}; and SN~Ia rates and lensing constraints \citep{friedmann18, rubin18}.

The See Change survey had the additional goal of testing the feasibility of performing a highly efficient space-based search at very high redshifts, beyond the previous \HST cluster search redshifts, where we have less detailed information on the expected cluster mass and star-formation rates.

In this paper, we describe the survey design and optimization, our pipeline, and the SNe we have discovered. Many of these elements that we introduced in this survey are likely to be useful in future space-based supernova searches, either similar cluster-based searches using the {\textit{James Webb Space Telescope}}, or the planned {\textit {Nancy Grace Roman Space Telescope}} ({\textit {Roman}}) SN survey \citep{SDT2015, hounsell18}. In a companion paper, we discuss our cosmological constraints and release the light curves \citep{companionpaper}. Section~\ref{sec:survey_design} presents the survey design and observing strategy. Section~\ref{sec:survey_execution} presents the data reduction pipeline, the search for SNe in each epoch, and the strategy for making follow-up decisions. Section~\ref{sec:ground} describes the ground-based follow-up program, and Section~\ref{sec:sn_sample} briefly describes the SN sample. Section~\ref{sec:conclusion} presents our conclusions.

\section{Survey Design}

\label{sec:survey_design}
In this section, we describe the design of the survey, including cluster selection, \HST filter selection, cadenced and follow-up \HST orbit strategies, and details of each individual \HST observation.

\subsection{Cluster Selection}

\begin{figure*}[h]
\begin{centering}
\includegraphics[width=0.95 \textwidth]{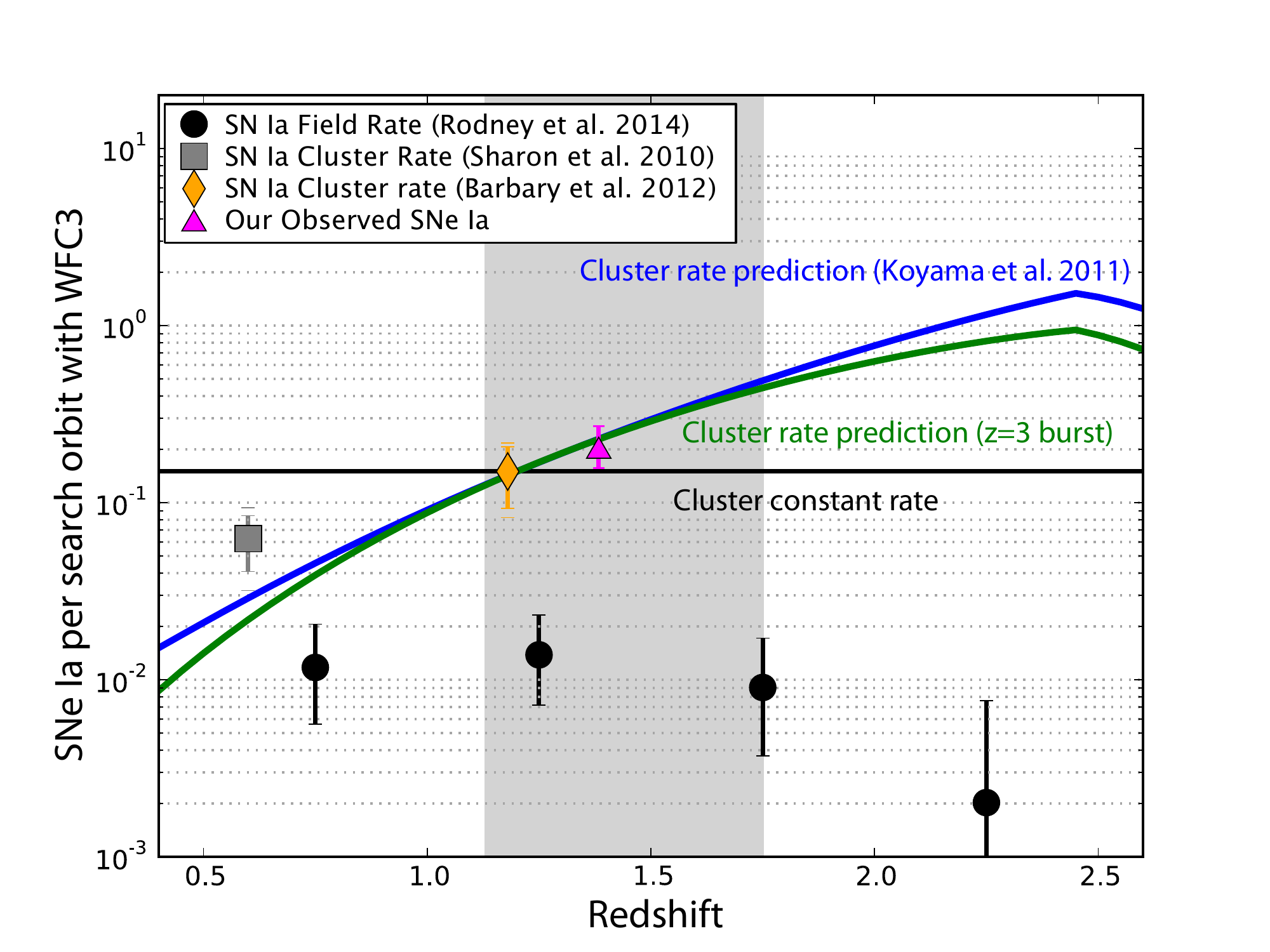}
\caption{Predicted and observed SNe per HST orbit for this program as a function of redshift. We expected (and found; magenta triangle) a much higher rate of SN in the clusters than in field galaxies. \orbitPercent of our orbits include a high-redshift SN~Ia within 22 rest-frame days of peak brightness, a highly efficient use of \HST for SN Ia characterization. }
\label{fig:snRates}
\end{centering}
\end{figure*}

\begin{figure*}[h]
\begin{centering}
\includegraphics[width=0.95 \textwidth]{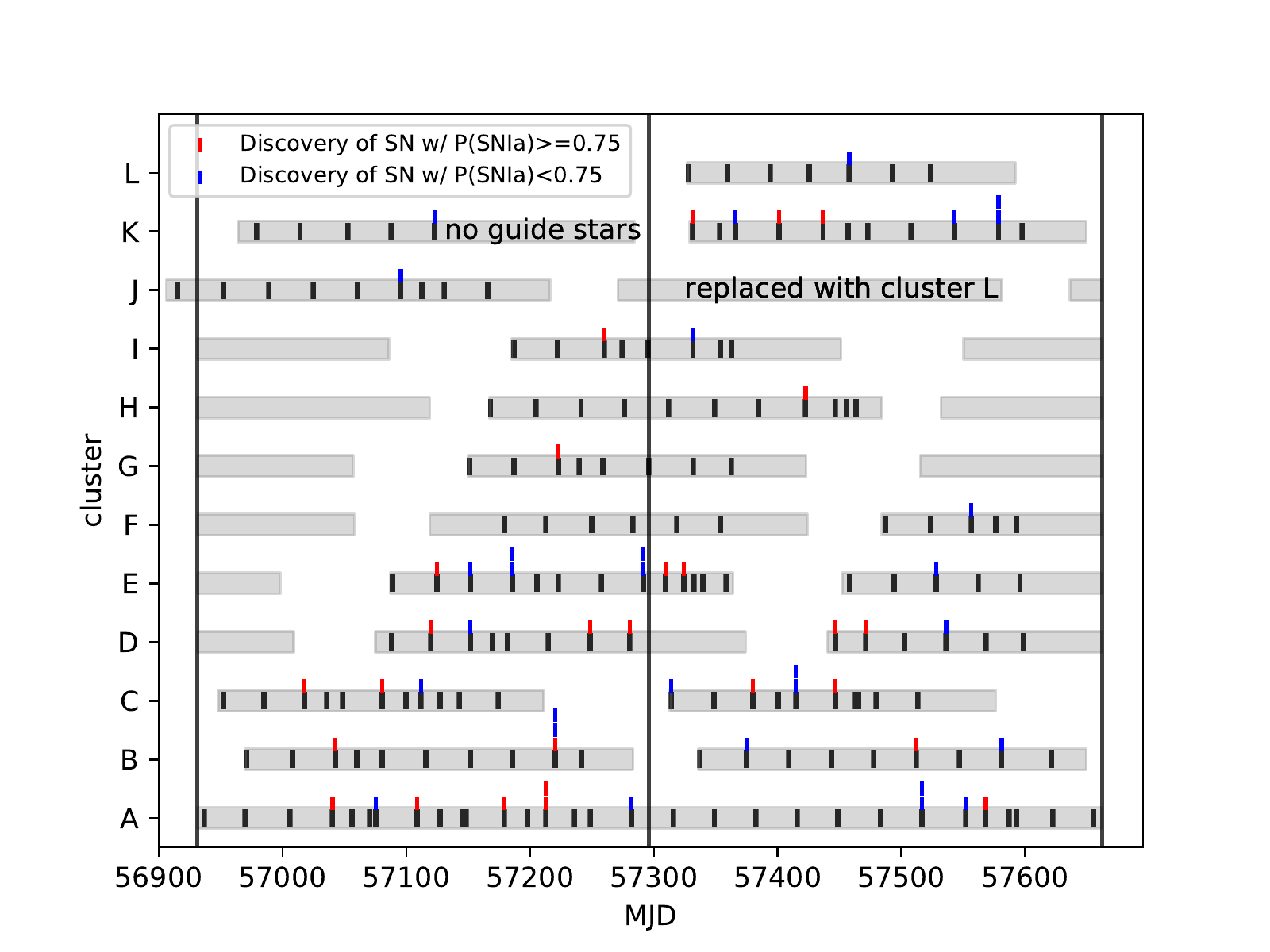}
\caption{\HST observations in the See Change program. Gray shaded bars are solar constraints as viewed from \HST. The rest-frame cadence is roughly 13-15 days. Vertical lines show \HST cycle boundaries. For cluster J, four orbits from program GO-13747 (PI: Webb) were cadenced for SN searching and are included here. A single, final reference orbit of cluster B was obtained in May 2017 and is not shown. Ticks spaced closely in time are the result of triggered followup layered on the cadenced search orbits. The discovery date of each transient discovered is shown by either a red ($\textrm{P(SNIa)}>=0.75$) or blue ($\textrm{P(SNIa)}<0.75$) tick mark. The typing probability is discussed in Section \ref{sec:sn_sample}.}
\label{fig:HSTObs}
\end{centering}
\end{figure*}

Cluster targets for See Change were selected to provide a SN~Ia sample at $z>1$ suitable for cosmology, as well as a sample of massive clusters for calibrating mass-observable relations (e.g., Sunyaev-Zel'dovich or X-ray). Potential targets were drawn from the South Pole Telescope survey (SPT, \citealt{bleem15}), the \textit{Spitzer} Adaptation of the Red-Sequence Cluster Survey (SpARCS, \citealt{muzzin09,wilson09,muzzin13}), the Massive and Distant Clusters of WISE Survey (MaDCoWs MOO clusters, \citealt{madcows}), the \textit{XMM} Cluster Survey (XCS, \citealt{xmmsurvey}), the \textit{Spitzer} Infrared Array Camera (IRAC) Distant Cluster Survey (IDCS, \citealt{stanford12}), and the IRAC Shallow Cluster Survey (ISCS, \citealt{iscs}), with the initial 
requirement to have a confirmed or very likely redshift above $1$ at the 
time See Change was first scheduled. As these cluster surveys employ a range of 
detection methods, we obtained a consistent ``richness" estimate for each 
potential cluster through the number of detected \textit{Spitzer} IRAC $3.6 
\mu\mathrm{m}$ sources within $2'$ of the reported cluster core. This aperture corresponds to $\sim 1$ Mpc between redshift 1 and 2, so this indicator should be consistent across the considered redshift range. This IRAC 
count is expected to correlate both with the number of galaxies (and therefore potential SN~Ia 
hosts), as well as the potential interest for cluster formation 
studies. We used a selection metric that scaled this number 
with the increased cosmological constraining power of more distant SNe and the 
expected increased SN rate closer to cluster formation \citep{sharon10,koyama11,barbary12}. The 
latter two effects both increase with redshift and compensate for 
the increased cluster richness at lower redshifts. Finally, clusters 
with poor \HST visibility were penalized, as the first visit of the second season to 
a target may contain old transients not suitable for follow-up. This 
evaluation led to the selection of $11$ clusters distributed in redshift 
($1.13<z<1.75$) and drawn from different detection surveys; an exceptionally massive MaDCoWs cluster was added halfway through the program after confirming the redshift with the Keck Multi-Object Spectrometer For Infrared Exploration (MOSFIRE) \citep{gonzalez15}.  During simulations of the program, we forecast $25 \pm 8$ SNe~Ia for the constant rate with redshift scenario, and $33 \pm 9$ SNe~Ia for an enhanced rate with redshift scenario (see Figure~\ref{fig:snRates}).
Table~\ref{tab:clusters} shows our selected clusters and the observed 
depth in each observed \HST bandpass, and Figure~\ref{fig:HSTObs} shows 
the \HST observing window and executed visits from our program for each 
cluster. We took cadenced observations every $\sim35$ days ($\sim$15 
rest-frame days at $z\sim1.3$), discussed further in 
Section~\ref{sec:filterselect}.

\begin{deluxetable}{lccllccccc}
\tablecaption{5$\sigma$ depths (AB mag) from \HST imaging in See Change \label{tab:clusters}}
\tablehead{\colhead{Cluster} & \thead{Internal \\ Designation} & \colhead{$z$} & \colhead{R.A.} & \colhead{Dec} & \colhead{$F814W$} & \colhead{$F105W$} & \colhead{$F125W$} & \colhead{$F140W$} & \colhead{$F160W$}}
\startdata
SPT0205\tablenotemark{a} & A & 1.32 & 02:05:46.27 & $-$58:29:06.7 & 27.9 & 28.2 & -- & 28.3 & 27.1 \\
ISCSJ-1432+3253\tablenotemark{b} & B & 1.40 & 14:32:18.31 & +32:53:07.8 & 27.4 & 28.0 & -- & 28.0 & 27.0 \\
MOO-1014\tablenotemark{c} & C & 1.23 & 10:14:07.6 & +00:38:22.4 & 27.6 & 28.0 & 27.1 & 27.9 & 26.6 \\
SPT-CLJ2106$-$5844\tablenotemark{d} & D & 1.13 & 21:06:04.94 & $-$58:44:42.4 & 27.4 & 27.9 & -- & 27.9 & -- \\
SPT2040\tablenotemark{e} & E & 1.48 & 20:40:59.59 & $-$44:51:36.7 & 27.3 & 27.9 & -- & 27.9 & 26.9 \\
SpARCS-J003550$-$431224\tablenotemark{f} & F & 1.34 & 00:35:49.7 & $-$43:12:24.2 & 27.4 & 27.8 & -- & 27.8 & -- \\ 
XMM44\tablenotemark{g} & G & 1.58 & 00:44:04.97 & $-$20:33:43.8 & 27.0 & 27.6 & -- & 27.6 & 26.8 \\
SpARCS-J033056$-$284300\tablenotemark{h} & H & 1.63 & 03:30:54 & $-$28:43:10 & 27.2 & 27.8 & -- & 27.8 & 26.9 \\ 
SpARCS-J022426$-$032331\tablenotemark{i} & I & 1.63 & 02:24:28.3 & $-$03:23:32.42 & 26.9 & 27.5 & -- & 27.6 & 27.0 \\ 
SpARCS-J104923+564033\tablenotemark{j} & J & 1.71 & 10:49:22.56 & +56:40:33.8 & 27.0 & 27.6 & -- & -- & 27.2 \\ 
IDCSJ1426\tablenotemark{k} & K & 1.75 & 14:26:32.87 & +35:08:24 & 27.4 & 27.9 & -- & 28.0 & 26.9 \\
MOO-J1142\tablenotemark{l} & L & 1.19 & 11:42:45.8 & +15:27:14 & 27.0 & 27.4 & -- & 27.4 & -- \\
\enddata
\tablecomments{Depths marked with -- indicate filters that were not used for that cluster. References: \textit{a}:\citealt{stalder13}, \textit{b}:\citealt{brodwin13}, \textit{c}:\citealt{brodwin15}, \textit{d}:\citealt{foley11}, \textit{e}:\citealt{bayliss14}, \textit{f}:\citealt{wilson09}, \textit{g}:\citealt{santos11}, \textit{h}:\citealt{nantais16}, \textit{i}:\citealt{muzzin13}, \textit{j}:\citealt{webb15a}, \textit{k}:\citealt{brodwin12}, \textit{l}:\citealt{gonzalez15}}

\end{deluxetable}

\subsection{Filter and Cadence Selection} \label{sec:filterselect}

\begin{figure*}
\begin{centering}
\includegraphics[width=0.95 \textwidth]{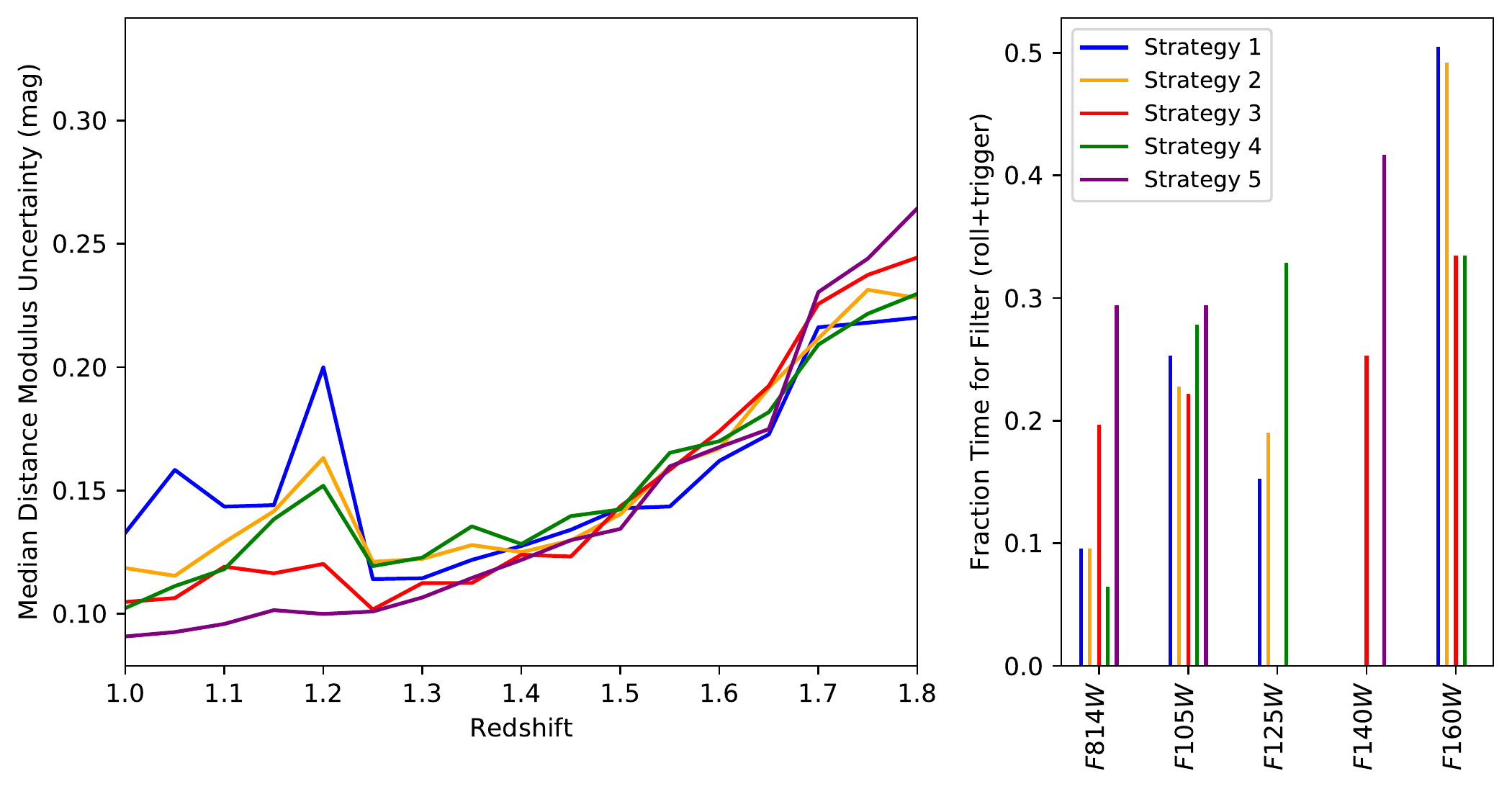}
\caption{The forecasted median distance-modulus uncertainty as a function of redshift for a small subset of surveys we simulated to optimize the strategy for See Change. For a fair comparison, all strategies shown here have a 35-day cadence and two orbits of triggered observations once a SN is discovered. The right panel shows the fraction of time in each WFC3 filter (combining one cadenced visit and the two-orbit trigger). The jump in distance modulus uncertainty at $z \sim 1.2$ occurs when the $F160W$ contains rest-frame wavelengths beyond the reddest described by the SALT2 model. At high redshift, $F160W$ becomes important (e.g., Strategy 1 showing smaller uncertainties than the others), despite the greater expense of $F160W$.\label{fig:surveyopt}}
\end{centering}
\end{figure*}

Before the survey began, we ran simulations of the survey for a wide range of filter choices, cadences, exposure times, and numbers of follow-up orbits. We optimized these to deliver small distance modulus uncertainties when the light curves were fit with SALT2 \citep{guy07,guy10,mosher14}. Figure~\ref{fig:surveyopt} shows examples of simulation outputs. With a sufficient cadence, our simulations showed $\sim 3/4$ of our orbits could be cadenced, with the rest reserved for triggered follow-up observations. Our optimum cadenced orbit included $\sim$900~s of WFC3-IR $F105W$, $\sim$1050~s of WFC3-IR $F140W$, and $\sim$400~s of WFC3-UVIS $F814W$. $F814W$ offered both information to reject low-redshift interlopers, as well as a second color for SNe to $z \sim 1.3$. $F105W$+$F140W$ provided a broad color baseline to enhance the distance measurement.\footnote{$F105W$+$F160W$ would provide an even wider baseline, but due to the  narrower width of the $F160W$ filter, the $F160W$ sensitivity is $\sim$ 0.5 mag (AB) less.} In our follow-up orbits, we also included WFC3-IR $F160W$ observations if they were predicted to significantly improve the SALT2 color estimate ($z \gtrsim 1.3$; see discussion of follow-up in Section~\ref{sec:followup_sim}).

\subsection{Exposure Details}
For each cadenced orbit, we executed seven exposures; three IR-$F105W$ exposures, three IR-$F140W$ exposures, and a single UVIS-$F814W$ exposure.\footnote{In the first orbits of the program, the exposures were split into four $F140W$ and two $F105W$ exposures. However, we found that with two dithers, several hundred unflagged bad pixels per visit in the $F105W$ dithers overlapped. This added significant workload to the human reviewers, and the 3$\times$3 pattern was adopted to improve bad pixel rejection. To improve the bad-pixel rejection when we were using two dithers, we downloaded several months of WFC3-IR imaging to derive an updated list of bad pixels. We flagged any pixels that were consistently discrepant with their surrounding pixels. An update to this work was presented in \citet{currie18}.} The IR exposures were dithered such that a pixel was subsampled 5 times across the two filters; to increase the signal-to-noise of faint transients, we combined subtractions across the two WFC3-IR filters, and this pattern improved sampling of the point spread function (PSF) in this combined image. The back-to-back exposures with a change in filter were observed with the same pointing to avoid a 25~s \HST pointing maneuver. For follow-up orbits of detected SNe, a similar dither pattern was used, but the filter selection was optimized based on simulations of the future light curve (using the detection epoch as a constraint; see Section~\ref{sec:followup_sim}). When the UVIS-$F814W$ was included in triggered \HST follow-up, a 2-point dither was used to reduce the possibility of losing an epoch to a cosmic-ray hit.

At the beginning of the program, the WFC3-IR ``blobs'' (small regions of $\sim 10\%$ lower sensitivity due to out-of-focus dust on the WFC3 channel select mechanism) were known \citep{pirzkal10}, but flat-field images were not yet available in the CALWF3 pipeline. To avoid the possibility of a SN observation landing fully in a blob, the dither patterns early in the program had large offsets to dither over the blobs. However, since the pixel scale is not uniform across the WFC3-IR detector, the large patterns caused regions of the image to be poorly sampled, leading to significant subtraction aliasing (we discuss our detection pipeline in Section~\ref{sec:SNsearch}). Once flat-field images for the blobs became available during program execution, the dither pattern was reduced to several pixels to improve sampling across the entire image.

The single, shallow WFC3-UVIS exposure in each cadenced visit used the UVIS-IR-FIX aperture, matching the geometric center of the IR detector. This would place the chip gap right in the center of each galaxy cluster, so each UVIS exposure had a Y-axis offset of $-15\arcsec$; combined with the variation in \HST roll-angle constraints over time, this avoided a ``hole'' forming in the stacked imaging at the middle of the cluster, and moved the chip gap off of the most likely SN host galaxies.

We used post-flash on the UVIS exposures to add electrons to the background, in order to avoid pathological charge-transfer-efficiency (CTE) losses \citep{anderson12}. The additional noise incurred ($< 10\%$) was well worth the trade to accurately measure SNe at these faint flux levels ($\sim 100$ e$^-$). The amount of post-flash varied based on the zodiacal background estimate of the Astronomer's Proposal Tool (APT) software, as we set the minimum post-flash for each visit that did not trigger an APT warning. Typically this was FLASH $\sim5-8$~e$^-$ . Despite the potential reduction in CTE losses that could have been achieved by placing the targets near an amplifier (by reducing the number of transfers), we generally did not choose to do this for UVIS follow-up imaging. There were several reasons that necessitated this decision:
\begin{enumerate}
  \item Roll-angle constraints were frequently tight, and sometimes the UVIS exposures would not have had sufficient overlap with previous imaging to allow alignment. The UVIS-$F814W$ exposures of these high-redshift clusters do not contain many bright objects.
  \item Similar to the above, if overlap with the cluster is lacking, new transients that might be discovered will have no UVIS coverage for those epochs. The UVIS fluxes or limits are important for our discovery-epoch estimate of the SN type (see Section~\ref{sec:single_epoch_typing}).
  \item The offset from the IR pointing to place a target near a UVIS amplifier frequently requires the acquisition of a new guide star, costing several minutes and carrying a (small) extra risk of acquisition failure.
\end{enumerate}

We took almost all IR data using the STEP50 readout pattern. This pattern performs four rapid reads at the beginning of the ramp, reducing read noise and providing good dynamic range for bright stars that would saturate with a longer read spacing. Using a large number of archival observations that used WFC3 IR, we verified that STEP50 offered the best performance in read noise.

We forced each visit to use a single guide star, rather than pairs, to save $\sim90$~s for science exposures. In pairs guiding, if the first star fails to acquire, guiding fails, so pairs guiding does not improve the likelihood of acquisition. We decided that the benefit to roll-axis stability in pairs guiding was less important than the extra exposure time. If \HST had rolled during single star guiding, we could align the individual WFC3 IR reads, though in practice, there does not appear to be significant roll drift in the sequence of images. 

\subsection{Elevated F105W Background Levels}
\label{sec:f105back}
\begin{figure*}[h!]
\begin{tabular}{cc}
\includegraphics[scale=0.56]{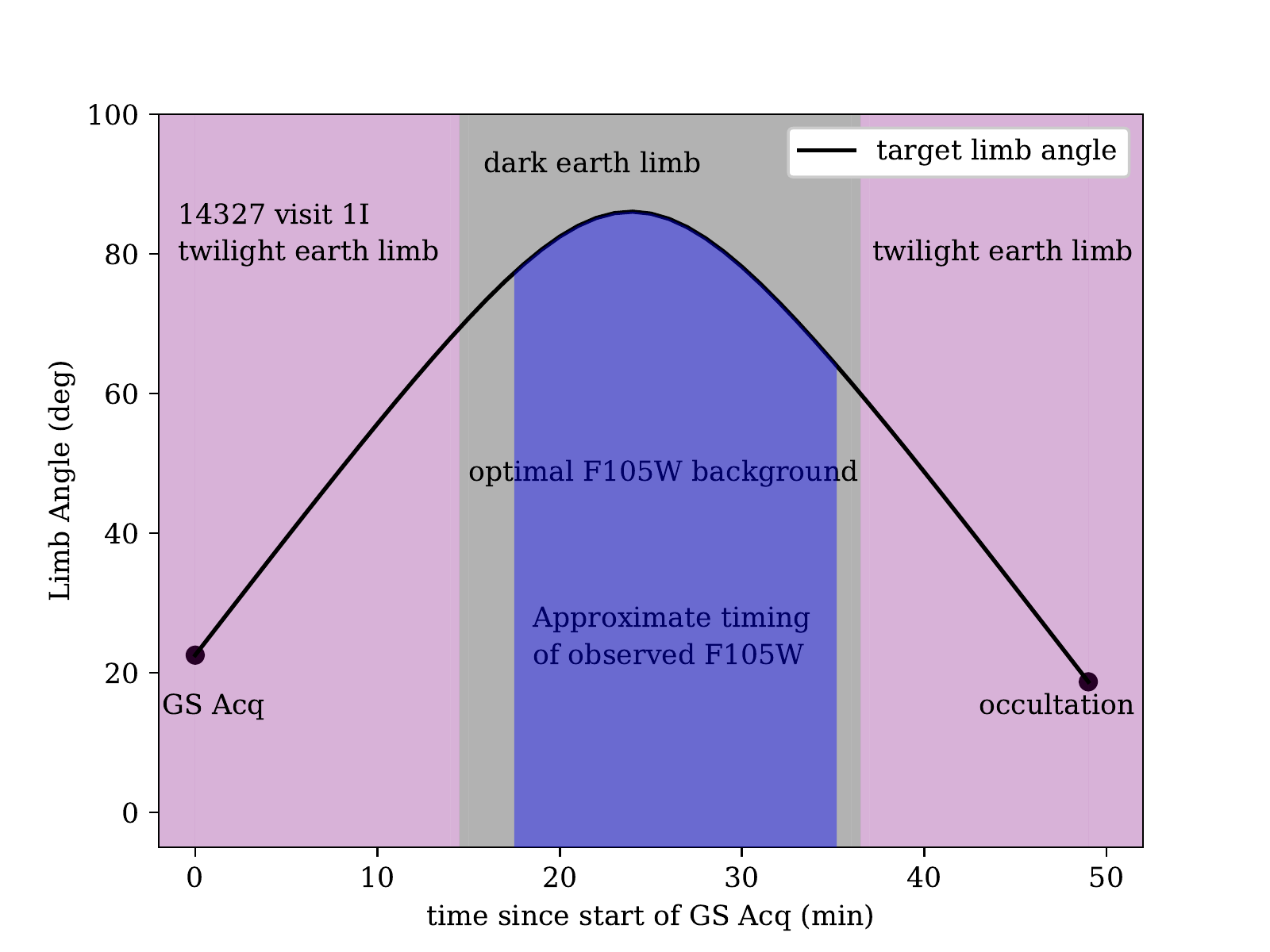} & 
\includegraphics[scale=0.56]{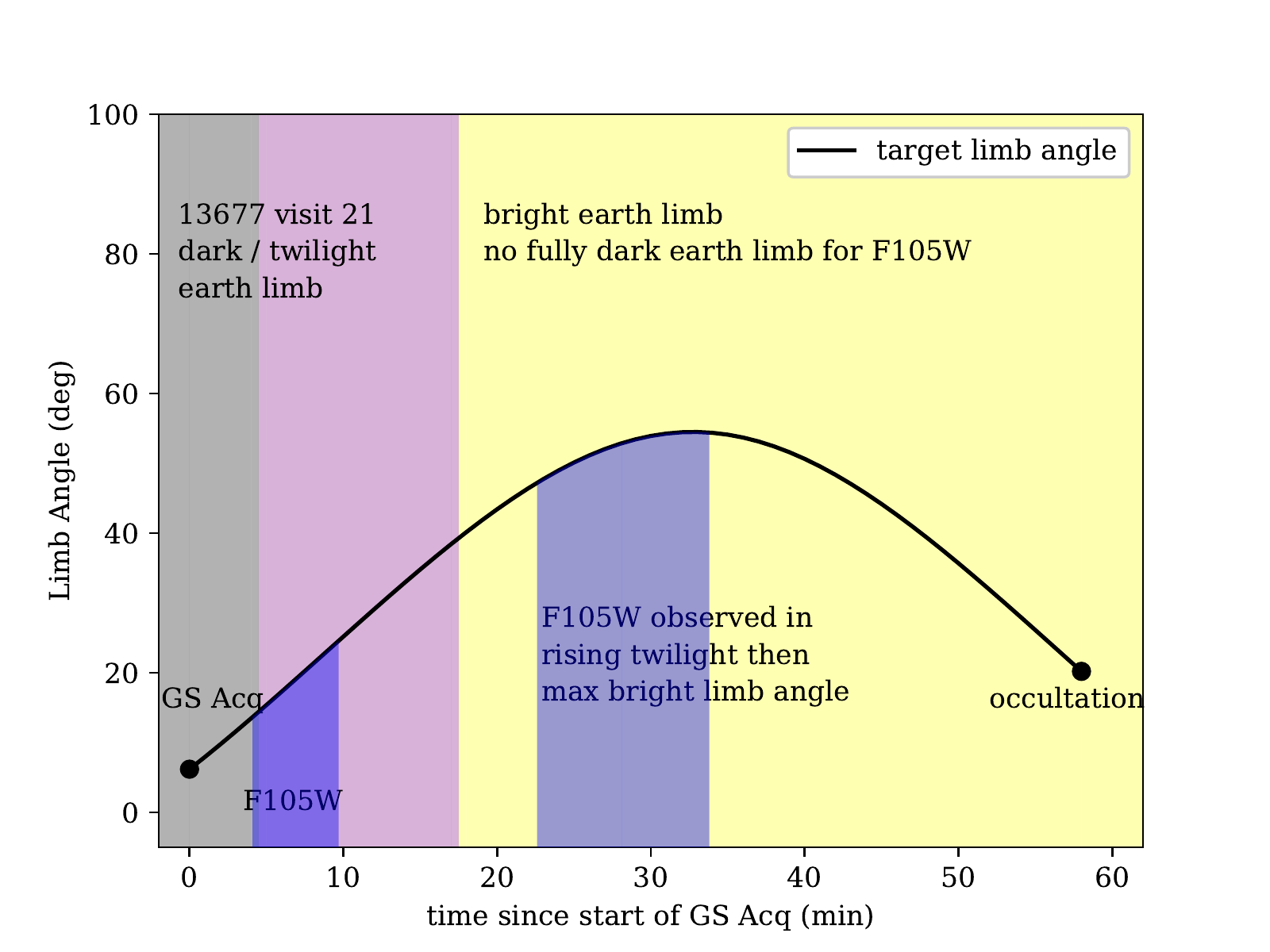}\\
\end{tabular}
\caption{Examples of placement of $F105W$ exposures (blue shaded regions) within $HST$ orbits to minimize pathological background levels due to He~{\sc ii} emission in bright Earth limb (yellow shaded). In both examples the timing of the exposures in the orbit were taken from the ``Visit Planner'' of the APT software used for \HST phase II submission, and the occultation reports were provided by our $HST$ Program Coordinator several weeks in advance of the expected observation. \textit{Left:} The observation occurs primarily in dark time (gray shaded), so we place the three F105W exposures in the middle of the orbit. \textit{Right:} The observation occurs primarily in bright time, however, it begins in dark time. In this case we place an $F105W$ exposure at the beginning of the orbit during rising twilight (purple shaded), and then place the other two exposures as near as possible to maximum Earth limb angle. Instructions and code for making exposure placement estimates for future visits can be found in the \code{SeeChangeTools v1.0.0} companion release \citep{seechangetools}.}
\label{fig:f105_background}
\end{figure*}

To minimize the impact of elevated $F105W$ background levels due to Helium~{\sc ii} emission \citep{brammer14}, we requested Earth occultation reports for each visit several weeks in advance from our program coordinator. We then placed the $F105W$ observations in the orbit with the following priority: 1) dark time, 2) maximum Earth limb angle, or 3) when the target is setting instead of rising over Earth limb (see Figure~\ref{fig:f105_background}). In order to ensure the timing of the $F105W$ exposures with respect to Earth limb, all visits were set up in the APT as non-interruptible sequences, forcing every visit to occur during one orbit of visibility rather than allowing the \HST scheduler to break up the exposures. Despite monitoring the occultation reports and strategically placing the $F105W$ exposures as described above, some exposures in the program had significantly elevated background levels. 

\section{Pipeline and Execution}

\label{sec:survey_execution}

\subsection{Data Reduction Pipeline and Supernova Search} \label{sec:SNsearch}
Our data reduction pipeline monitored a fast-track remote directory at STScI for new observations. The fast-track directory and automated monitoring removed at least an hour from the process of archiving the data and then retrieving it through standard methods (data were archived on average $\sim 5$ hours after observation). The time after data processing with our pipeline to final candidate review was also $\sim 5$ hours, so the typical timescale from data acquisition to identification of transients was $\sim10$ hours. 

Once data were ingested by our pipeline, the pipeline started the WFC3-UVIS charge transfer efficiency (CTE) correction.\footnote{The entire survey was executed before CTE-corrected UVIS exposures started to be routinely produced by CALWF3.} For each filter, dithered exposures were aligned inter-visit, and a combined image was produced by \code{AstroDrizzle}. We measured the sky background in all images using \code{SEP}\footnote{https://sep.readthedocs.io/en/v1.0.x/} \citep{sourceextractor,sep:paper, sep:zenodo}, and we used inverse variance map (IVM) weighting in \code{AstroDrizzle} with uncertainties derived from the measured sky background. This improved the combined image stack on the occasions where the $F105W$ background levels varied significantly from dither to dither. The combined image was then aligned to a reference image, and the World Coordinate System (WCS) offsets were propagated to the individual flat-fielded images (FLT's) with \code{TweakBack}. The final combined image was then generated with \code{AstroDrizzle} from these aligned FLT's.

To continually monitor our reduction pipeline and human searchers, the pipeline planted simulated SNe in each image. For maximum accuracy, the human searchers did not know which were simulated SNe and which were real, making this a blinded search. Section~\ref{sec:simSNe} provides more detail on the simulated-SN planting procedure. In short, we selected objects to plant simulated SNe on with \code{SEP}, and planted following the second-moment matrix so that the rate of simulated SNe mimics the light. The pipeline planted in a uniform magnitude distribution from 25 to 27 AB (as discussed further in Section~\ref{sec:simSNe}). We planted directly in the FLT images (before stacking) so that the whole pipeline was tested (minus image alignment).

To generate subtractions, we used \code{AstroDrizzle} to create a reference image from all previous exposures (including archival data from other programs where available) on the same pixel grid as the new image. We then directly subtracted these two images. We did not attempt to match the PSFs of the two images as the PSF of \HST is relatively stable between orbits. We did find artifacts when new and reference images were taken at different rotation angles since the PSF of WFC3 is asymmetrical. These artifacts limit our sensitivity near especially bright objects (the cores of galaxies or stars), but they were easily identified (both by eye and by the random forest classifier described in Section~\ref{sec:autoclass}) and unlikely to have been interpreted as candidates.

We used \code{SEP} to detect candidate transients in the generated subtractions with a convolution filter and kept any group of pixels that had a total increase in flux of 0.5 counts per second in at least one of the IR channels. We performed aperture photometry using \code{SEP} with a range of radii between $0.1\arcsec$ to $2.0\arcsec$ for each of the candidates on both the reference and new images in every filter. After candidates were identified and their basic properties measured, their random forest probability (of being a real transient) in each IR filter was estimated (Section~\ref{sec:autoclass}). The total processing time was 2-3 minutes per epoch per filter when running on 4 cores (not including the runtime of the CTE correction for UVIS images, which was approximately 60 minutes).

For candidates above a minimum random forest threshold (determined from review of human detection efficiency curves of simulated SNe), we immediately started forward-modeling PSF photometry \citep{nordin14} on a local computing cluster running the Sun Grid Engine. The PSF photometry took $\sim$ 2-3 hours for each candidate, and was started as early as possible to be ready for review as soon as human searching was completed. With 48 cores/96 threads available, all photometry jobs for most epochs were able to run simultaneously. 

Once candidates were identified, their properties estimated, and PSF photometry started, human searchers received an e-mail notifying them that a search was available. A \pkg{TKinter} GUI allowed each searcher to rate every candidate in a given epoch on a 0--3 scale, with 3 being the best transient candidates. The random forest ratings were used to automatically rate the lowest probability candidates with a 0, so they did not need to be reviewed by humans. This rejected dozens of spurious artifacts in each subtraction (cores of bright objects, detector artifacts, etc), saving a lot of human search time. In the beginning of the program, before the random forest was finished, each epoch was reviewed by three searchers. When the random forest was implemented, it was treated as a human searcher and the number of humans was reduced to two. 

After human candidate review, and after all PSF photometry was completed, every candidate went through a final review by an experienced searcher. A web interface was used to interact with our database and all data products, to present the reviewer with the PSF photometry, known photo-$z$ and spec-$z$ constraints, PSF photometry, human search ratings, and random forest probabilities. Each candidate was then assigned a final `yes' or `no' classification as a real transient. After this final review, the simulated SNe were unblinded and we determined if there were any real SNe in the image. The location of simulated SNe was checked in the pristine subtraction to ensure no simulated SNe were planted on real transients.

Once a transient was identified, we made \HST and ground-based trigger decisions. The fluxes in each filter were compared to a Monte Carlo simulation including SNe~Ia and core collapse SNe, separated into~Ib/c and type~II (see Section~\ref{sec:single_epoch_typing} for a detailed explanation of this procedure). When a likely SN~Ia was identified, we simulated its future light curve in various follow-up scenarios to finalize the trigger decision (see Section~\ref{sec:followup_sim}). Finally, once valuable targets were identified and a follow-up plan was determined, \HST triggers were activated, usually for one, two, or three orbits depending on planned future cadenced observations.

\subsection{Random Forest Candidate Classifier}
\label{sec:autoclass}
Large-area transient searches generate a volume of candidate transients that is daunting for humans to search, even with traditional rejection based on simple cuts. The vast majority of these candidates are subtraction artifacts, and machine learning has been employed as an initial rejection step before a cleaned list of candidates is presented to humans. \citet{bailey07} first developed machine learning techniques to identify transients. In the Nearby Supernova Factory SN search to which it was applied, this was able to decrease the number of SN candidates that were shown to humans by a factor of 10 compared to a cuts-based approach. \citet{bloom12} and \citet{brink13} built on this work and developed random forest classifiers for the Palomar Transient Factory. \citet{goldstein15} then implemented a random forest classifier for the Dark Energy Survey (DES) which decreased the number of candidates shown to humans by a factor of 13.4 compared to previous DES methods while only losing 1.0\% of artificially injected Type~Ia SNe. All of these efforts share certain characteristics: they all use ground-based data, all have good sampling ($\gtrsim 2$ samples per PSF full width at half maximum), and all use the observer-frame optical.

The main challenge present in the See Change dataset that has not been previously addressed with machine learning techniques is the undersampling of the PSF. Because of this undersampling, subtractions of \HST images typically have large residuals near the cores of galaxies that can mask the presence of neighboring bright transients \citep{rodney14}, potentially introducing biases into subsequent analyses. While these candidates are difficult to discern by eye, they do represent a large amount of light in the image and can be identified with machine learning techniques. We therefore implemented a random forest algorithm to search for SN candidates in the See Change dataset. The features used for this random forest were inspired by the features of \citet{bailey07}, \citet{bloom12}, and \citet{goldstein15}, although several new features were introduced. A description of these features can be found in Table~\ref{tab:randomforest}.

The random forest was trained on a set of previously taken \HST images (roughly the first half of the See Change HST orbits). 10,000 synthetic SNe were injected into these images, and the images were processed through the analysis software described previously resulting in the detection of 30,000 artifacts along with the synthetic SNe. The features described in Table~\ref{tab:randomforest} were measured on patches around each synthetic SN and artifact. We fit a random forest model to these features using \texttt{scikit-learn} \citep{pedregosa11}, using 10-fold cross-validation to evaluate the process and avoid over-tuning. The output of the random forest was mapped to produce 0--3 integer ratings for each candidate just as the human searchers were doing. The levels of these 0--3 labels were tuned to match the levels that a human would typically assign to each candidate. Using the average rating of the brighter fake SNe as a metric, we found that the random forest performed almost as well as the best human searchers on this dataset, and significantly better than the average human searcher. As a result, it was used to replace one of the three human searchers for each epoch, and all candidates ranked $0$ by the random forest were automatically rejected. This amounted to $\sim 30\%$ of candidates in each search epoch. By removing a human searcher and rejecting poor candidates, the random forest cut the amount of human effort in our search by a factor of $\sim 2$. 

\begin{figure*}[h]
\begin{centering}

\includegraphics[width=0.32 \textwidth]{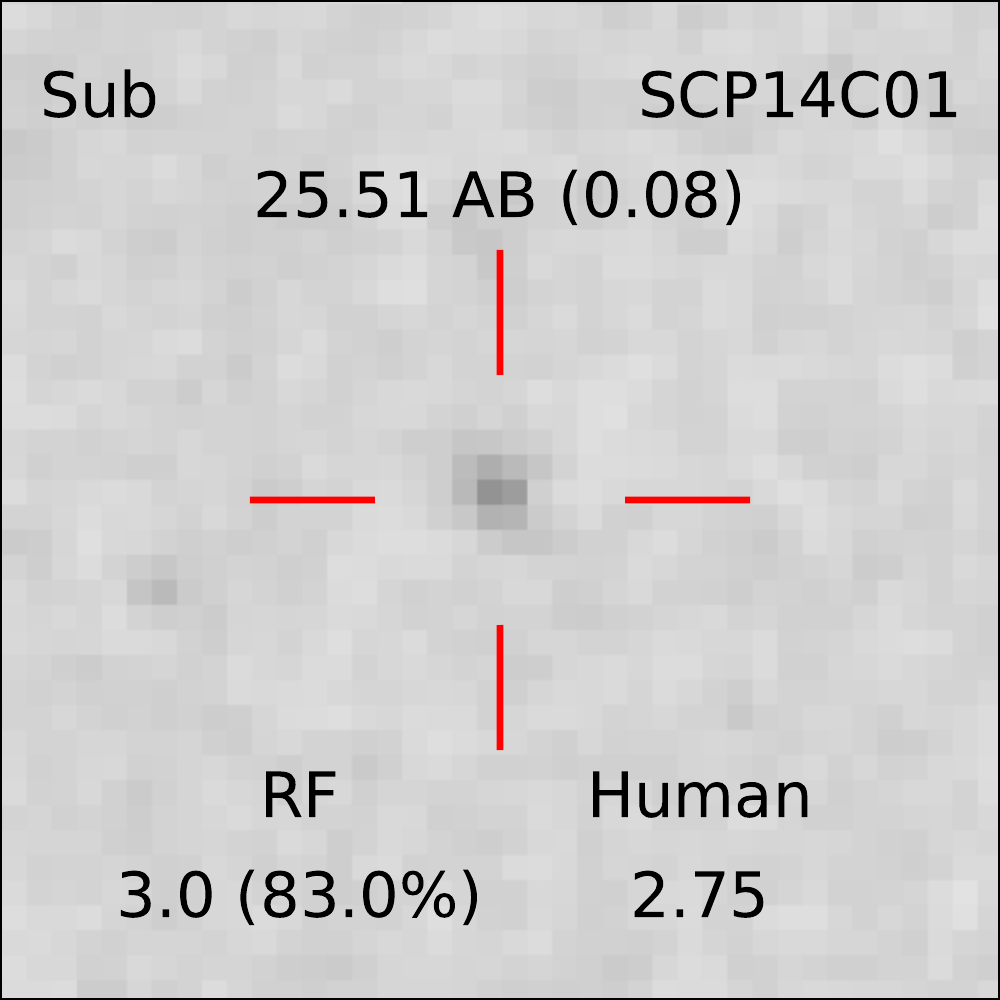}
\includegraphics[width=0.32 \textwidth]{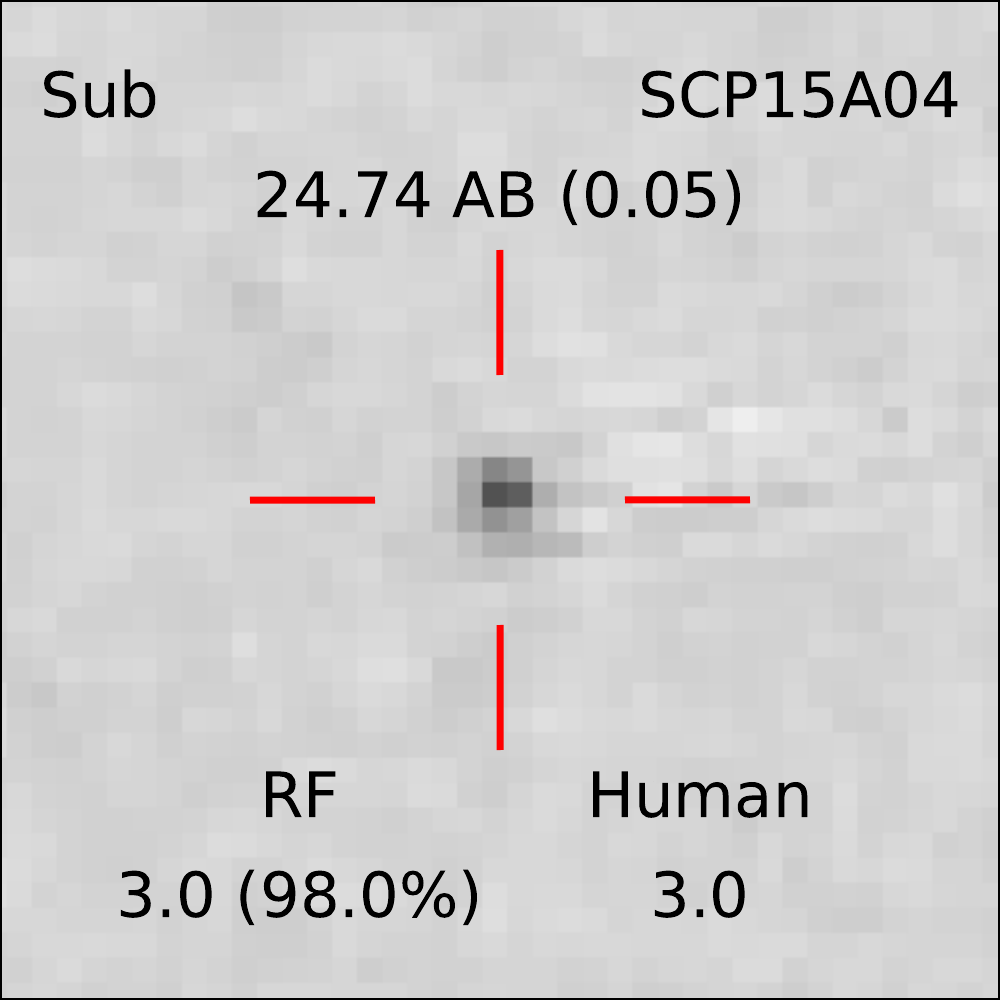}
\includegraphics[width=0.32 \textwidth]{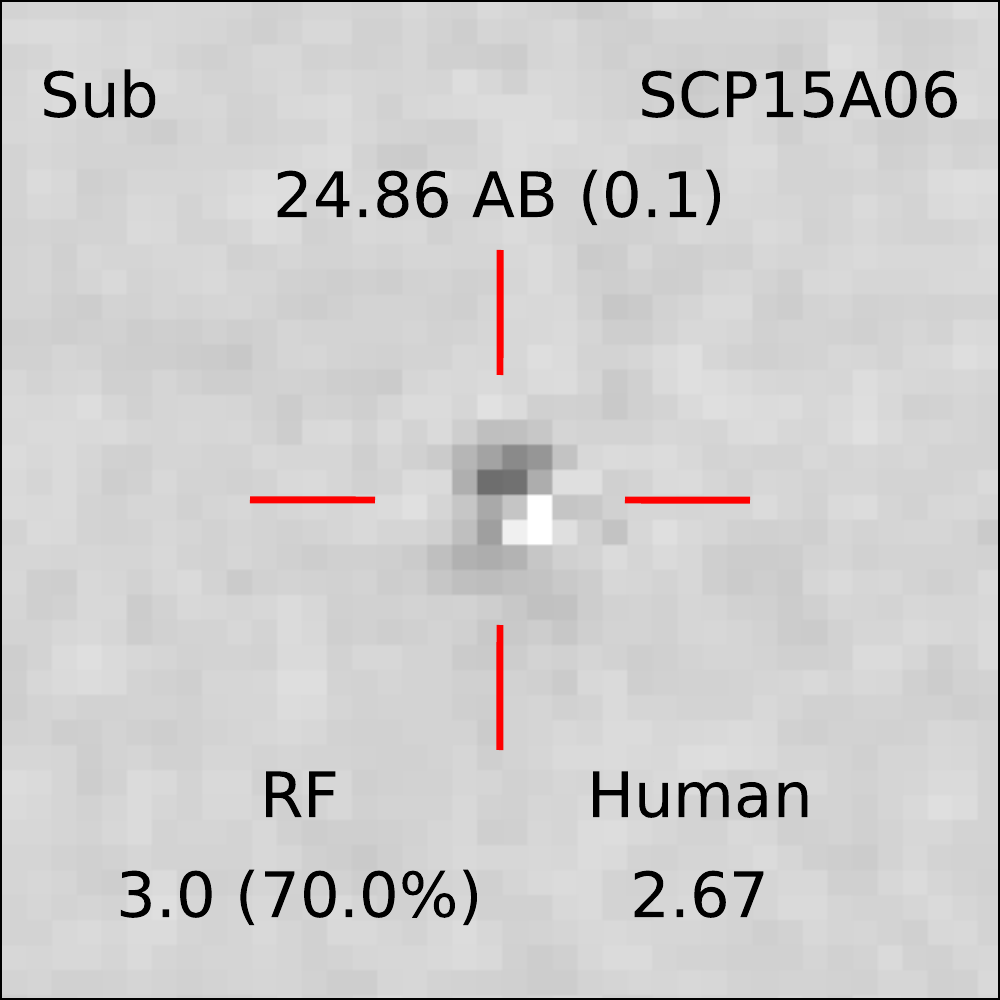}

\includegraphics[width=0.32 \textwidth]{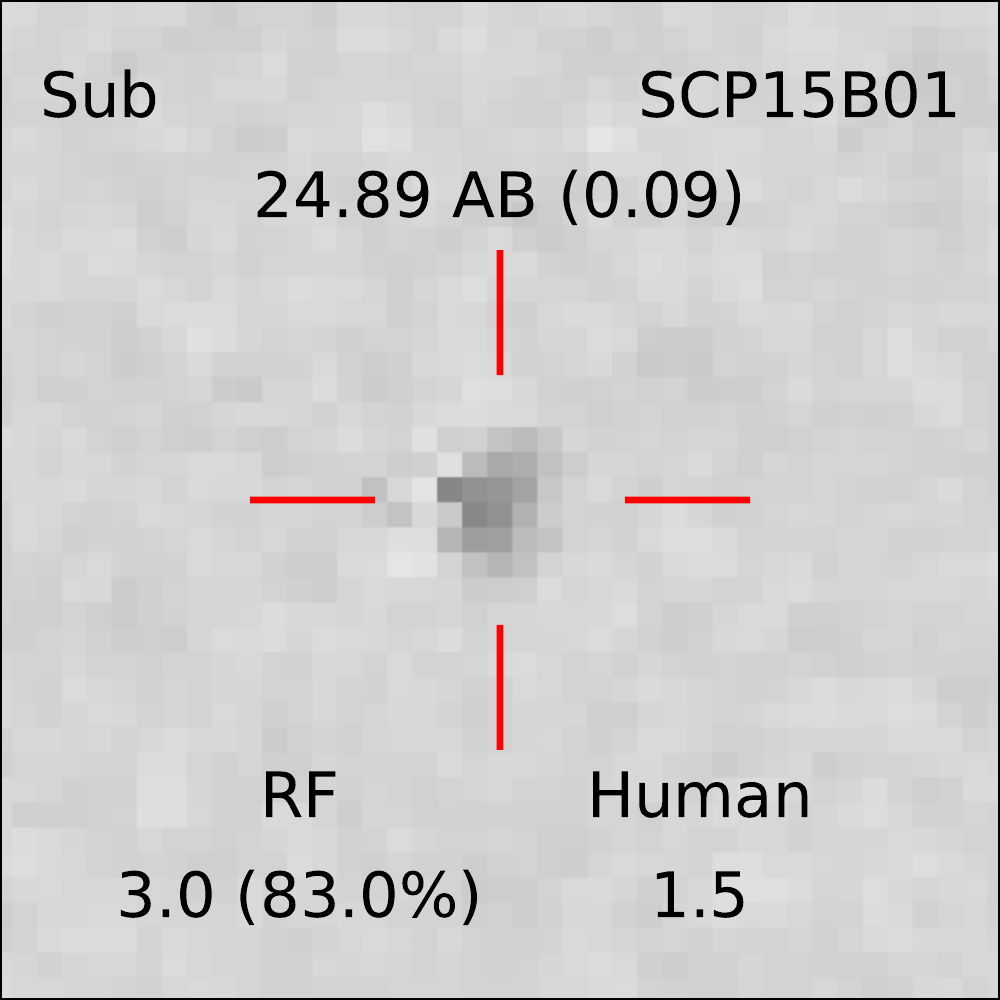}
\includegraphics[width=0.32 \textwidth]{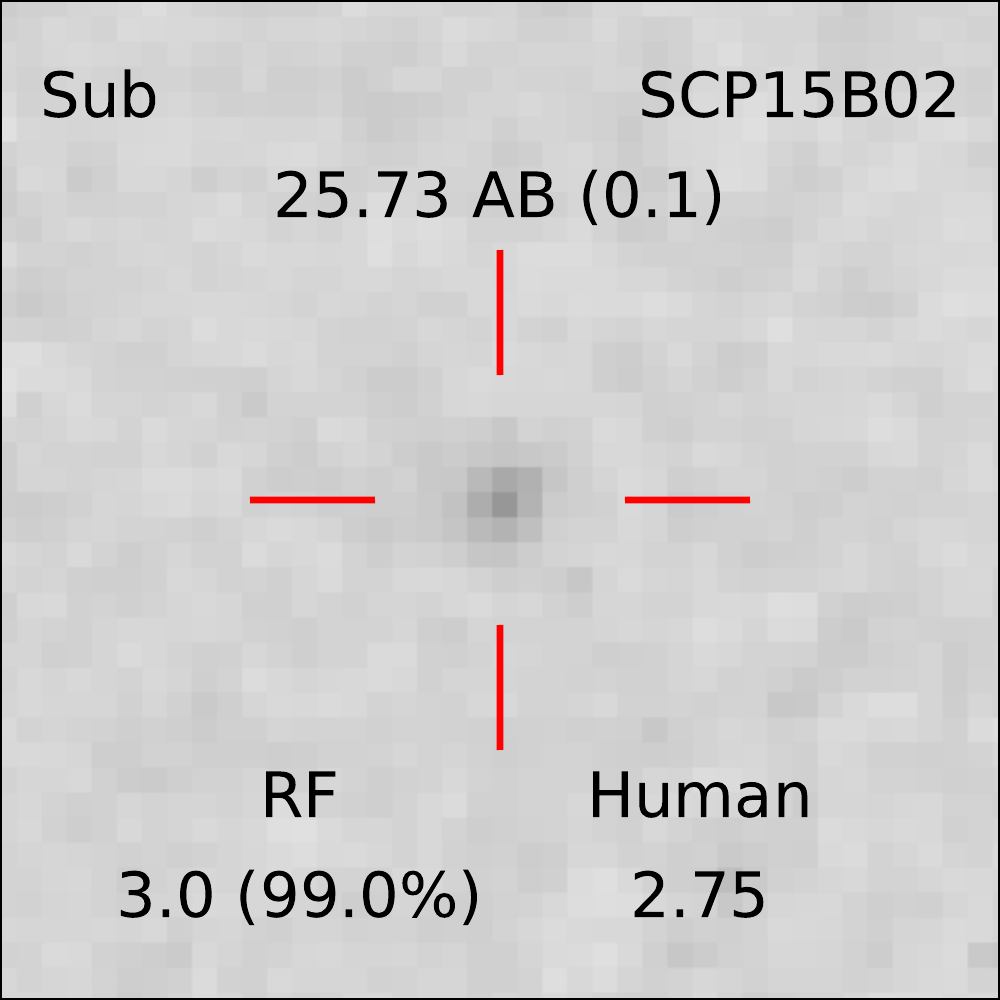}
\includegraphics[width=0.32 \textwidth]{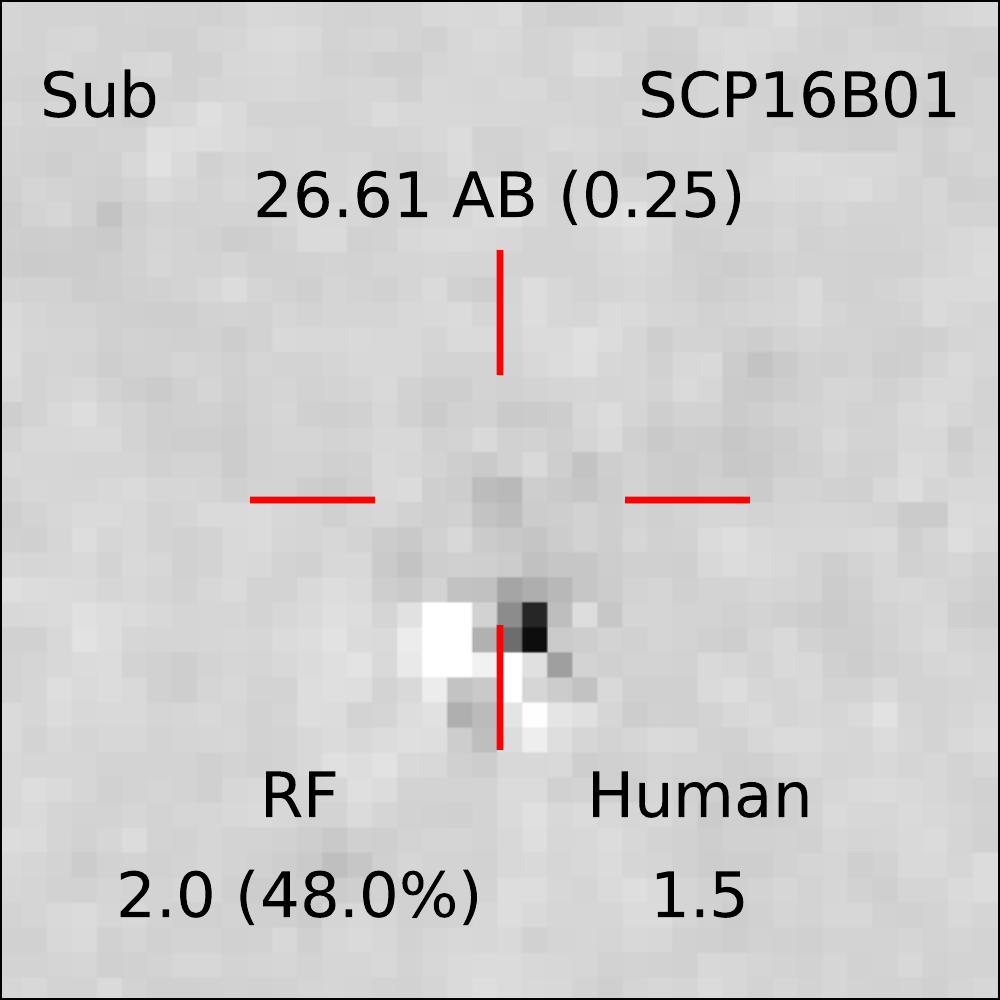}

\includegraphics[width=0.32 \textwidth]{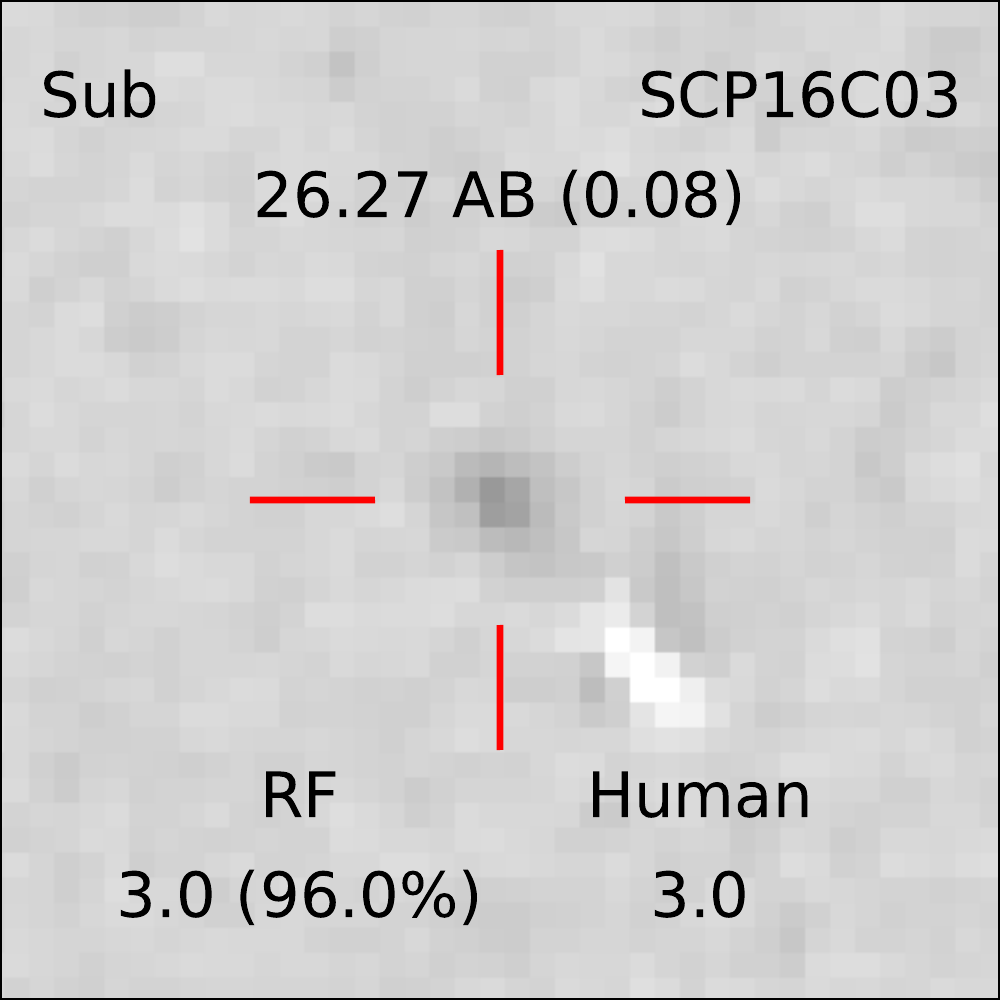}
\includegraphics[width=0.32 \textwidth]{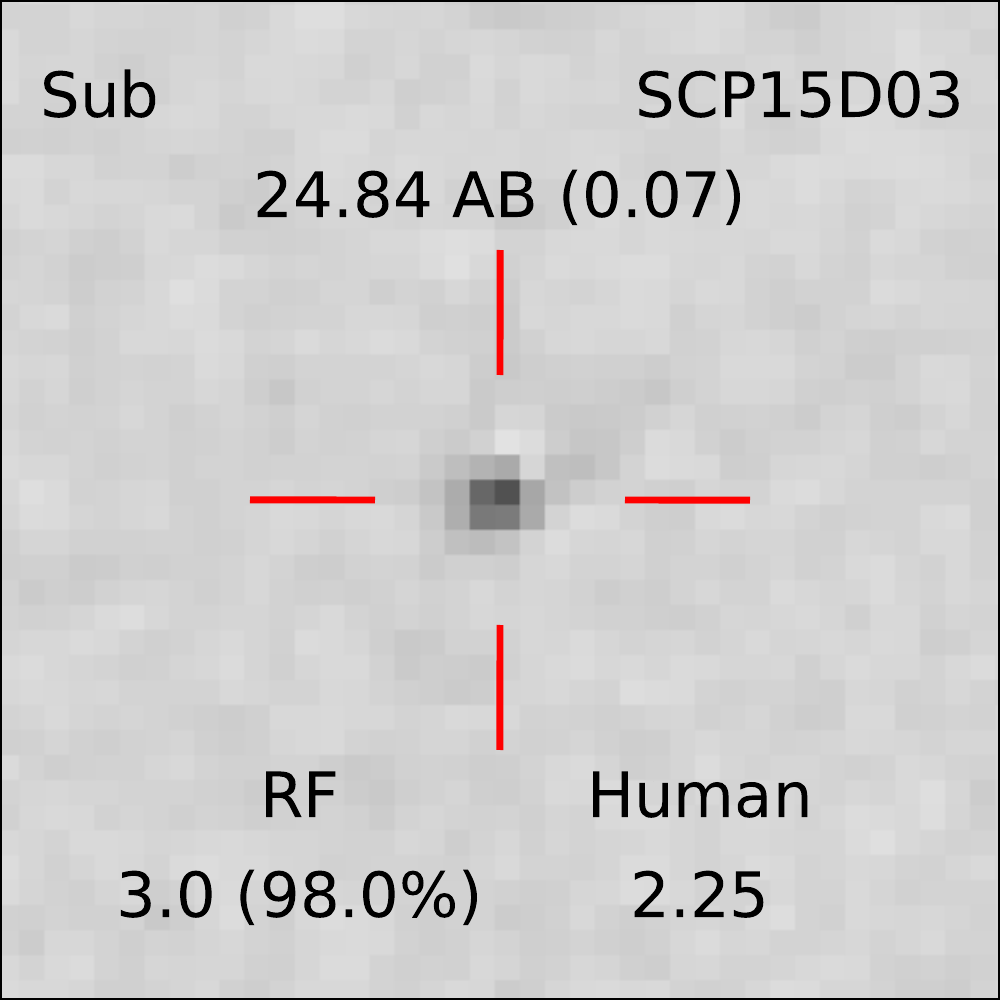}
\includegraphics[width=0.32 \textwidth]{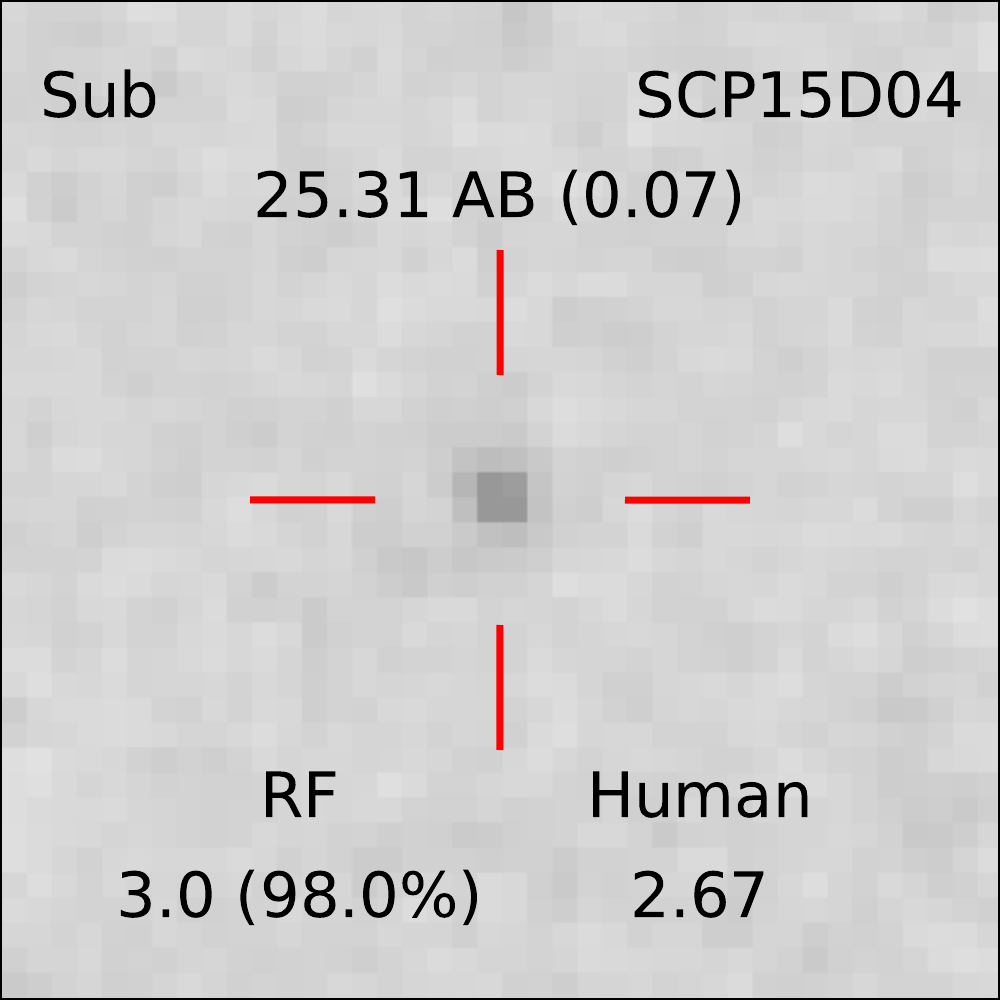}

\includegraphics[width=0.32 \textwidth]{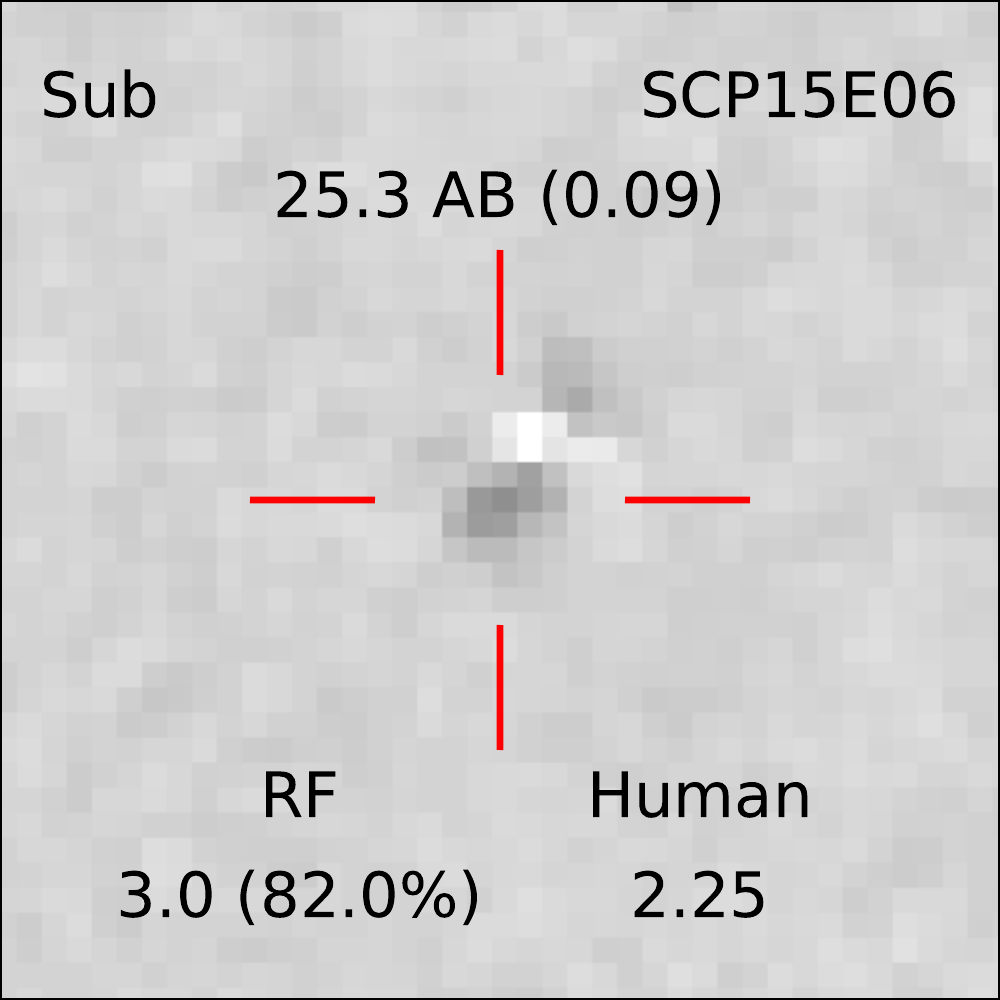}
\includegraphics[width=0.32 \textwidth]{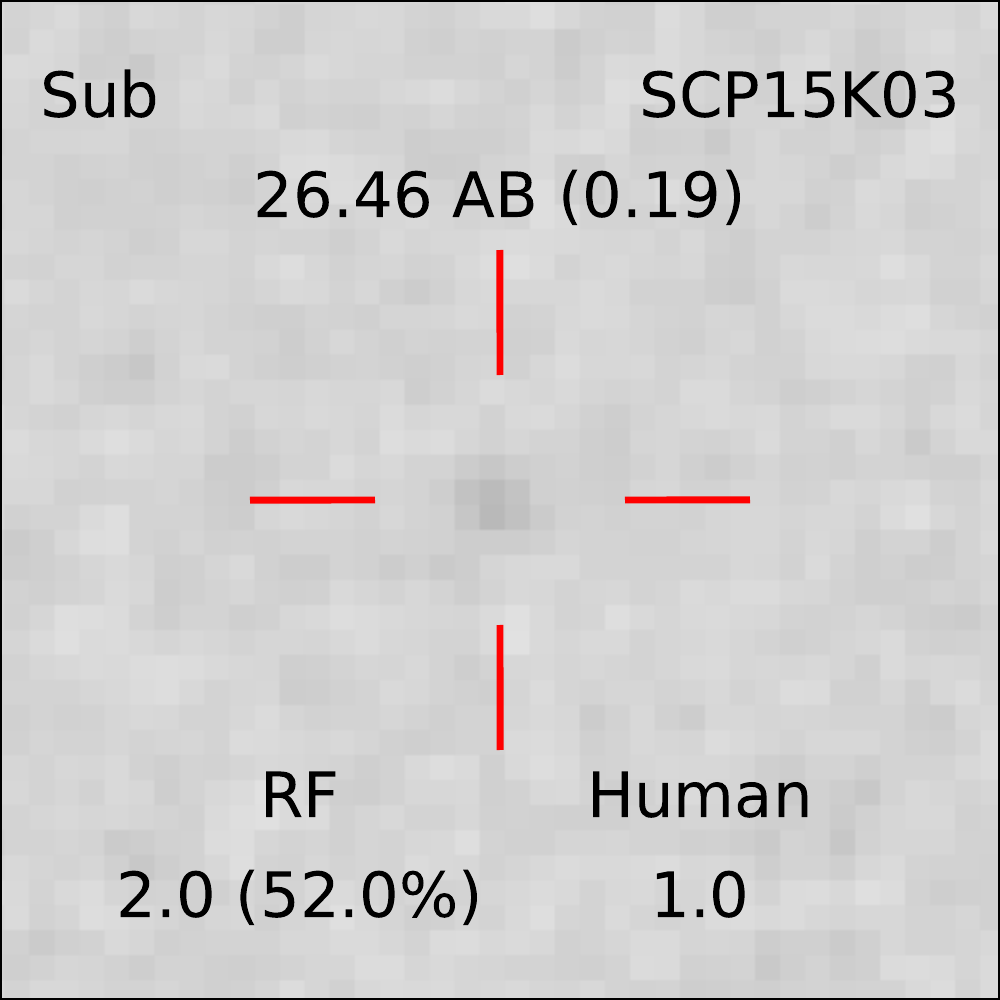}
\includegraphics[width=0.32 \textwidth]{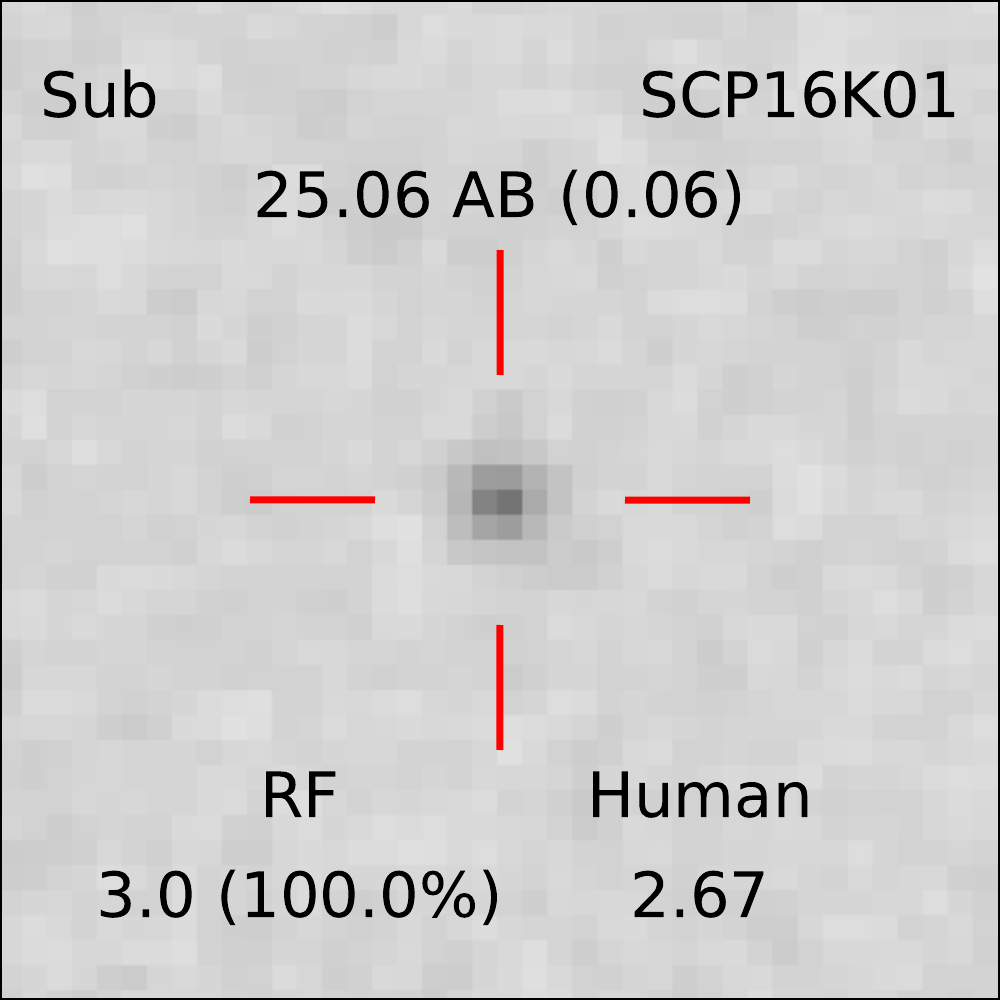}

\caption{\label{fig:rf_cutout_examples} Subtraction image cutouts for 12 randomly selected transients among the most likely SNe Ia. Each cutout is from the epoch of discovery, and the magnitude at discovery is shown above the red cross (centered on the candidate). The random forest probability and equivalent candidate rating is shown, as well as the average rating from the human searchers. This random sampling demonstrates a good range in human and random forest confidence.}

\end{centering}
\end{figure*}

Figure \ref{fig:rf_cutout_examples} shows cutouts of each subtraction that was searched for 12 randomly-selected SNe Ia at the epoch of discovery, with the AB magnitude, random forest rating and probability, and average human rating shown. This includes a good range of human and random forest confidence for a simple qualitative comparison.

\begin{figure*}[h!]
\includegraphics[scale=0.98]{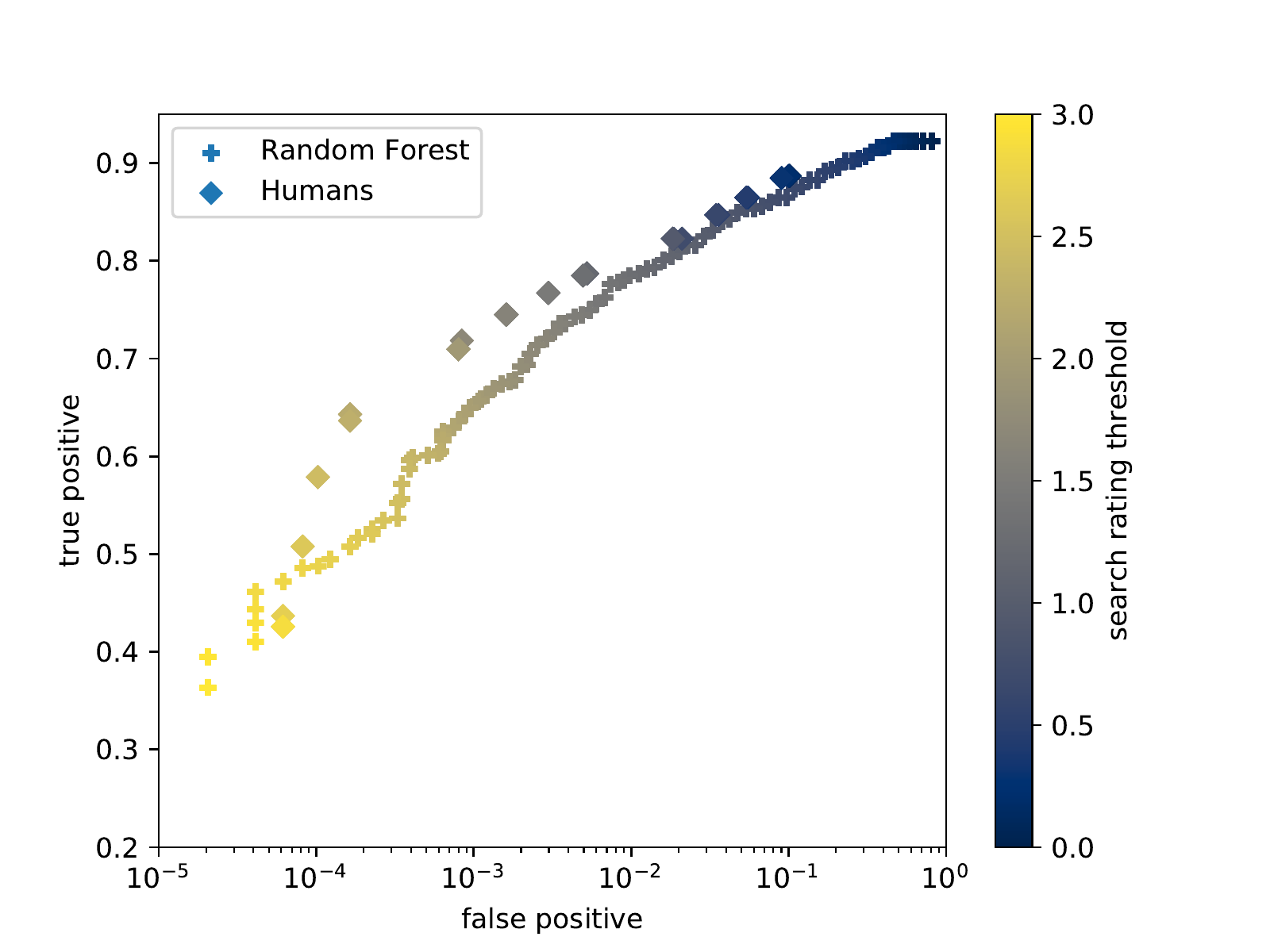}
\caption{ROC curves for the random forest classifier (plusses) and human searchers (diamonds), using the search rating on a 0-3 scale as the threshold (colorbar). The true positive rate is estimated using ratings of the blinded, fake SNe that were planted during each search epoch. The false positive rate is estimated from the real candidates in the survey (by excluding the 57 candidates marked as real). This means the false positive rate is an upper limit, as it is possible that some of the candidates that were rejected are actually real. The true positive rate peaks at 92\% because our candidate detection algorithm missed 8\% of fake SNe, which were not presented to the humans or the random forest for evaluation. The two search methods begin to diverge at a $\sim 1\%$ false positive rate. This is an encouraging sign for future space-based transient surveys, as a RF classifier using a similar model would allow for efficient, automated followup without the need for time-intensive human inspection.}
\label{fig:ROC}
\end{figure*}

Figure \ref{fig:ROC} demonstrates the comparability of the random forest and human searchers via receiver operating characteristic (ROC) curves. The true positive rate was estimated by using the ratings of each fake SN that was planted in real time (451 fake SNe in this set); the false positive rate is estimated from the actual candidates in the survey, minus the 57 that were marked as real transients (48,773 candidates in this set). The threshold for each data point is the average rating of all humans for the human curve, and $3\times$ the random forest probability for the RF curve. We chose not to force the random forest probabilities into integer values like we did in the search, simply for a more informative curve. The human and random forest curves are very similar down to a false positive rate of about $0.5\%-1\%$, corresponding to an average candidate rating of around 1.5 on our 0-3 scale. For future large surveys where human review of every candidate is impossible, this is an encouraging sign; a random forest classifier model such as this could be used to make automated followup decisions with a high true positive and low false negative rate, with confidence that the classifier performs almost identically to the average of several human searchers.

\begin{deluxetable*}{ccp{9cm}c}
 \tablecaption{Features (applied to small regions centered on the candidate) used by the random forest classifier to identify transients}
 \tablehead{  \colhead{Feature Name} & \colhead{Importance} & \colhead{Description} & \colhead{Source}}
 \startdata
\texttt{ref\_flux} & 0.14 & Aperture (r=1.5 pixels) flux in reference image & New \\
\texttt{avg\_raw\_flux} & 0.12 & Average flux in 5$\times$5 box in each raw image & New \\
\texttt{percent\_increase} & 0.11 & Percent increase in flux between new and reference image & New \\
\texttt{resi\_flux} & 0.10 & (squared residual of subtraction image after psf-fit) / psf-fit flux & New \\
\texttt{avg\_std} & 0.10 & Standard deviation of average flux in 5$\times$5 box in the raw images & New \\
\texttt{colmeds} & 0.07 & Max of median of pixel values of each column in 51$\times$51 pixel box around object in subtraction & G15 \\
\texttt{max\_var} & 0.07 & Maximum width of 2D Gaussian fit to subtraction image & B12 \\
\texttt{min\_var} & 0.05 & Minimum width of 2D Gaussian fit to subtraction image & B12 \\
\texttt{std\_radial\_meds} & 0.04 & Standard deviation of median of pixel values on radially oriented lines passing through center & G15 \\
\texttt{max\_patch\_size} & 0.04 & Number of 4-connected negative pixels in 9x9 box in subtraction & New \\
\texttt{diff\_sum} & 0.04 & Flux in 5$\times$5 box in subtraction image & G15 \\
\texttt{annulus\_flux\_2\_4} & 0.02 & Flux in annulus of inner radius 2, outer radius 4 in subtraction image & New \\
\texttt{r\_aper\_psf} & 0.02 & The sum of a 5-pixel circular aperture flux and a PSF-fit flux divided by the PSF-fit flux in subtraction image & G15 \\
\texttt{max\_raw\_flux\_diff} & 0.02 & Max flux difference in 5$\times$5 box amongst raw images & New \\
\texttt{std\_avg} & 0.01 & Average standard deviation of flux in raw images & New \\
\texttt{min\_dist} & 0.01 & Distance from candidate to nearest edge in raw image & B12 \\
\texttt{n2sig5} & 0.01 & Number of pixels less than $-2$T in 5$\times$5 box in subtraction image where T is the median absolute deviation of the pixels in the box & G15 \\
\texttt{num\_neg} & 0.01 & Number of negative pixels in 7$\times$7 box in subtraction image & G15 \\
\texttt{n2sig3} & $<$0.01 & Number of pixels less than $-2$T in 3$\times$3 box in subtraction image & G15 \\
\texttt{n3sig5} & $<$0.01 & Number of pixels less than $-3$T in 5$\times$5 box in subtraction image & G15 \\
\texttt{n3sig3} & $<$0.01 & Number of pixels less than $-3$T in 3$\times$3 box in subtraction image & G15 \\
 \enddata
\tablecomments{The importance column indicates the relative weight of each feature used in the classifier, as determined by the training of the classifier. The source column indicates the first reference that included each feature, with B12 corresponding to \citet{bloom12}, G15 corresponding to \citet{goldstein15}, and New corresponding to new features introduced for this analysis.}
\label{tab:randomforest}
\end{deluxetable*}

\subsection{Simulated-Supernova Planting and Detection Efficiency}
\label{sec:simSNe}
To evaluate our pipeline and search procedure in real-time, we planted blinded, simulated SNe in each search epoch, in each IR filter. In each epoch, we detect objects suitable for simulated SN host galaxies as follows,
\begin{enumerate}
	\item npix $>3$ and peak pixel $>5\times\textrm{RMS}_\textrm{sky}$,
    \item npix $>10$,
    \item peak pixel $<20\times\textrm{RMS}_\textrm{sky} $
\end{enumerate}
where npix is the number of connected pixels, and the final combined cut is performed as (1 and/or 2) and 3. These cuts reject objects that are very small unless they are somewhat bright, and rejects high surface-brightness objects which are not suitable as simulated host galaxies for SNe at $z>1$. 

We then generated an average of 10 SNe per search epoch, with AB magnitudes ranging from 25.0 to 27.0 in steps of 0.2 in $F105W$ and $F140W$; the average SN~Ia in this redshift range will have color$\sim0$ at peak brightness in the IR filters, which map fairly well to rest-frame $B$ and $V$ bands. In addition, for the first HST cycle, a random $15\%$ of simulated SNe were generated as faint, red SNe corresponding to $z\sim 2.2$ to train both human and machine searchers to watch for SNe near our detection limit in redshift. The location of each simulated SN is drawn from the ellipse moments returned by \code{SEP}.\footnote{For about the first half of the program, our galaxy cuts were less-restrictive, and the simulated-SN locations were drawn from a multi-variate normal centered on each host. After learning from this sample of simulated SNe, improvements were made to the planting procedure, and the detection efficiency reported here is for this second, more realistic set, representing about half of the SN survey. The first set of simulated SNe typically landed in regions of lower surface brightness, shifting the detection efficiency falloff to fainter magnitudes.} The SNe are planted in each individual FLT using the same PSF as in \citet{rubin18}, and then processed with the same pipeline that generated the stacked images for each epoch and IR filter. 

The primary human and random forest search was performed on these images with simulated SNe. As discussed in Section~\ref{sec:SNsearch}, the locations of the simulated SNe were blinded until after all candidates had gone through final human review to determine whether the object was a real transient valuable enough for an \HST trigger. After unblinding, the location of simulated SNe was checked in the pristine images, to verify that no simulated SNe were planted on real SNe. 

\begin{figure*}[h]
\begin{centering}
\includegraphics[width=0.95 \textwidth]{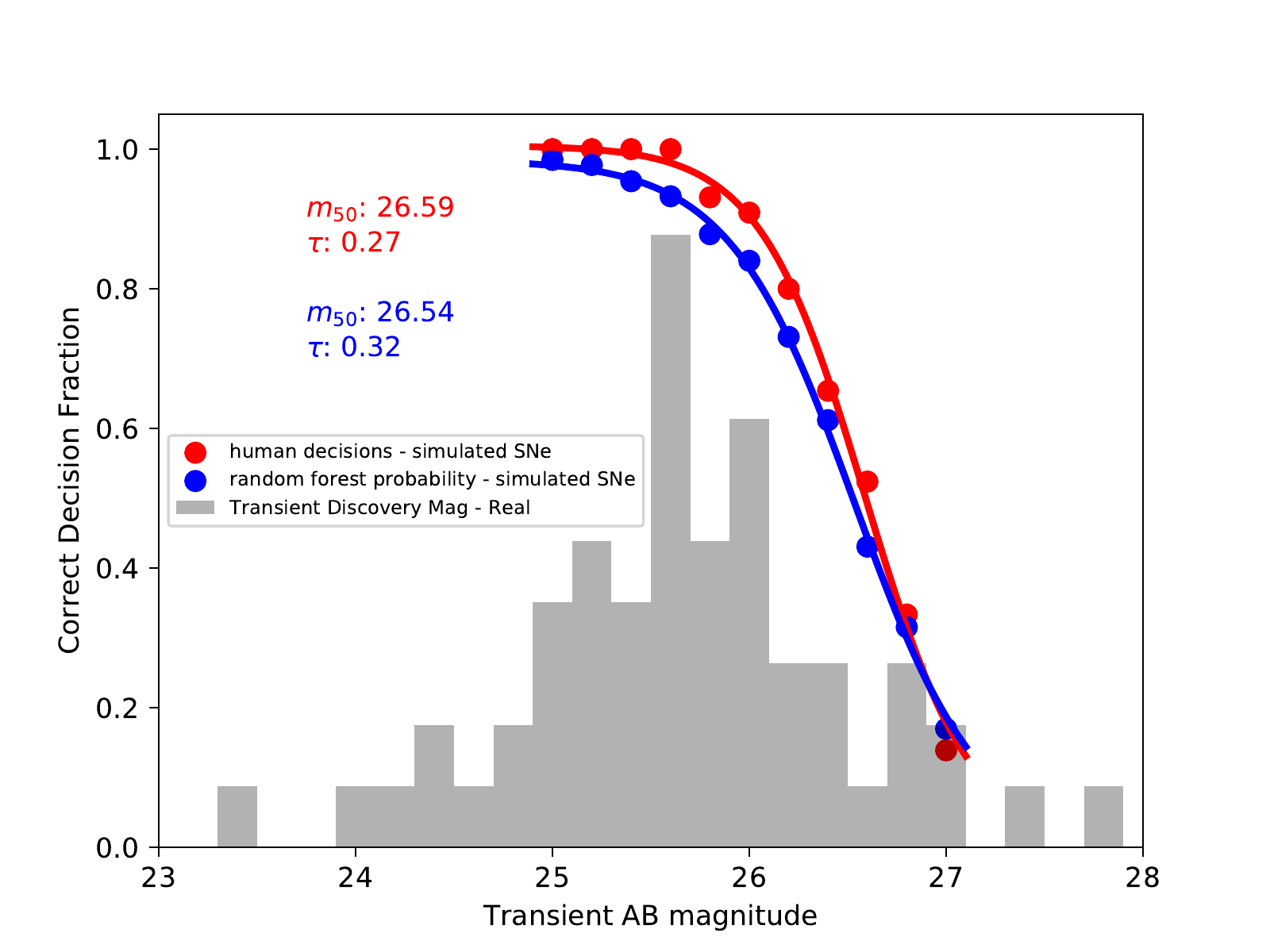}
\caption{Detection efficiency curves for simulated SNe, for both humans and the random forest classifier. The density distribution of transient magnitudes in $F105W$ at the epoch of discovery is shown in gray. During the execution of the SN survey we planted simulated SNe in real-time, and only unblinded after a final decision was made on the status of every candidate transient. For a single $HST$ visit this procedure typically cost $5-10$ person-hours, not counting the software development. Once we had some statistics on the operational detection efficiency of the random forest classifier, we removed a human searcher ($1-2$ person-hours), but in effect, we could have removed all human searching for a small cost in detection efficiency; this capability will be crucial for detecting SNe in larger future SN surveys with an under-sampled PSF, such as \textit{Roman}.}
\label{fig:deteff}
\end{centering}
\end{figure*}

Figure~\ref{fig:deteff} shows the detection efficiency curve from this search procedure, for human searchers, and for the average probability returned by the random forest classifier. We parameterize the detection efficiency with the same form as Equation 1 of \citet{rodney14}, i.e.,
\begin{equation}
\eta_{\mathrm{det}}\left(m\right)=\eta_0\times\left(1+\mathrm{exp}\left(\frac{m-m_{50}}{\tau}\right)\right)^{-1}
\end{equation}
The magnitude of 50\% detection efficiency ($\mathrm{m}_{50}$) for the full search and review procedure is 26.59 AB mag, while the magnitude of 50\% average random forest probability is 26.54 AB mag. 

\subsection{Single-Epoch Typing}
\label{sec:single_epoch_typing}
\begin{figure*}[h]
\begin{centering}
\includegraphics[width=0.95 \textwidth]{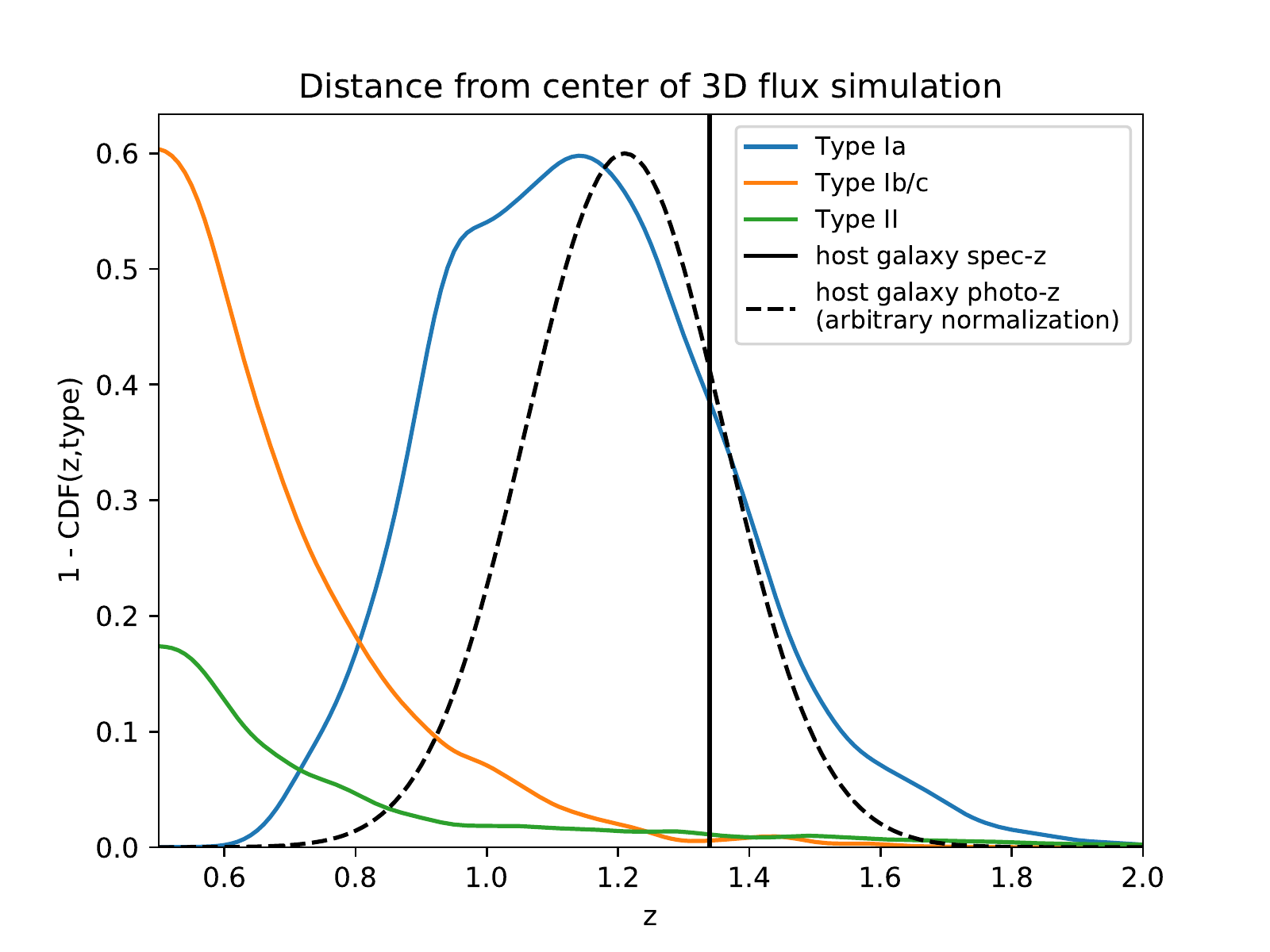}
\caption{Example of single-epoch typing constraints for SCP15A03. In this example, the photo-$z$ was known during the survey, but the spec-$z$ was determined after the \HST survey finished. Combining the photo-$z$ constraint with the color and flux constraints taken from the Monte Carlo strongly implies SCP15A03 is a SN~Ia. Including the final spec-z, the SN~Ia probability is $\sim98\%$; also including light curve shape constraints (not shown), the SN~Ia probability is well over $\sim99\%$. The Ia, Ibc, and II curves were lightly smoothed for plotting purposes.}
\label{fig:survivalfunction}
\end{centering}
\end{figure*}

Once a transient was identified, we had to make \HST and ground-based trigger decisions. The fluxes in each \HST filter were compared to a Monte Carlo simulation including SNe~Ia and core collapse SNe, separated into types Ib/c and type II. A WMAP9 flat $\Lambda$CDM cosmology \citep{wmap9} was assumed, with peak absolute magnitudes for each SN type drawn from a normal distribution \citep{richardson14}. For SNe~Ia, we also drew from $x_1$ and color, which introduces some redshift-color correlation. The core collapse SNe were reddened by drawing a host $E(B-V)$ from a $\mathcal{U}[-0.1,0.65]$ distribution, assuming $R_V$ of 3.1. To be conservative, the core collapse models were always simulated at peak brightness, regardless of possible phase constraints (as these SNe are typically fainter than SNe~Ia). The SNe~Ia were simulated for phase ranges of $\mathcal{U}[-10,-5],\ \mathcal{U}[-5,0],\ \mathcal{U}[0,5],\ \mathcal{U}[5,10]$; for a given object, a heavily constrained SALT2 fit to preliminary photometry provided a rough estimate of which SN~Ia phase simulations to include. A Monte Carlo simulation was used to produce 1000 SNe of each type at each redshift for redshift from 0.5 to 2.0 in steps of 0.05. We calculated the distance from each candidate's 3D flux measurement (one dimension for each of $F814W$, $F105W$, and $F140W$) to the center of the simulated distributions, thereby constructing the survival function, $\mathrm{SF}(z)=1-\mathrm{CDF}(z)$, for each SN type. This distribution was calculated by binning the Monte Carlo fluxes in 3D, determining which bin the real SN 3D flux would fall into, and calculating the percentile rank of that bin within the distribution of all bins; $\mathrm{SF}(z)=1$ is then the modal bin of the distribution. We did not normalize this distribution because a value of $1$ usefully indicates the SN is directly in the modal bin of the distribution. Measurement error was accounted for by adding noise to the Monte Carlo fluxes, using $\mathcal{N}[0,\sigma_\mathrm{SN,band}]$ to effectively spread out the Monte Carlo distribution. This SF($z$) was then combined with any known redshift constraints to provide a constraint on the SN type (an example of this procedure is shown in Figure~\ref{fig:survivalfunction}). 

Generally, any candidate that was a possible SN~Ia in the redshift range of the survey was triggered on. Occasionally, for the faintest candidates, we would wait until another cadenced epoch had been taken before triggering. These candidates are either very early detections of SNe~Ia, possible SNe~Ia well past the targeted redshifts, or faint CC SNe. In all of these cases, there was little harm in waiting for the next visit.

\subsection{Finalizing follow-up Decisions}
\label{sec:followup_sim}
\begin{figure*}[h]
\begin{centering}
\includegraphics[width=0.95 \textwidth]{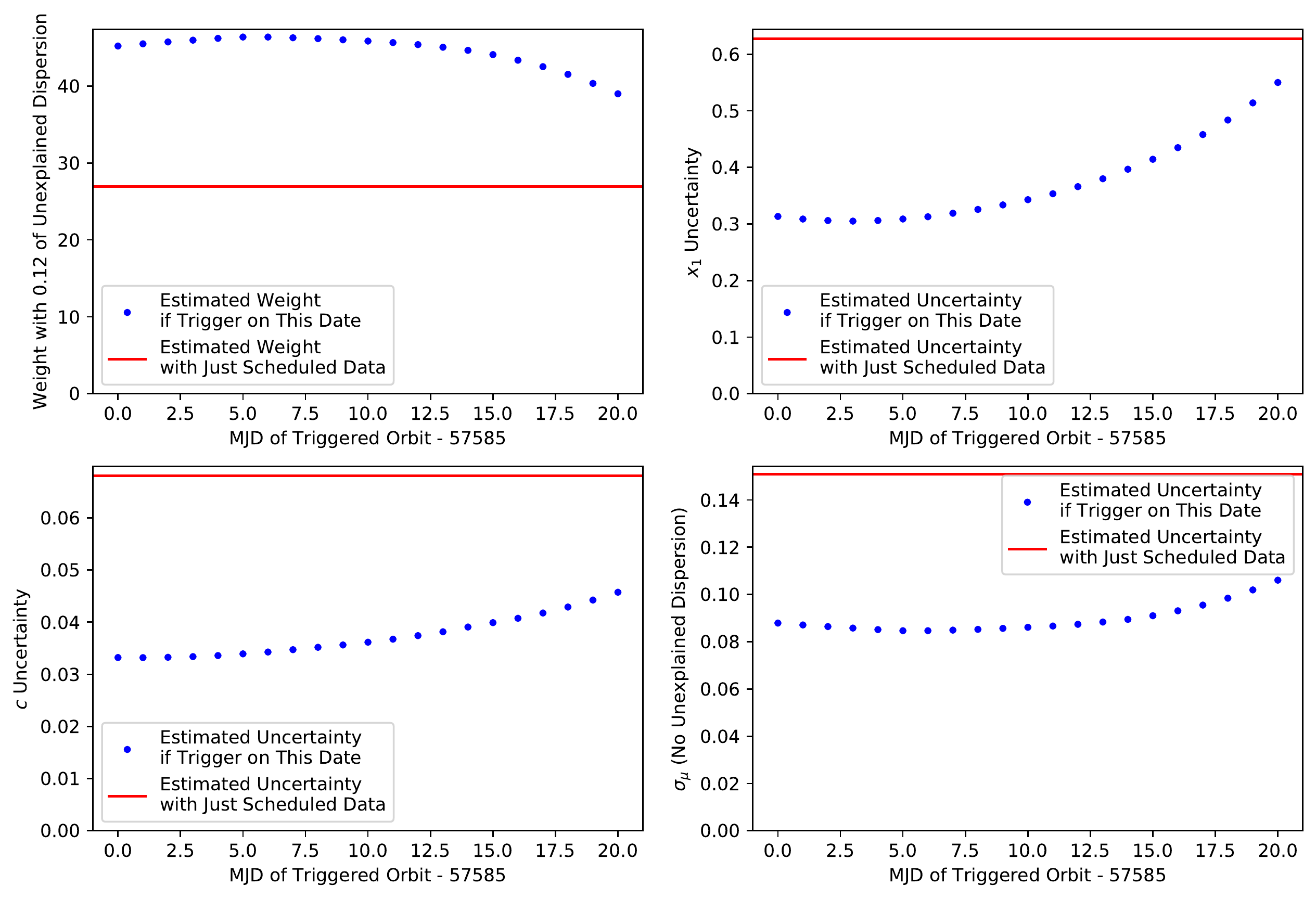}
\caption{Example of SN forecasting for \HST follow-up decisions. The SN, SCP16A04, was discovered on MJD 57568. We show the Hubble diagram weight (upper left) ($1/\sigma^2_\mu$, including 0.12 mag of unexplained dispersion), the SALT2 $x_1$ uncertainty (upper right), the SALT2 color uncertainty (lower left), and the measured $\sigma_\mu$ assuming fiducial values for the SALT2 standardization parameters (lower right). The solid red line is the expected measurement from existing photometry and future cadenced visits. The blue points are the expected measurements if a ToO is activated to observe the SN on the given date. These forecasts depend on a constrained SALT2 fit to the existing photometry, usually only one or two pre-maximum epochs.  The actual trigger date was executed near the start of the plotted window.}
\label{fig:forecast}
\end{centering}
\end{figure*}

After candidates were identified as likely SNe~Ia, we simulated the distance modulus uncertainty for a variety of \HST follow-up scenarios to finalize our ToO decision. We first performed a SALT2-4 fit, using priors on $x_1$ and color ($x_1 \sim \mathcal{N}[0,1]$, color $\sim \mathcal{N}[0,0.15]$) so that the date of maximum could be constrained with only one or two epochs of SN light. The best-fit model was then used as a template to generate future observed photometric points, using the same uncertainties as the existing photometric data. We iterated over potential observation dates starting 2-3 weeks from the next \HST scheduling window, performing a SALT2-4 fit to each simulated light curve, and evaluating the $x_1$, color and standardized distance modulus uncertainties (Figure~\ref{fig:forecast} illustrates this procedure). Generally, we expected to find a SN~Ia in every $\sim$5th \HST orbit, and our ToO decisions from these forecasts weighed the increased Hubble diagram weight against the chance of finding a new SN. In borderline cases, we tended to lean towards triggering on discovered SNe rather than saving orbits for potential future ones. 

\section{Ground-Based Program}
\label{sec:ground}

The aim of the ground-based program was to determine spectroscopic redshifts of host galaxies, cluster galaxies, and (where possible) live transient classifications (particularly for SNe in faint host galaxies). Observations were split between the Keck Low Resolution Imaging Spectrograph (LRIS) \citep{oke95} and MOSFIRE \citep{mclean12}, the Subaru Faint Object Camera and Spectrograph, \citep{kashikawa02}, the Very Large Telescope (VLT) FOcal Reducer/low dispersion Spectrograph 2 (FORS2) \citep{appenzeller98} and X-shooter \citep{vernet11}, the Gran Telescopio CANARIAS (GTC) Optical System for Imaging and low-Intermediate-Resolution Integrated Spectroscopy (OSIRIS) \citep{osiris98}, and the Gemini Multi-Object spectrograph (GMOS) \citep{gmos04}. The primary sources of host galaxy redshifts in the program were VLT and Keck MOSFIRE, though every telescope and instrument contributed at least one redshift. Spectra that contributed to transient redshifts are available in the \code{Seechange Galaxy Spectra} companion release \citep{seechangeSpectra}.

\subsection{Mask design and alignment of catalogs to the \HST images}
For the ground-based program, wherever possible we performed multi-object spectroscopy. At the beginning of the program, we hoped to obtain redshifts of cluster galaxies and future transient host galaxies, while at the end, most clusters had several transient hosts that needed redshifts. In general, each cluster had different \textit{a priori} information available, from existing spectroscopic redshifts, to photo-$z$ constraints, to astrometric quality, to photometric coverage. To weight targets for mask design software, we used a combination of photo-$z$ constraints (preferring targets near the cluster redshift), red color in the \HST filters (which in many cases were not used in the pre-existing photo-$z$ catalogs), and brightness. 

The small \HST FoV, combined with a small number of specific transient hosts of increased value, made selection of alignment stars and alignment of target catalogs both challenging and critically important. Some of the ground-based imaging of the clusters was 10+ years old, and we found that the proper motions for alignment stars from the original catalogs were not sufficiently accurate to ensure high value targets would land on slits (determined by comparing to the new \HST imaging). For alignment star selection, we heavily prioritized stars that were visible in the \HST FoV. For fixed-mask instruments like LRIS and FORS2, this can crowd out some science targets. 

For target selection, our catalogs contained a combination of \HST targets, \textit{Spitzer} targets, and ground-based targets from a variety of different surveys. These combined catalogs contained targets aligned to different stellar catalogs and were affected by the proper motion of potential alignment stars; when compared to the \HST imaging, targets were frequently offset by amounts similar to the half-width of a standard slit ($\sim$ tenths of an arcsecond). To ensure as many science targets land on slits as possible, we developed a procedure that determines a fully affine transformation from either a catalog or a FITS image with WCS to the \HST imaging. This affine transformation implicitly accounts for potentially unknown distortion coefficients in the ground-based catalogs and imaging. 

For each catalog, we used the WCS of the \HST imaging to predict the $x$-$y$ coordinates of each object from their reported RA, Dec coordinates. We then detected objects in the image with \code{SEP}, and matched them to these predictions by a combination of the closest object and a total angular offset limit\footnote{This procedure did require that the alignment was accurate enough that the closest object was likely to be the right one.}. With these two sets of $x$-$y$ coordinates, we determined the full affine transformation from the catalog coordinates into the \HST coordinates using OpenCV's\footnote{https://docs.opencv.org/2.4/index.html} \code{estimateRigidTransform} procedure. We performed an iterative outlier rejection procedure, rejecting the biggest outlier greater than 3 times the normalized median absolute deviation (NMAD) from the predicted location after the affine transformation, until no 3-NMAD outliers remained. With the transformation matrix estimated, we then transformed any object from the other catalog or image into \HST coordinates without needing to run an alignment procedure like \code{TweakReg} or \code{SCAMP}, which are generally less flexible about data format. This explicitly ensured that any uncertainty in distortion or the accuracy of astrometry did not affect our ability to get the object into the center of a slit. This algorithm is available in the \code{SeeChangeTools v1.0.0} companion software release \citep{seechangetools}.

\subsection{VLT}
VLT observations were obtained using FORS2 \citep{appenzeller98} and X-shooter \citep{vernet11} under proposals 294.A-5025, 095.A-0830, 096.A-0926 (PI: Hook), 097.A-0442 (PI: Nordin) and 100.A-0851 (PI: Williams). The data were reduced using a combination of \code{ESOReflex} \citep{2013A&A...559A..96F}, \code{IRAF}\footnote{IRAF is distributed by the National Optical Astronomy Observatory, which is operated by the Association of Universities for Research in Astronomy (AURA) under a cooperative agreement with the National Science Foundation.} and custom routines. We utilized the deployable slits in MOS mode for FORS2 using the GRIS\_300I+11 grism, which gives a wavelength coverage of 6000 -- 11000\,\AA\ and a resolution of $R=660$ at the central wavelength of 8600\,\AA. The southern and equatorial clusters at redshift $z<1.5$ were targeted with FORS2. At higher redshifts, the [O~{\sc ii}] 3727\,\AA\ and Ca~{\sc ii}~H\,\&\,K features become difficult to detect with FORS2, as they move further to the red end of the spectrum where sensitivity is low, therefore X-Shooter (and in particular the NIR arm), was used for the hosts of SN~Ia candidates occurring in the higher redshift clusters. In some cases X-Shooter ToO time was used to quickly obtain a redshift and help us decide whether to trigger further {\it HST} orbits on a given SN~Ia candidate. \citet{vltpaper} fully discuss the observation setup, data reduction, and present the reduced spectra and associated host properties. Figure~\ref{fig:SNSpectrum} presents a live SN spectrum that, considered with the light curve, supports a cluster (MOO-1014, $z=1.23$) SN~Ia classification.

\subsection{Keck}
\subsubsection{LRIS}
Keck LRIS \citep[][]{lris1995} was used in multislit mode for the lower-redshift northern and equatorial See Change clusters. Masks were designed using the \texttt{AUTOSLIT3} software.\footnote{Developed by J. Cohen and P. Shopbell} Slit widths were set to 1\arcsec\xspace with lengths in the range 6-9\arcsec, with minimum separations of 1.5–2\arcsec.  Given the high redshifts of our targets, our results derive from the red arm of LRIS, which is equipped with a fully-depleted LBNL CCD \citep{holland2003, rockosi2010}, making the gold-coated 600 line/mm grating blazed at 1~micron combined with the 6800~\AA\xspace dichroic the most effective configuration. With this grating and slit width the full width at half maximum of a line in the dispersion direction from an extended source spans 7.4~pixels or 5.9~\AA. We chose to not bin the detector readout, as the thickness of the CCD results in larger numbers cosmic rays and binning would have increased the level of interference with our spectra. In addition, binning in the spectral direction would have resulted in undersampling of the image of the slit on the detector, making sky subtraction more difficult. Multiple exposures, each of 1200--1800~seconds duration, were taken in order to obtain sufficient depth, while improving cosmic ray rejection and allowing realignment to compensate for mechanical flexure.

We processed the LRIS spectroscopic data with {\tt BOGUS}, which is an IRAF-based pipeline
designed for two-dimensional processing of multislit data, written by D.~Stern, A.~Bunker, and S.~A.~Stanford.\footnote{Available
upon request} After gain and
overscan correction of the raw two-dimensional images, {\tt BOGUS} 
splits the mask into individual slitlets and processes
each slitlet using standard optical longslit techniques. After
flattening the spectrum with domeflats taken after the observations, 
cosmic rays are identified from unsharp masking
of the images, sky lines are subtracted using a low order polynomial
fit to each column, and images are shifted by integer pixels in the
spatial and dispersion directions and recombined. An additional step of pair-wise image subtraction
improves fringe subtraction at long wavelengths. As a final step,
{\tt BOGUS} shifts each of the slitlets to roughly align them in
the wavelength direction which simplifies wavelength calibration.

After the two-dimensional processing provided by {\tt BOGUS}, we
extracted the spectra using the {\tt APALL} procedure within IRAF. 
We extracted arc lamps in an identical manner and used them to do a first pass wavelength
calibration of the data. This typically relied on a fifth order
polynomial wavelength solution, providing a calibration accuracy of $\approx 0.1$~\AA\ RMS. 
As a final step in the wavelength calibration,
the spectra were linearly shifted in wavelength based on the sky lines, and we
conservatively estimate that the wavelength solutions are robust
to better than 1~\AA.

\subsubsection{MOSFIRE} \label{sec:mosfire}
For MOSFIRE, at these redshifts a choice must usually be made between targeting Ca H \& K and the 4000\AA \xspace break in $Y$ band or H$\alpha$ in $H$ band. To inform this decision, we estimated emission-line fluxes based on galaxy colors by comparing to the MPA-JHU SDSS DR7 catalogs\footnote{https://wwwmpa.mpa-garching.mpg.de/SDSS/DR7/} described in \citet{Kauffmann03,Brinchmann04,Salim07}. We ran the estimated emission-line fluxes through the MOSFIRE exposure time calculator to estimate the feasibility of searching for an emission line. We also estimated the signal-to-noise ratio (SNR) of the spectrum in the $Y$ band, knowing from simulations using the weighted cross-correlation (described in Section \ref{sec:crosscorr}) that we needed an average SNR $\sim1$ per wavelength pixel in order to confirm an absorption-feature redshift. 

During observing, we used the ABA$'$B$'$ dither pattern, with recommended sampling and exposure times from the MOSFIRE documentation\footnote{https://www2.keck.hawaii.edu/inst/mosfire/home.html}. For most masks, we monitored mask alignment, seeing, background, and cloud obscuration in real-time by including a faint star in the science mask and estimating its position in each dither with a custom \code{python} script. 

On most nights, we observed at least one A0V star to be used as a telluric standard. In practice, we combined and normalized the A0V observations into a single instrumental response curve for each MOSFIRE grating/filter combination, and used the \HST photometry of the host galaxy for absolute flux calibration. 

To process the data, we used the MOSFIRE Data Reduction Pipeline (DRP), 2018 release version\footnote{https://github.com/Keck-DataReductionPipelines/MosfireDRP, release version Jan 5, 2018}, which reduced the raw data into individual, wavelength-calibrated 2D FITS files for each configurable slit in each mask. To extract one-dimensional (1D) spectra for use in the weighted cross-correlation, we collapsed each 2D spectrum in the wavelength dimension\footnote{The DRP-reduced data are well-rectified and the drift over the wavelength range of each filter is typically less than a pixel.}, fitting a Gaussian to estimate the spatial weighting profile. The 2D data are then collapsed in the spatial direction using the weights from the DRP and the Gaussian profile determined above. 

\subsection{GTC}
The set of spectroscopic longslit data taken  at the GTC were obtained under programs GTC11-14B, GTC27-15A, GTC13-15B, GTC8-16A (PI: Ruiz--Lapuente). In GTC11-14B, we used the OSIRIS grism R300R aiming at the wavelenghth range 4800--10000 {\AA}  with a resolution of  $R=348$. After that program,  we moved to grism R2500I that covers the spectral range 7330---10000 {\AA} with a resolution of $R=2503$ and would allow better sky subtraction and higher sensitivity to narrow host galaxy spectral features for redshift determination. We processed the GTC/OSIRIS raw data using IRAF$^{11}$ and custom routines. Twilight flat fields were created, and were used along with the bad pixel masks to reduce the science frames. The frames were sigma clipped and combined  to eliminate cosmic ray events. The spectra have been wavelength calibrated using comparison arc-spectra of Hg+Ar, Ne and Xe  lamps, and flux calibrated using spectrophotometric standard stars. The 1D spectra were extracted  using the APALL procedure within IRAF. The individual spectra were median combined into a resulting 1D spectrum. The final spectra were linearly shifted  based on the positions of four sky lines (5577{\AA}, 6300{\AA}, 6863{\AA} and 7276{\AA}) to correct for instrument flexure. 

\begin{figure*}[h]
\begin{centering}
\includegraphics[width=0.49 \textwidth]{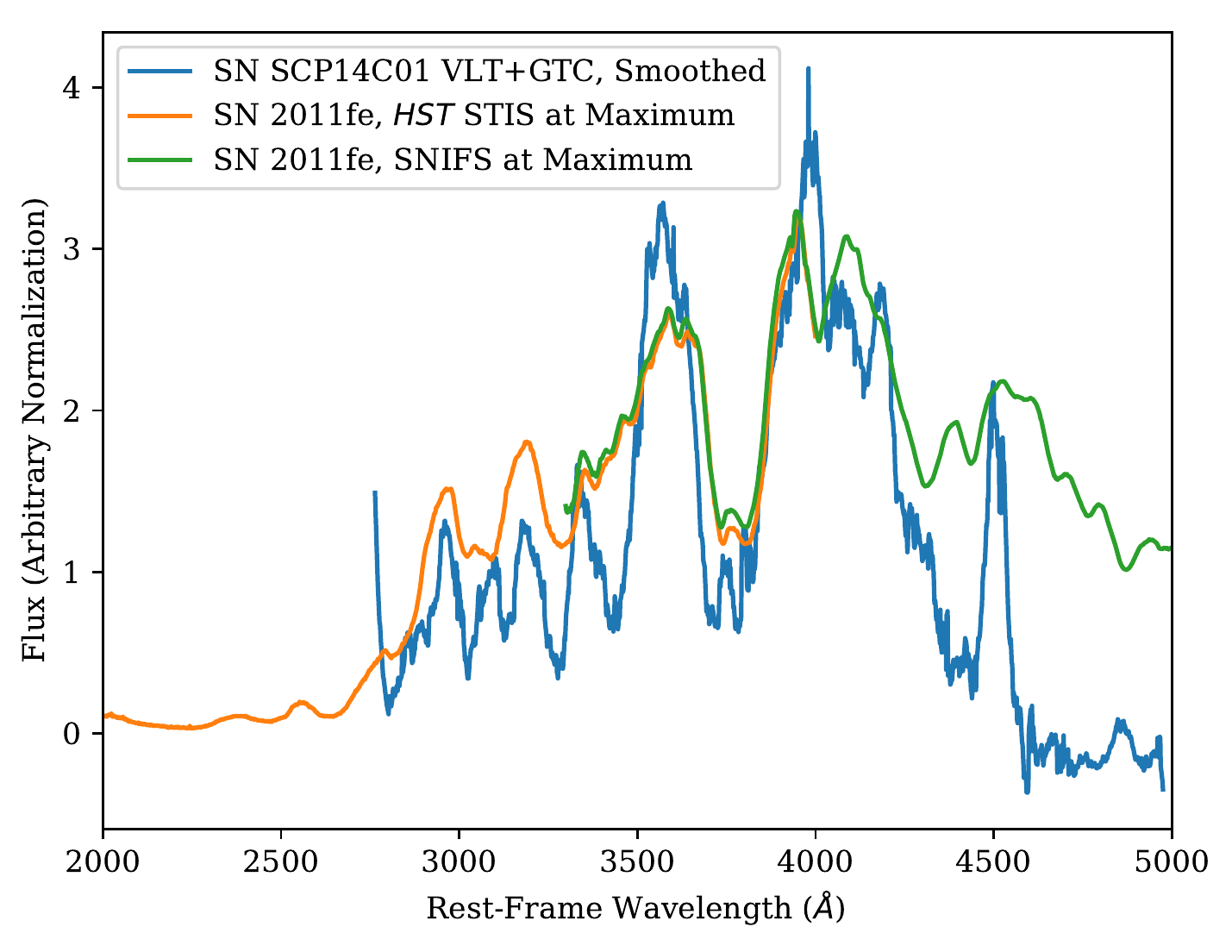}
\includegraphics[width=0.49 \textwidth]{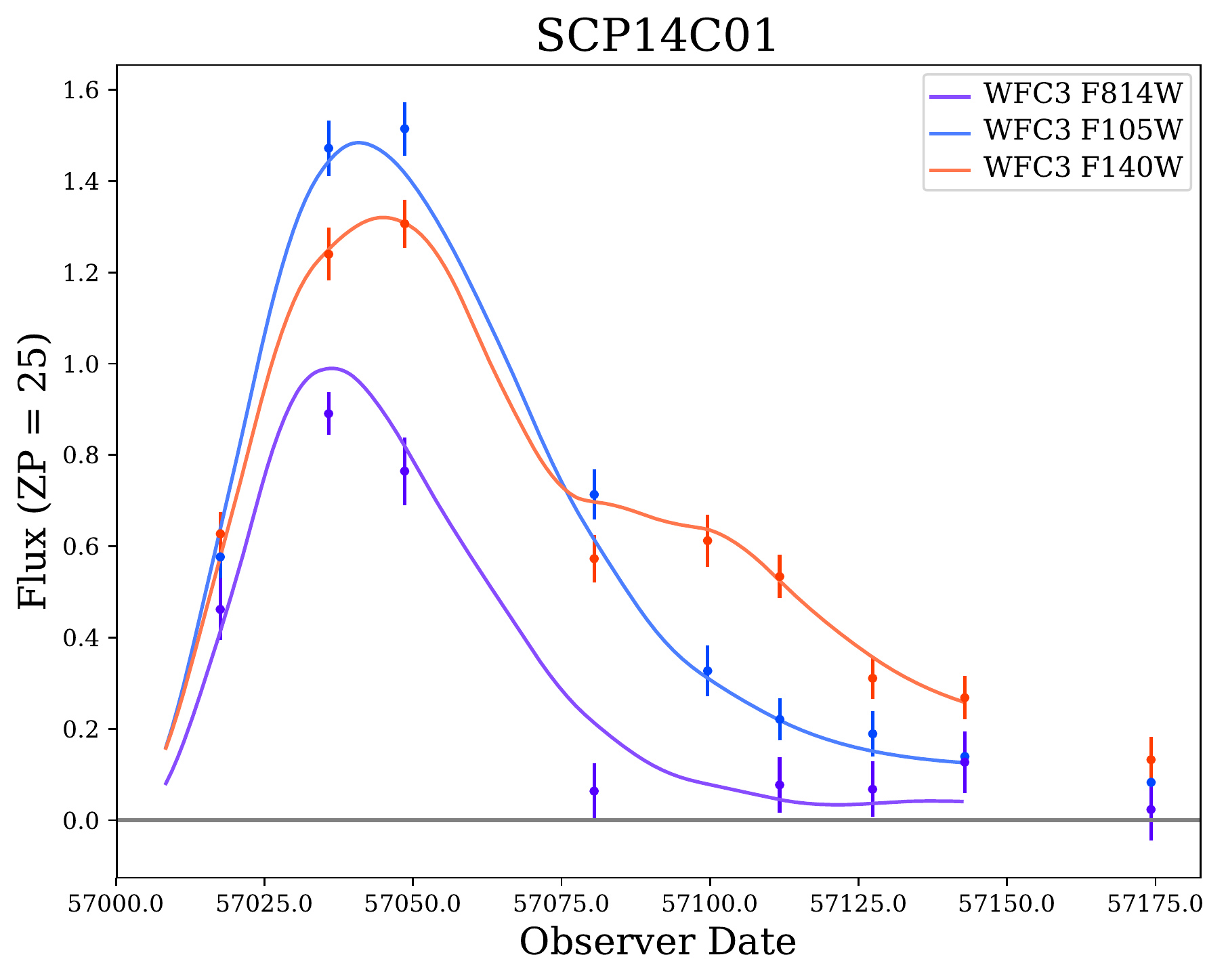}
\caption{Combined spectra of SN SCP14C01 from VLT and GTC, smoothed and de-redshifted. We see a reasonable match to the SN~Ia SN 2011fe \citep{pereira13, mazzali14}. Combined with the light curve, we have high confidence that this is a SN~Ia at the redshift of cluster MOO-1014, $z=1.23$.}
\label{fig:SNSpectrum}
\end{centering}
\end{figure*}

In Figure~\ref{fig:SNSpectrum}, we show a combined GTC and VLT spectrum of SCP14C01, a SN Ia with no detected host galaxy in MOO-1014. These spectroscopic data, along with the well-sampled light curve, were crucial for confirming the redshift and type of this SN. 
 
\subsection{Weighted Cross-Correlation}
\label{sec:crosscorr}
For host galaxy spectroscopic redshifts and uncertainties, we implemented a weighted cross-correlation algorithm \citep{kelson00}.\footnote{\url{http://code.obs.carnegiescience.edu/Algorithms/realcc/at_download/file}} This allowed us to easily combine spectra across the different telescopes and instruments, with telluric features naturally de-weighted. We also used the algorithm in simulations to determine exposure times (especially for Keck MOSFIRE observations), constrained by HST photometry and spectroscopy from any prior runs. As mentioned in Section~\ref{sec:mosfire}, for MOSFIRE observations on any given night it is important to determine which instrument configuration has the highest likelihood of providing a spectroscopic redshift. We applied simulated noise to a typical elliptical galaxy spectrum to determine the required SNR for the weighted cross-correlation to detect a redshift at $5\sigma$. For the MOSFIRE $Y$-band spectrograph configuration, an average SNR of $\sim 1$ per wavelength pixel over the filter is required. This routine is included in the \code{SeeChangeTools v1.0.0} companion release \citep{seechangetools}.

\subsection{Ground-based summary}
The ground-based program provided spectroscopic host-galaxy redshifts for $\sim 75\%$ of the SNe Ia in the cosmology sample. It also provided host-galaxy redshifts for many of the low-redshift transients discovered, as well as confirming SN Ia host-galaxy redshifts up to $z=2.29$. We also were able to confirm the type of a SN Ia with no detected host galaxy in MOO-1014 at $z=1.23$ with GTC OSIRIS and VLT FORS. The spectra used for confirmation of spectroscopic redshifts are available via the \texttt{Seechange Host Galaxy Spectra} companion release \citep{seechangeSpectra} in a format that is compatible with the weighted cross-correlation routine included in \code{SeeChangeTools v1.0.0} \citep{seechangetools}.

\begin{figure*}[h]
\begin{centering}

\includegraphics[width=0.32 \textwidth]{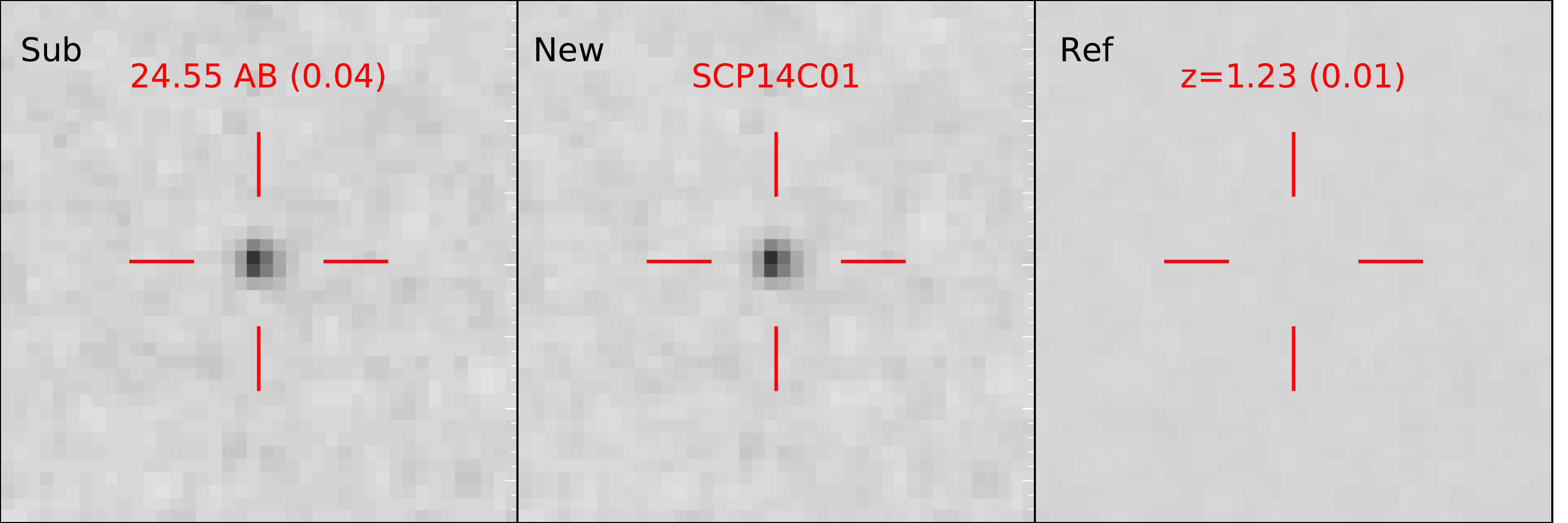}
\includegraphics[width=0.32 \textwidth]{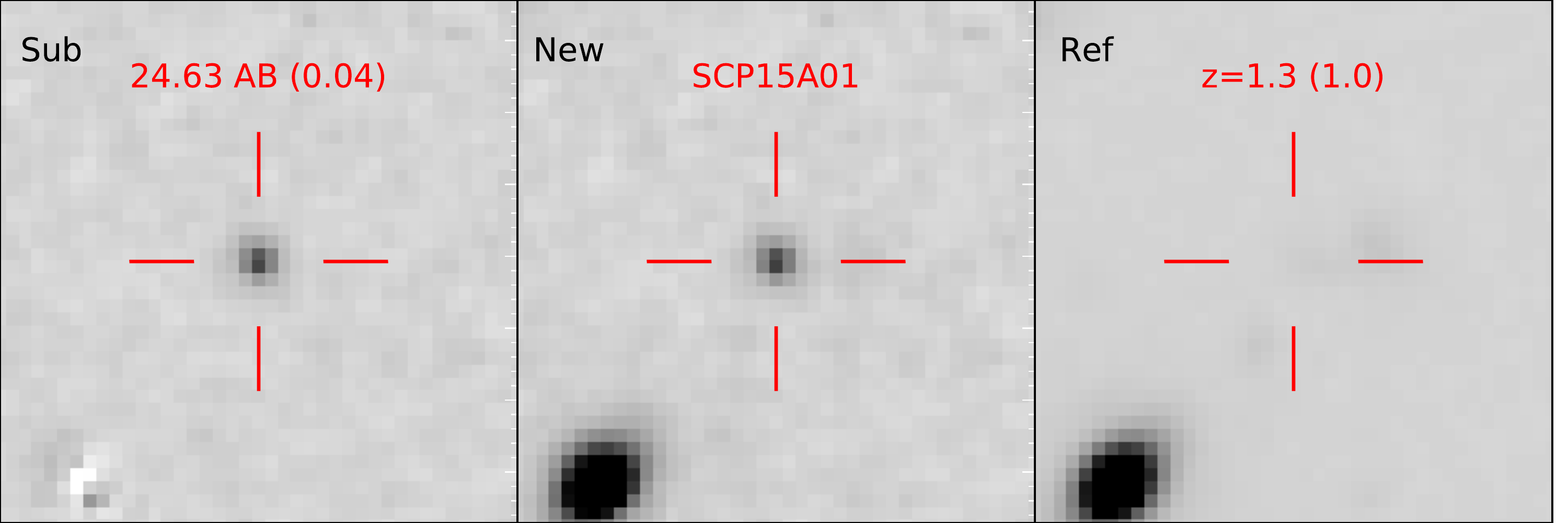}
\includegraphics[width=0.32 \textwidth]{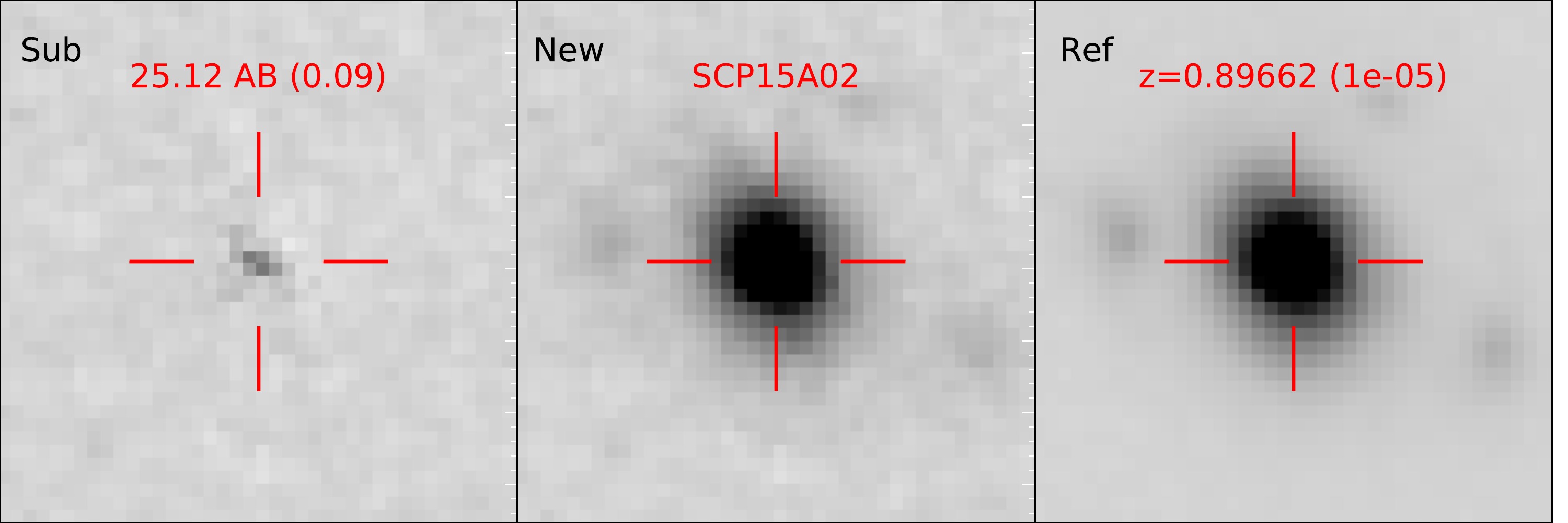}

\includegraphics[width=0.32 \textwidth]{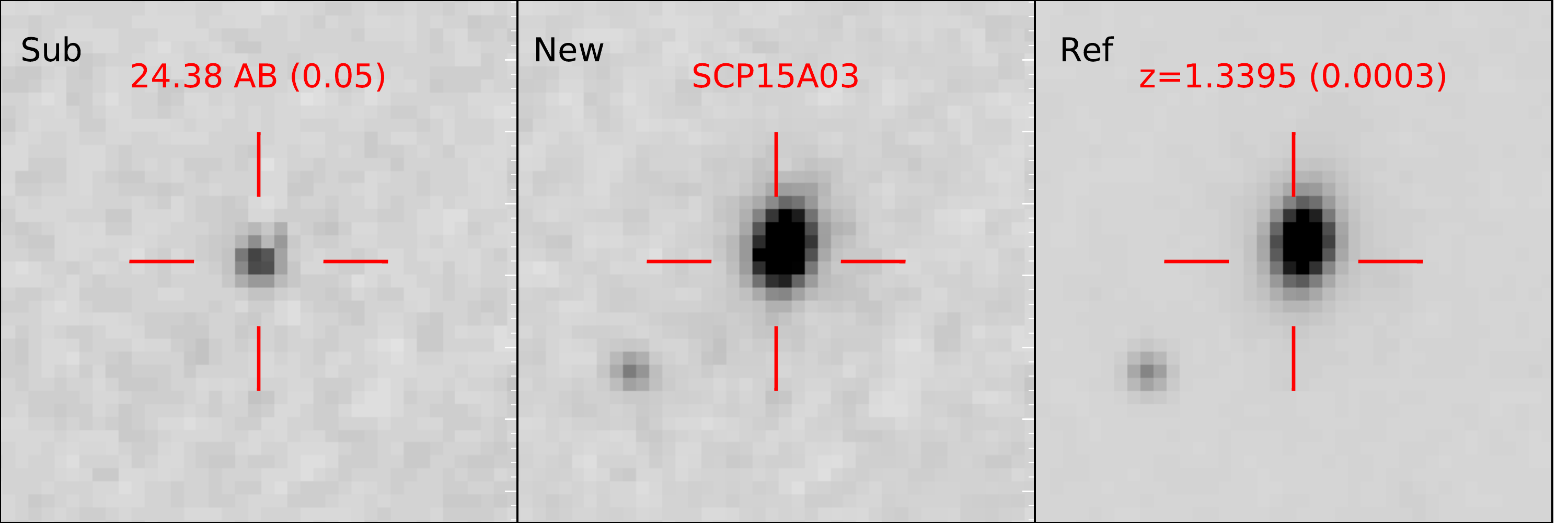}
\includegraphics[width=0.32 \textwidth]{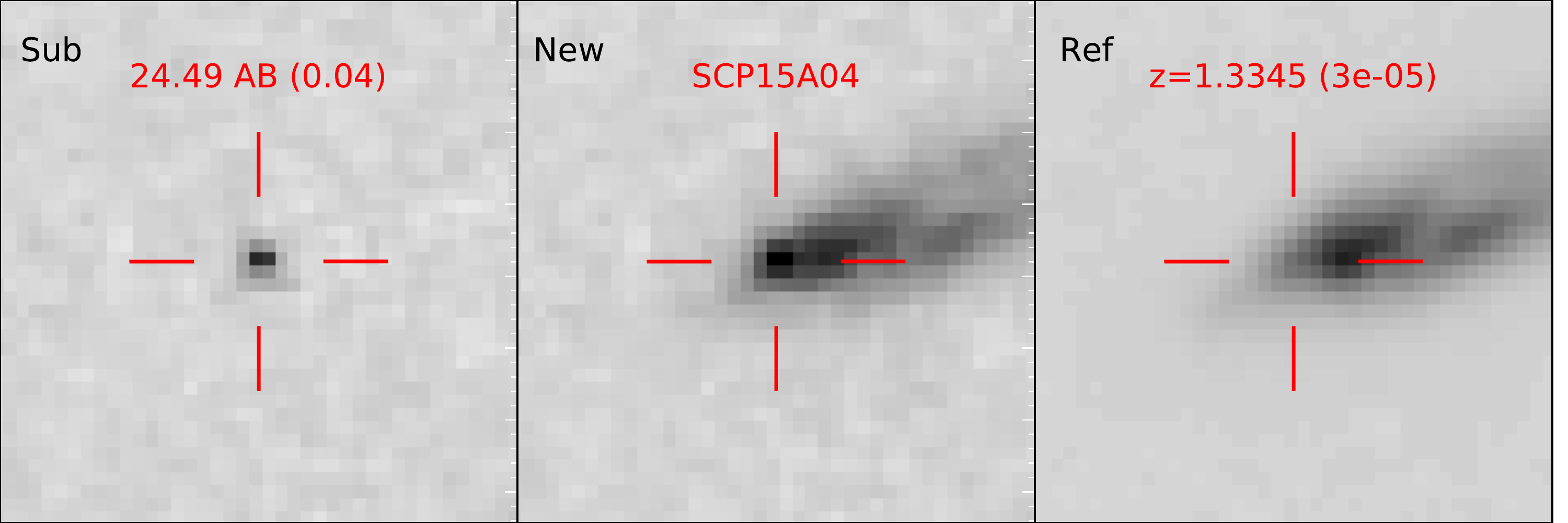}
\includegraphics[width=0.32 \textwidth]{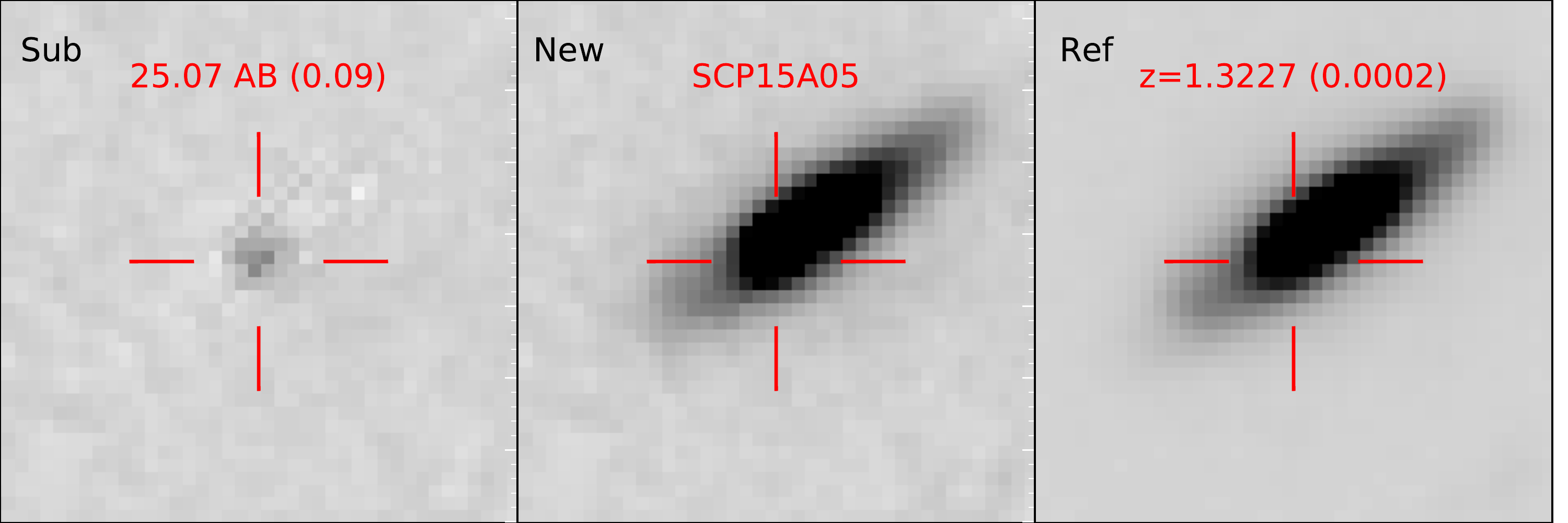}

\includegraphics[width=0.32 \textwidth]{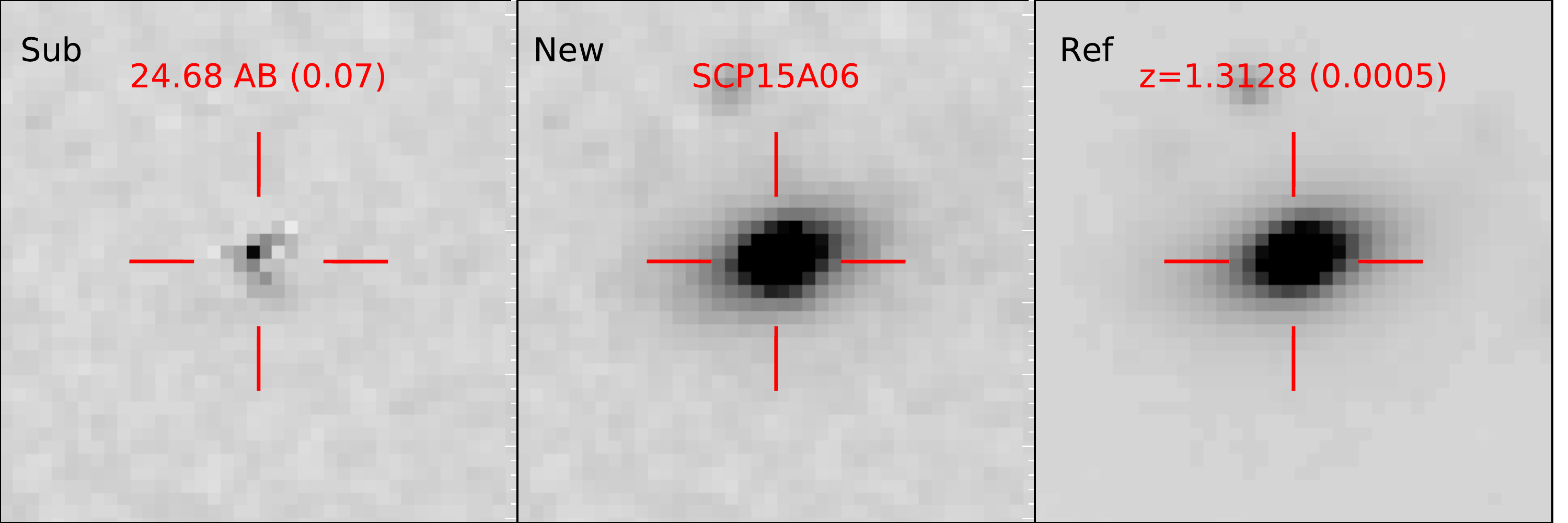}
\includegraphics[width=0.32 \textwidth]{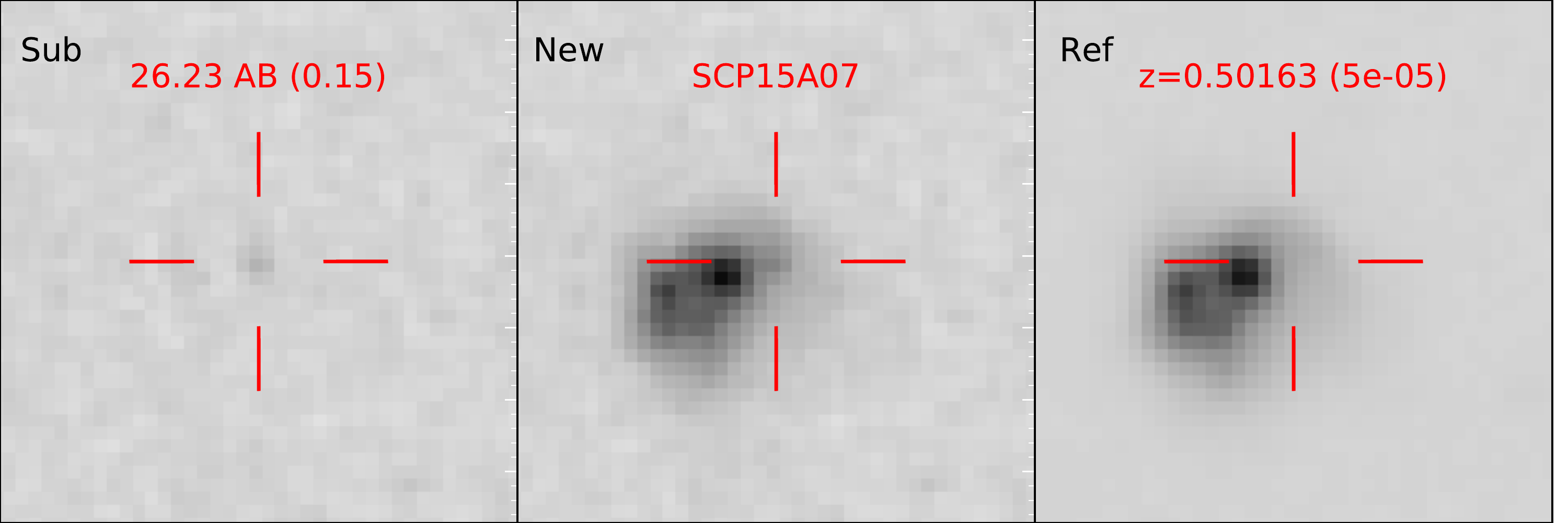}
\includegraphics[width=0.32 \textwidth]{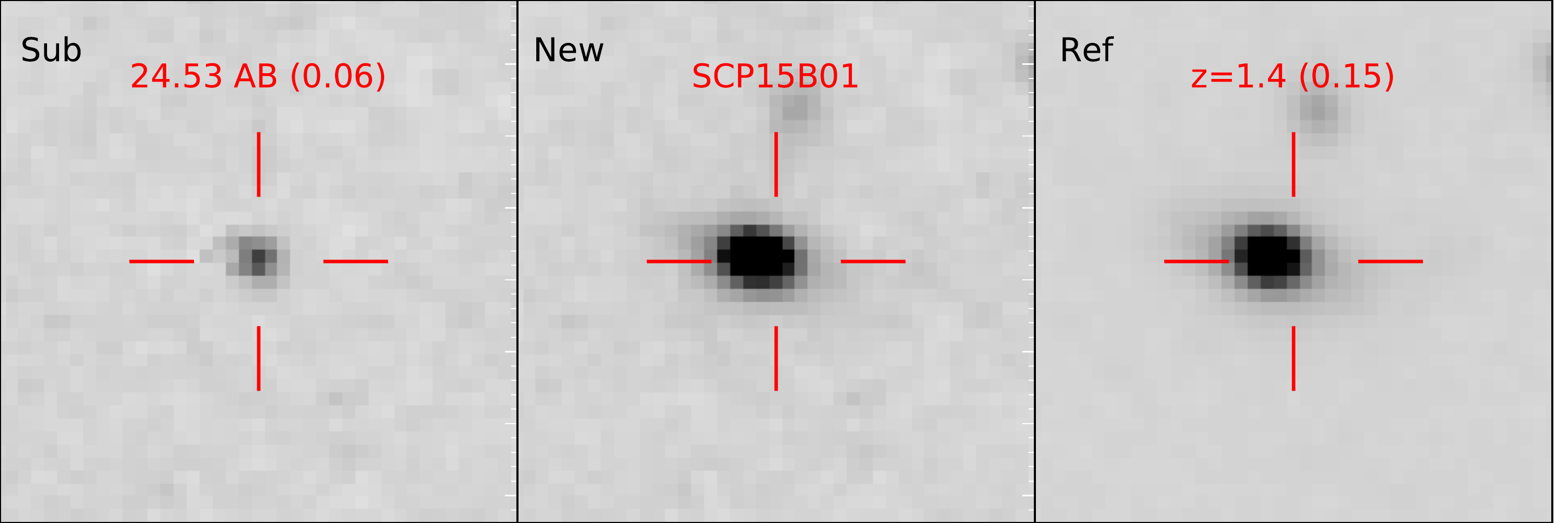}

\includegraphics[width=0.32 \textwidth]{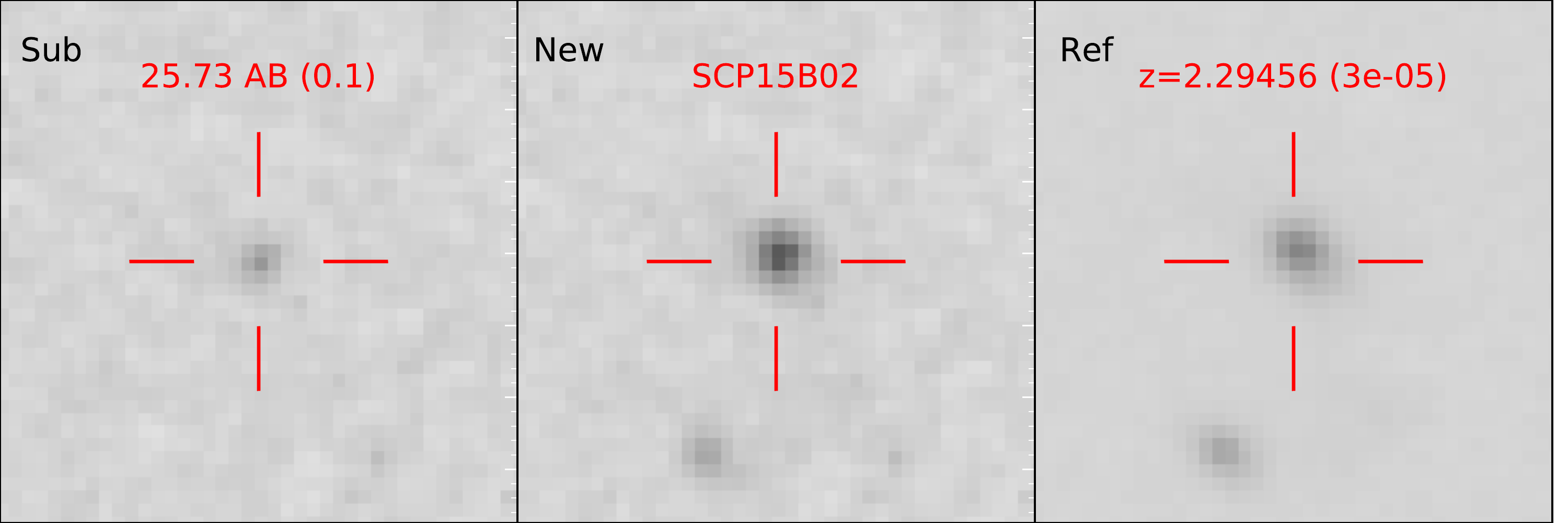}
\includegraphics[width=0.32 \textwidth]{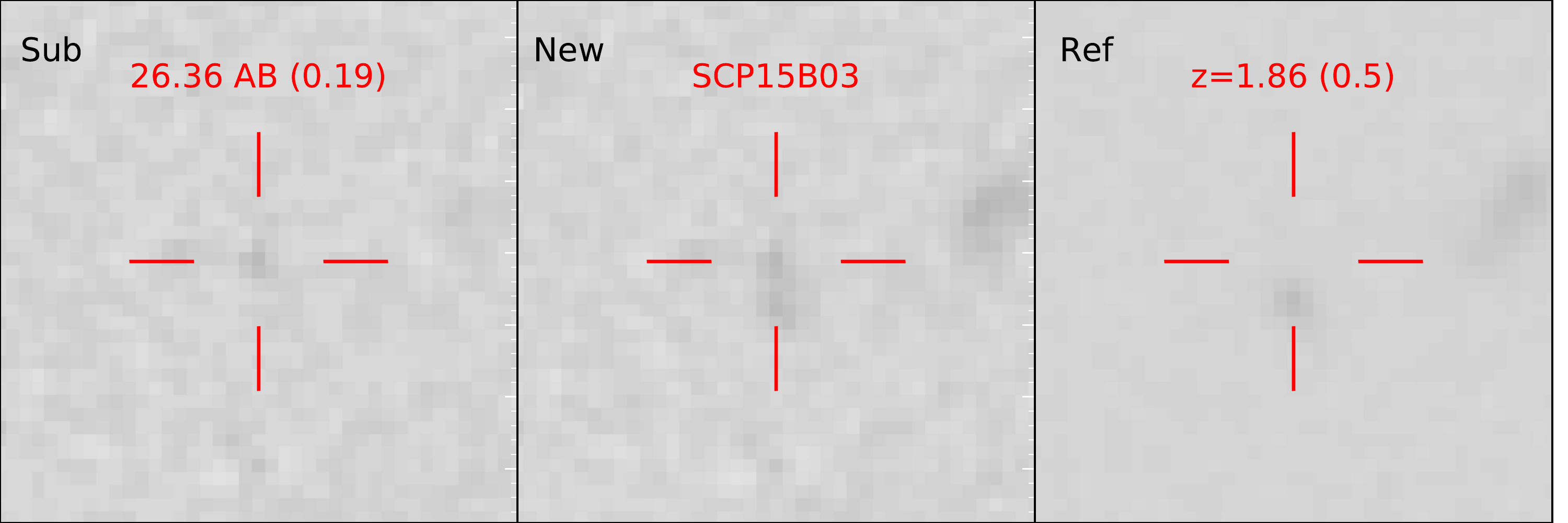}
\includegraphics[width=0.32 \textwidth]{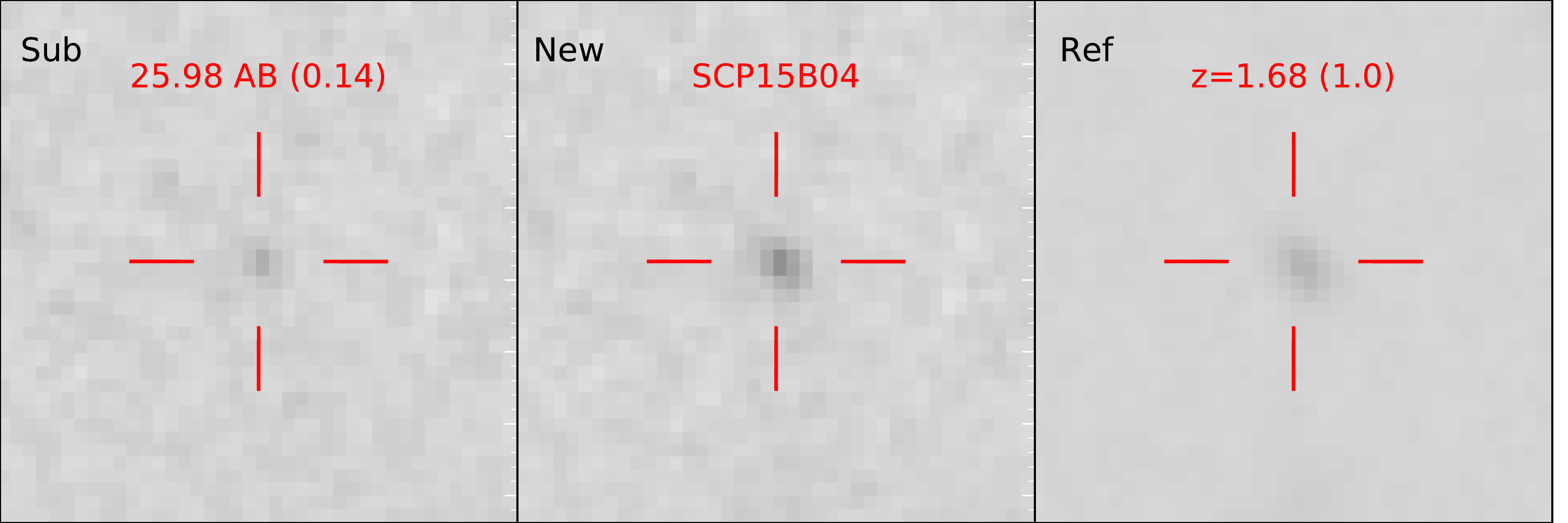}

\includegraphics[width=0.32 \textwidth]{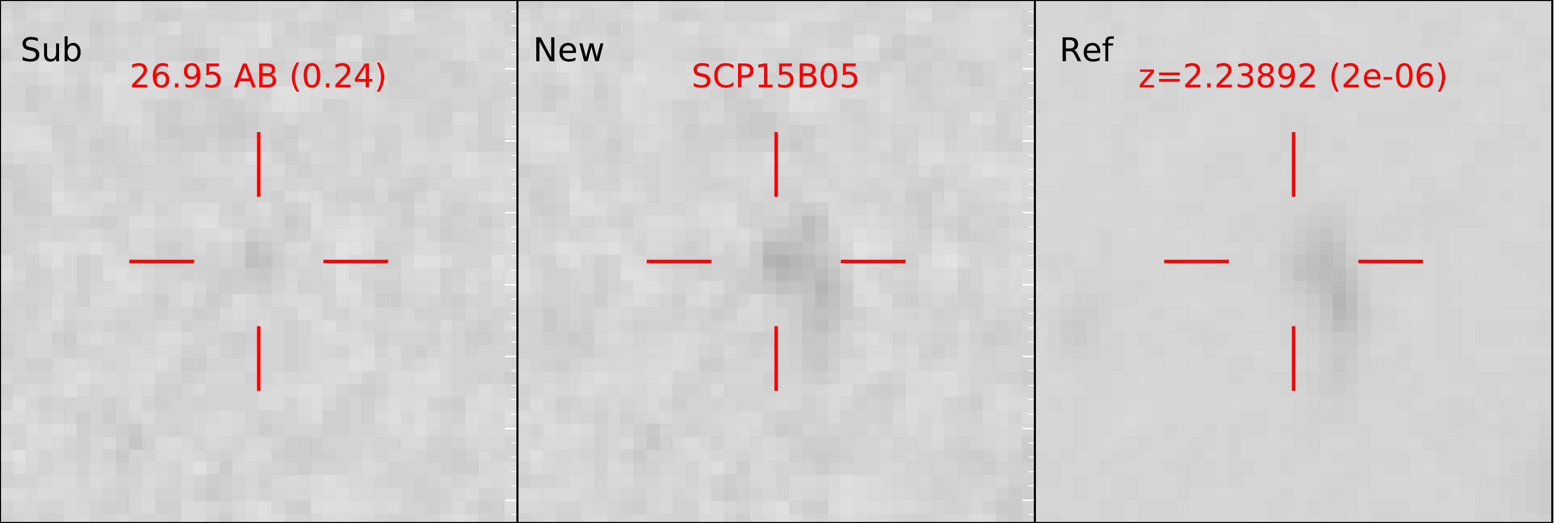}
\includegraphics[width=0.32 \textwidth]{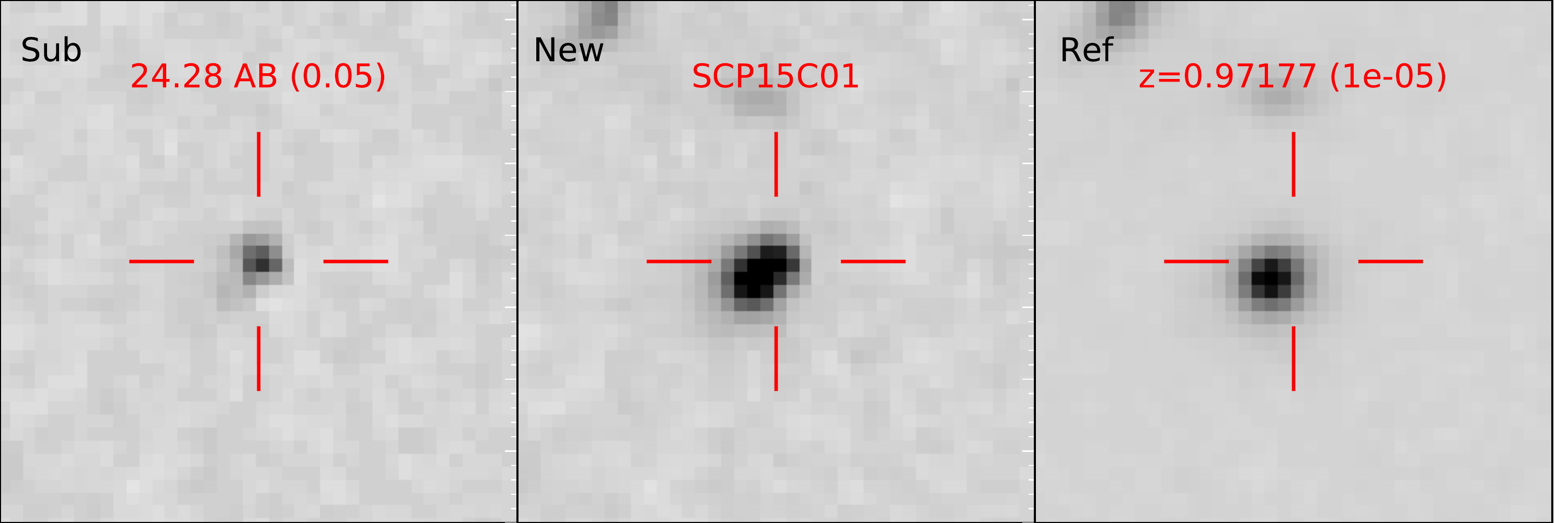}
\includegraphics[width=0.32 \textwidth]{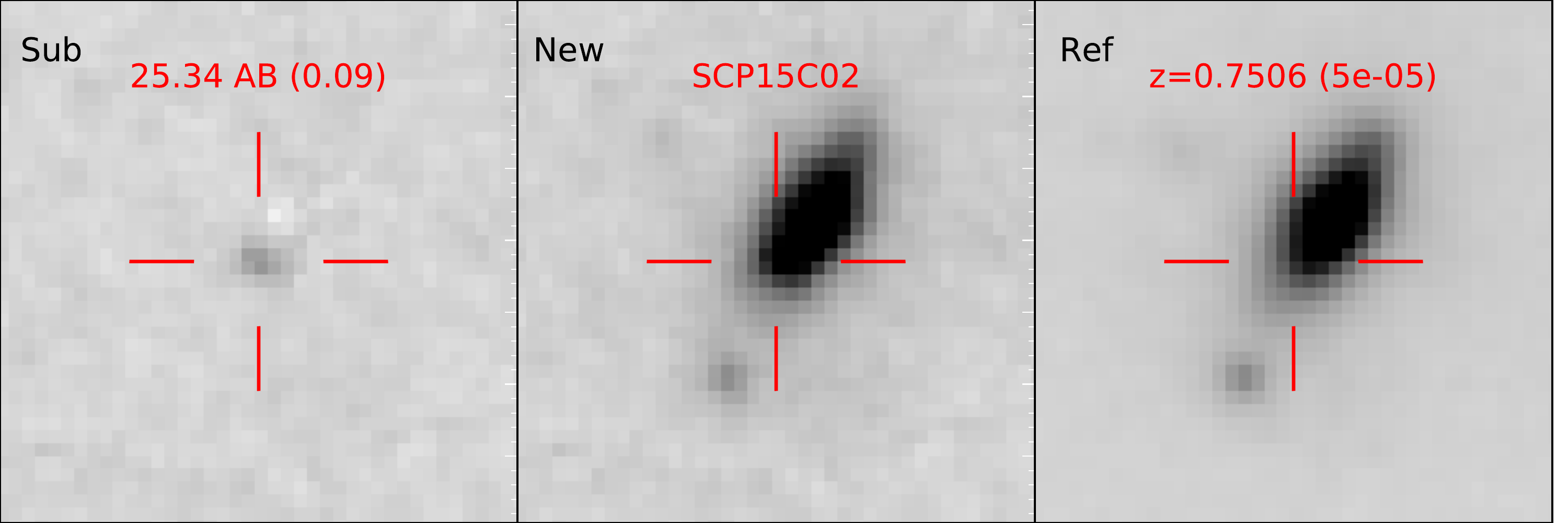}

\includegraphics[width=0.32 \textwidth]{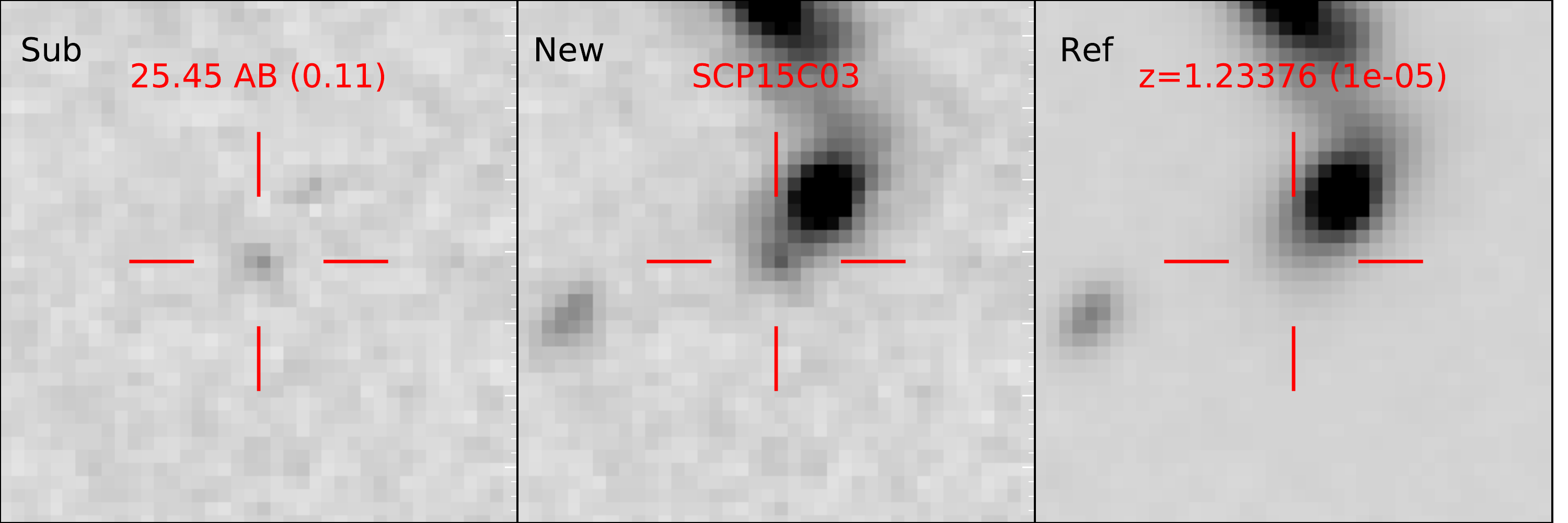}
\includegraphics[width=0.32 \textwidth]{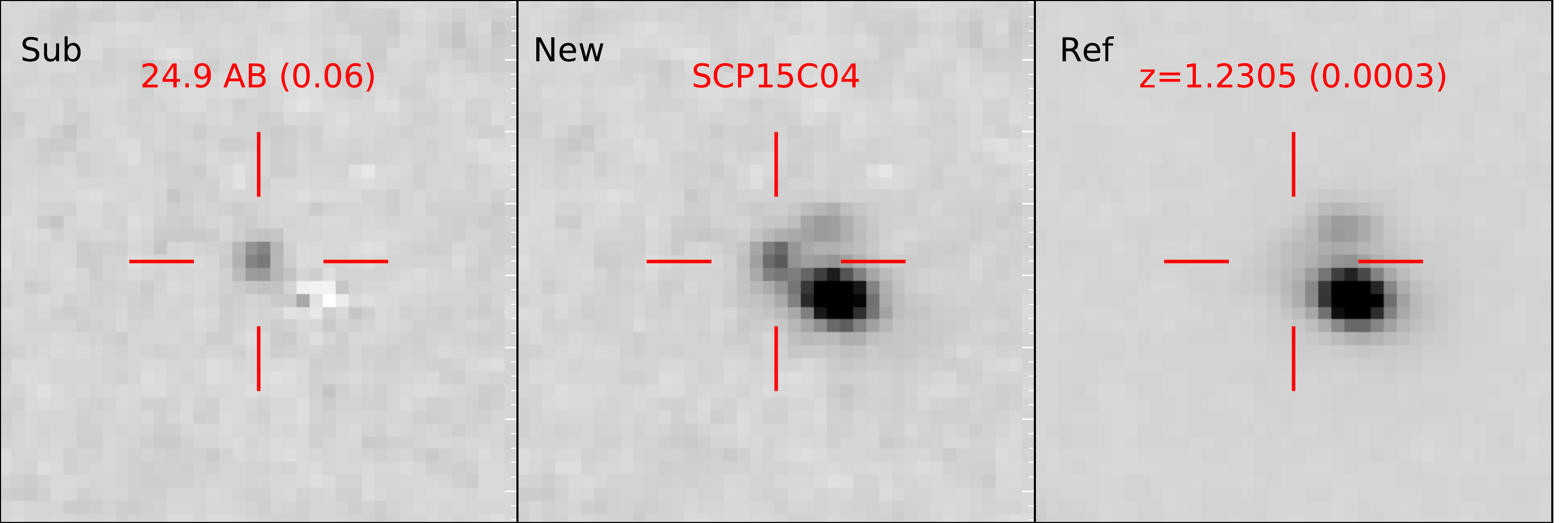}
\includegraphics[width=0.32\textwidth]{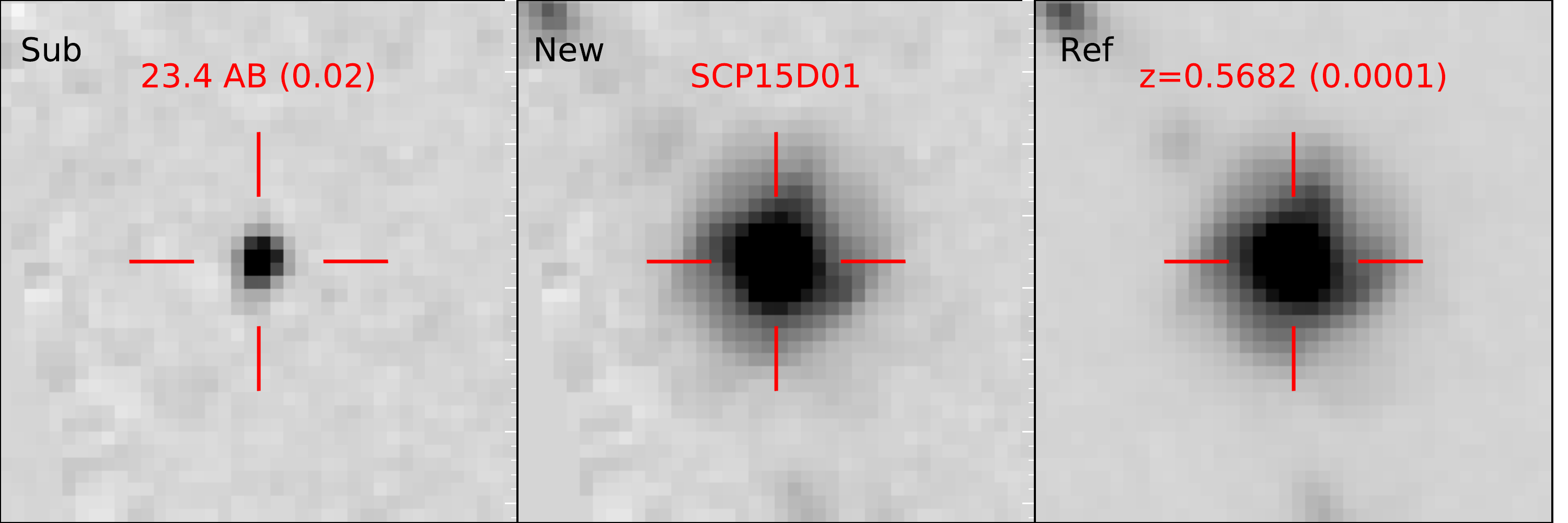}

\includegraphics[width=0.32\textwidth]{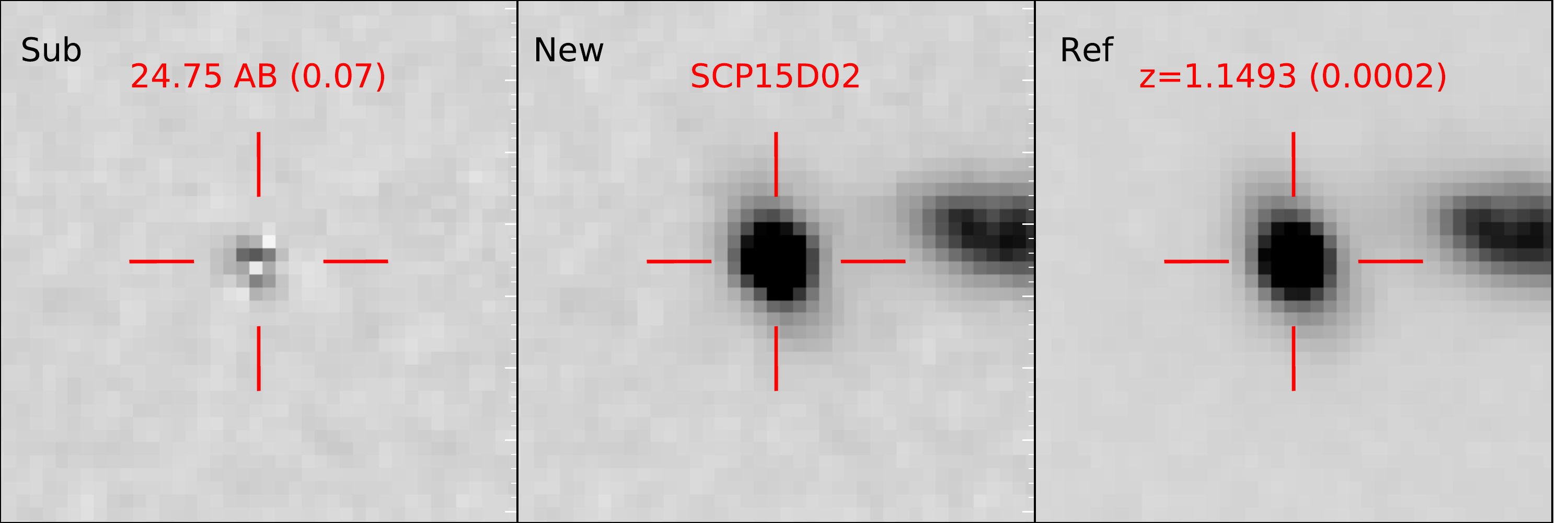}
\includegraphics[width=0.32\textwidth]{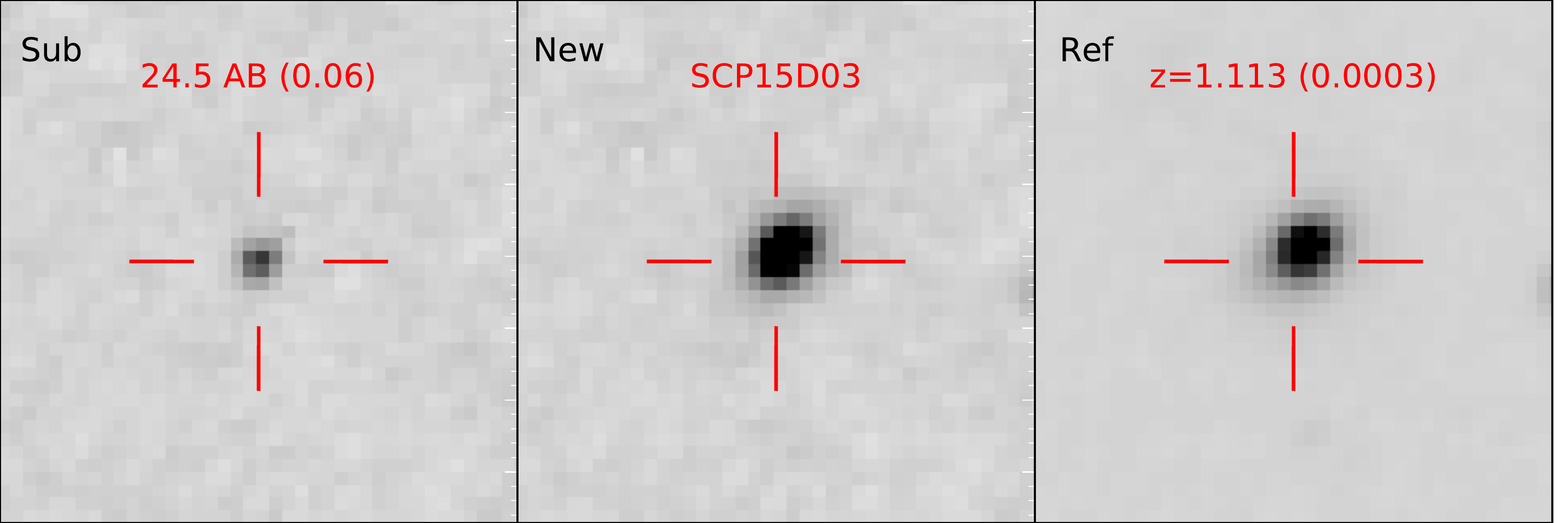}
\includegraphics[width=0.32\textwidth]{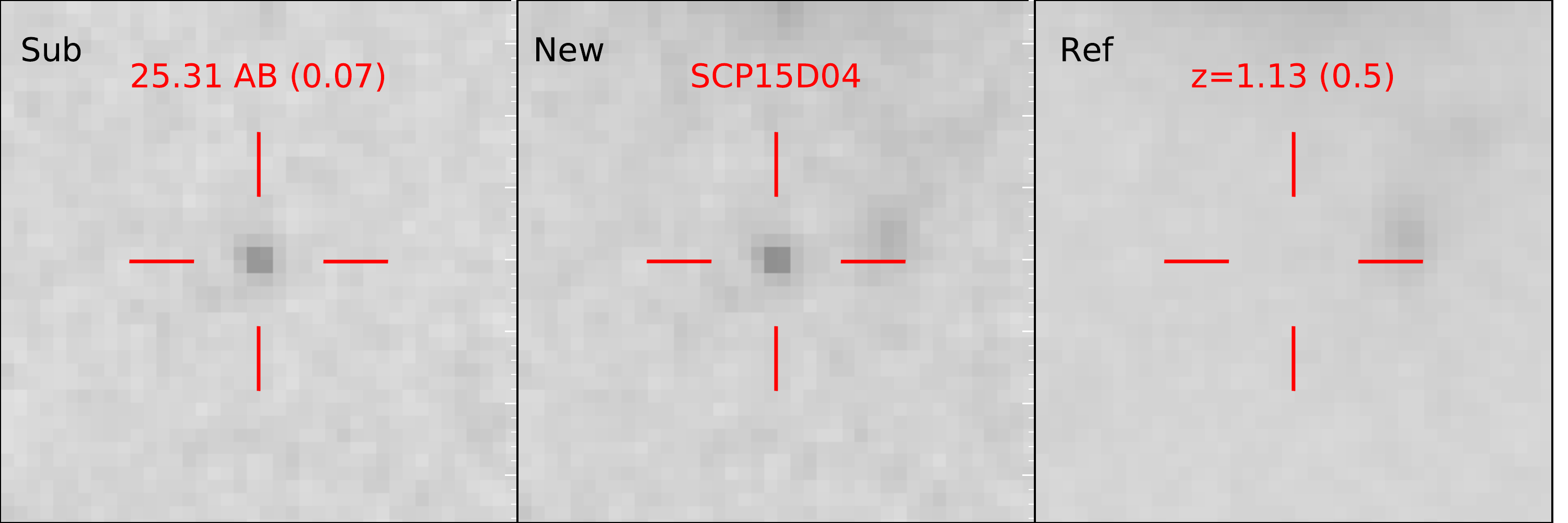}

\includegraphics[width=0.32 \textwidth]{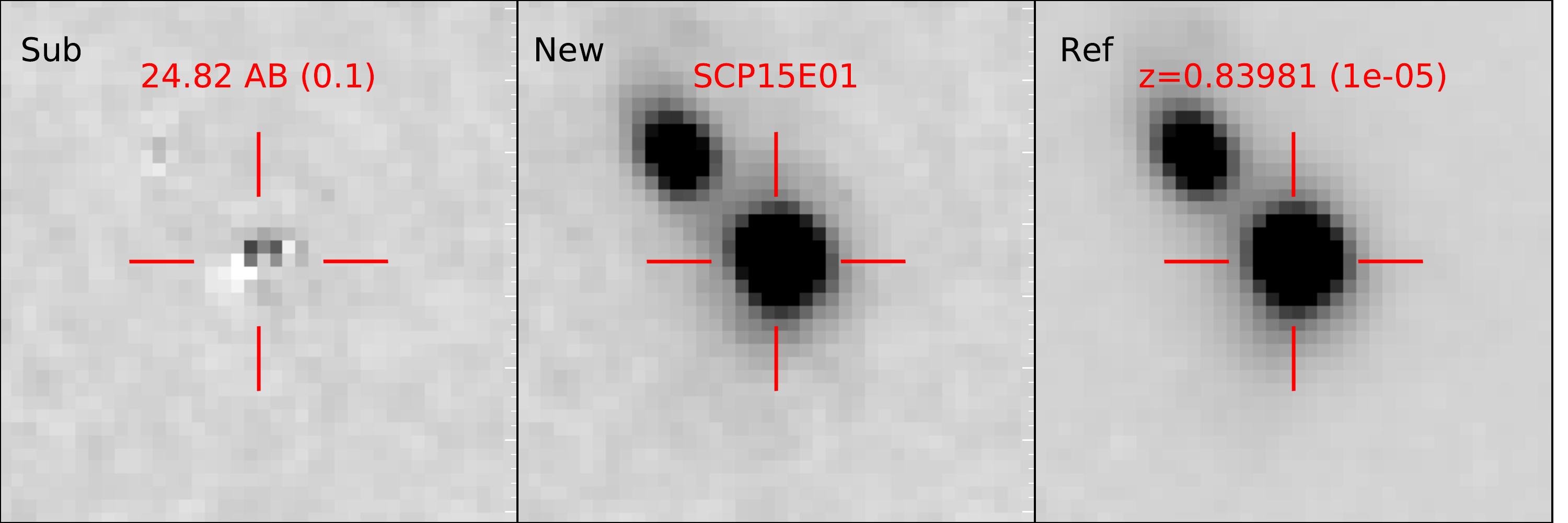}
\includegraphics[width=0.32 \textwidth]{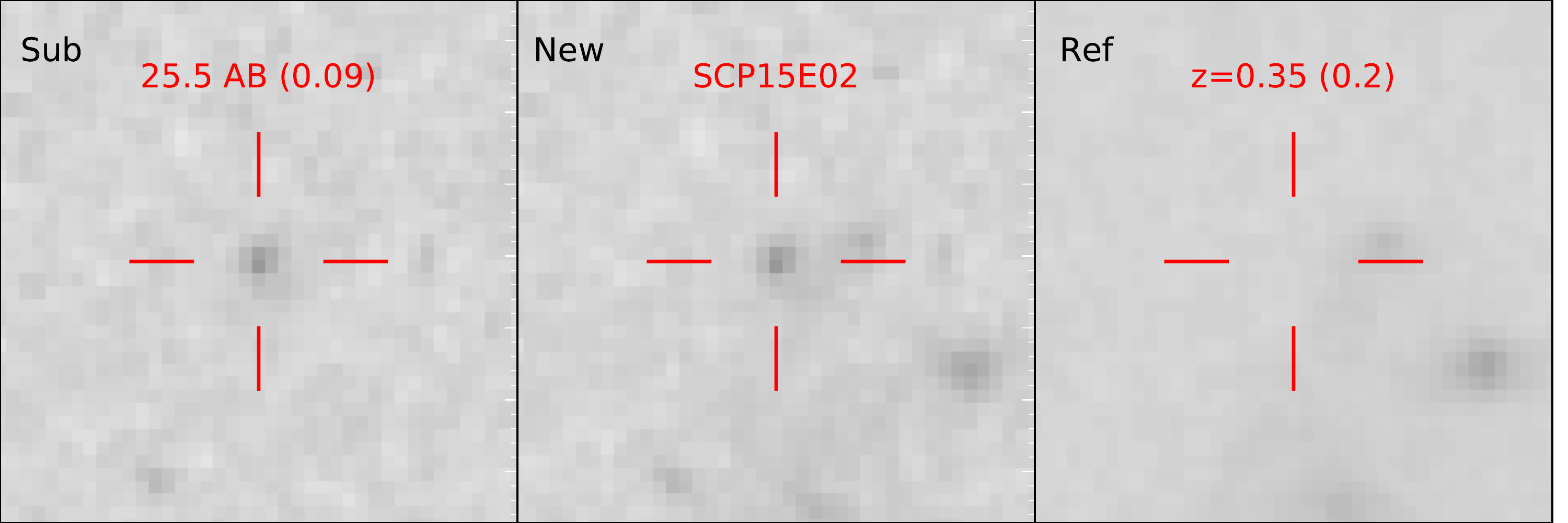}
\includegraphics[width=0.32 \textwidth]{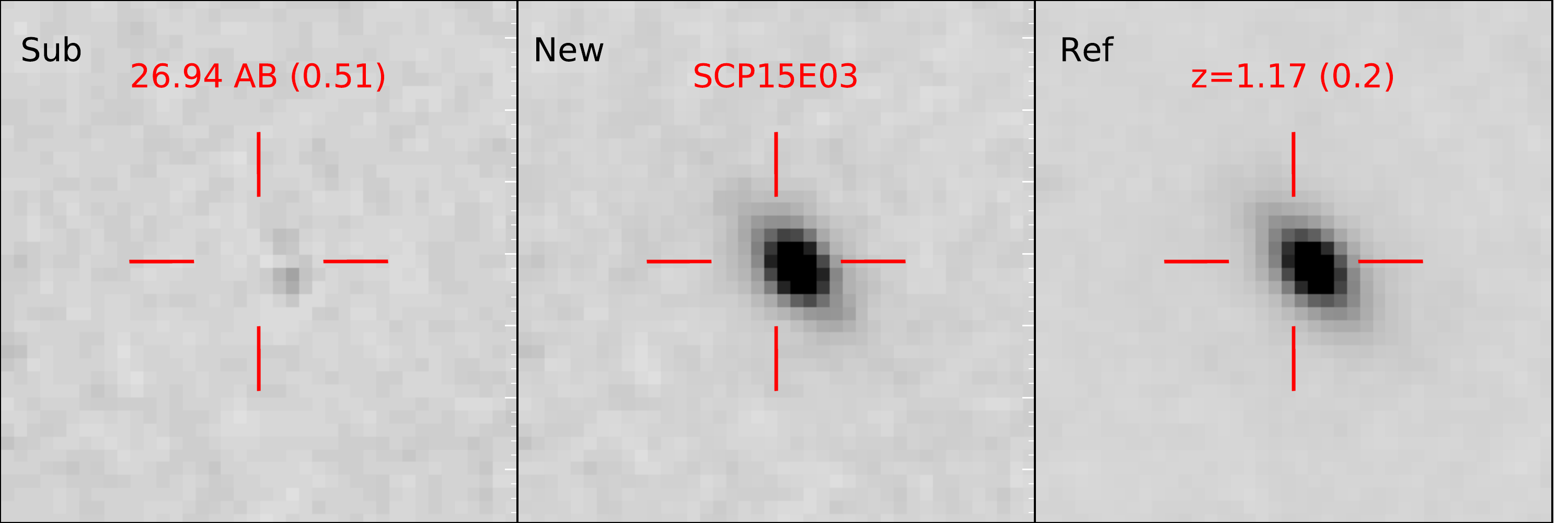}

\includegraphics[width=0.32 \textwidth]{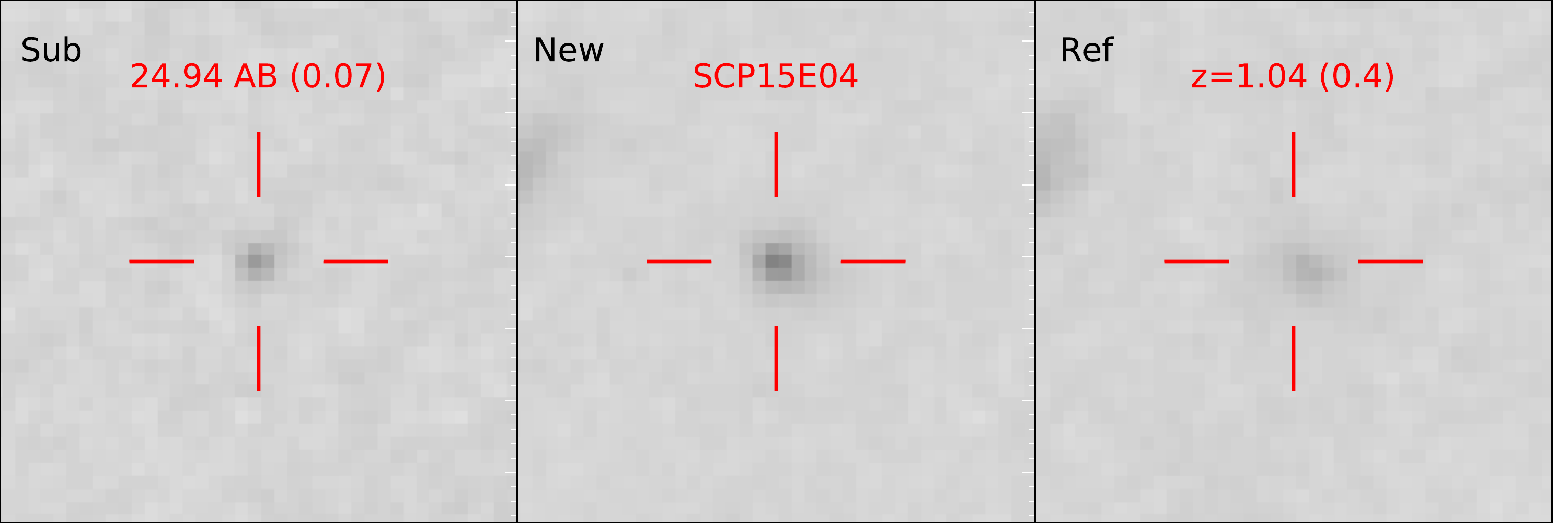}
\includegraphics[width=0.32 \textwidth]{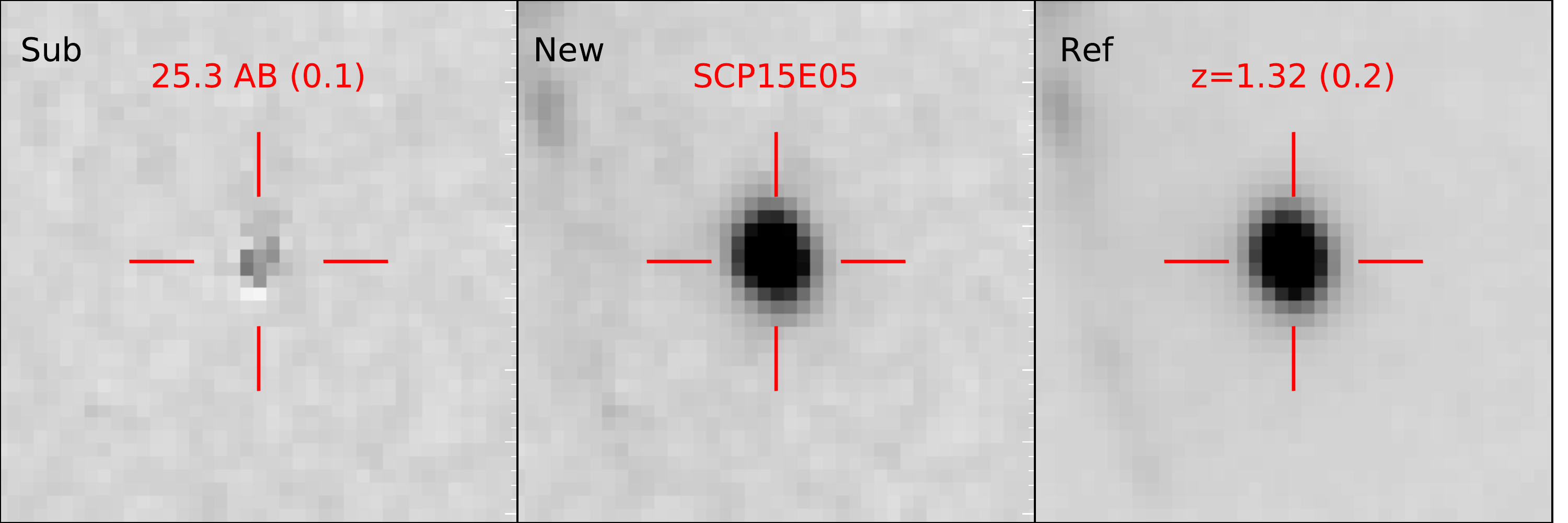}
\includegraphics[width=0.32 \textwidth]{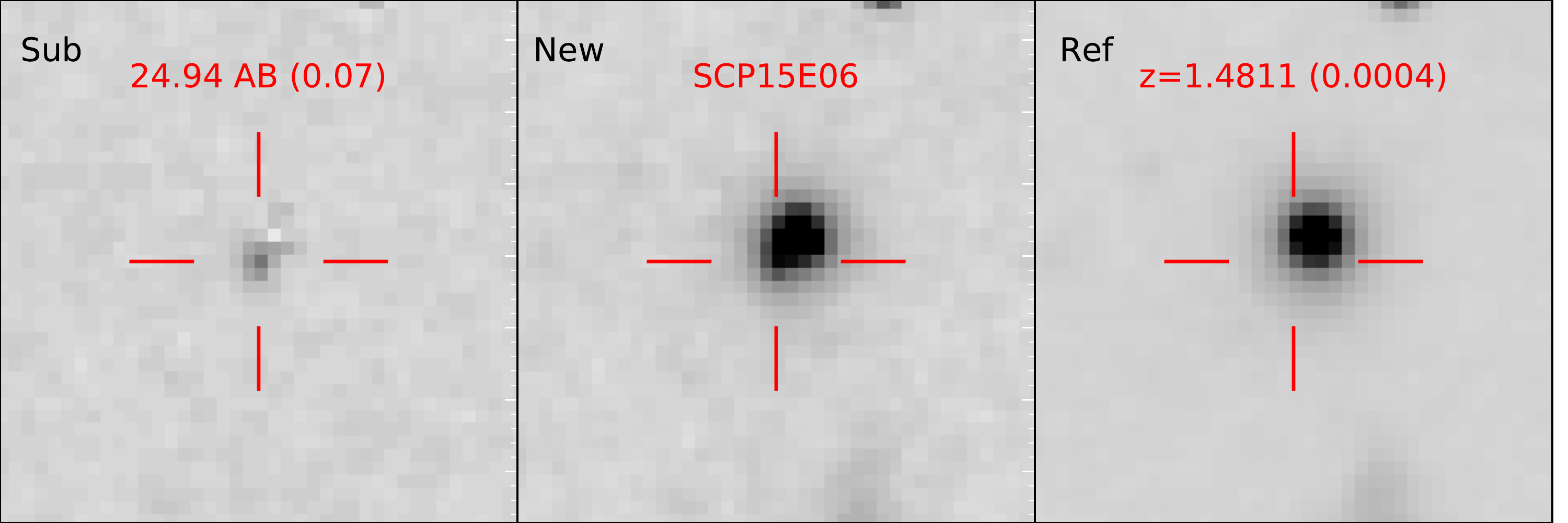}

\includegraphics[width=0.32 \textwidth]{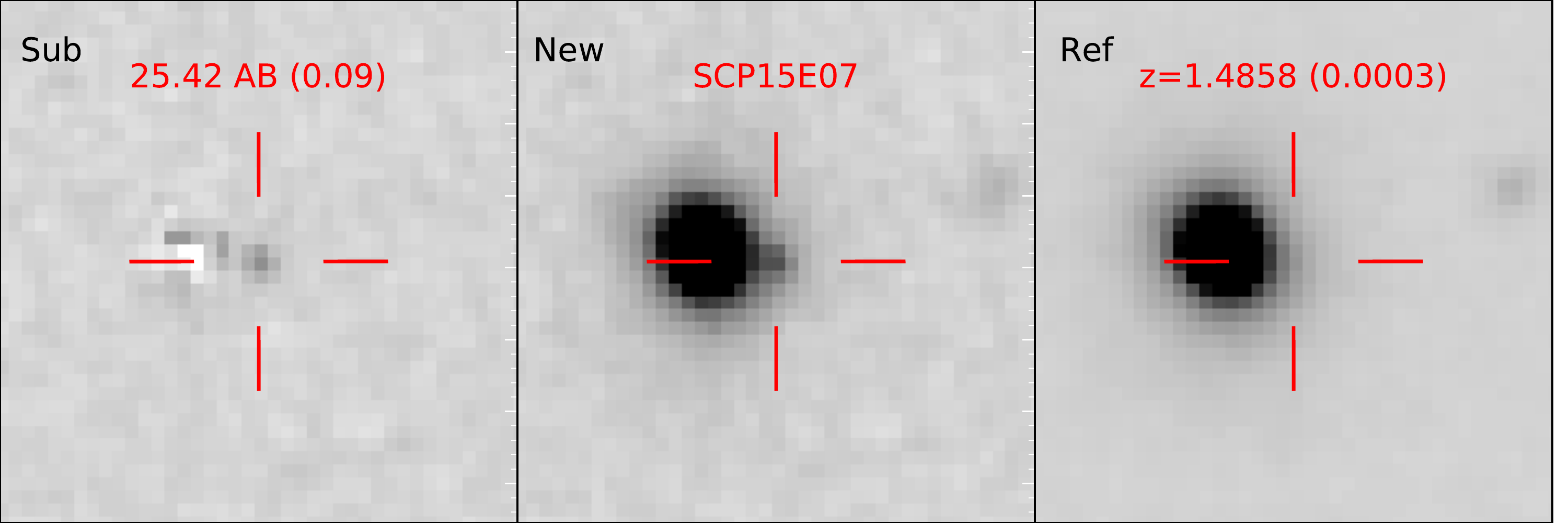}
\includegraphics[width=0.32 \textwidth]{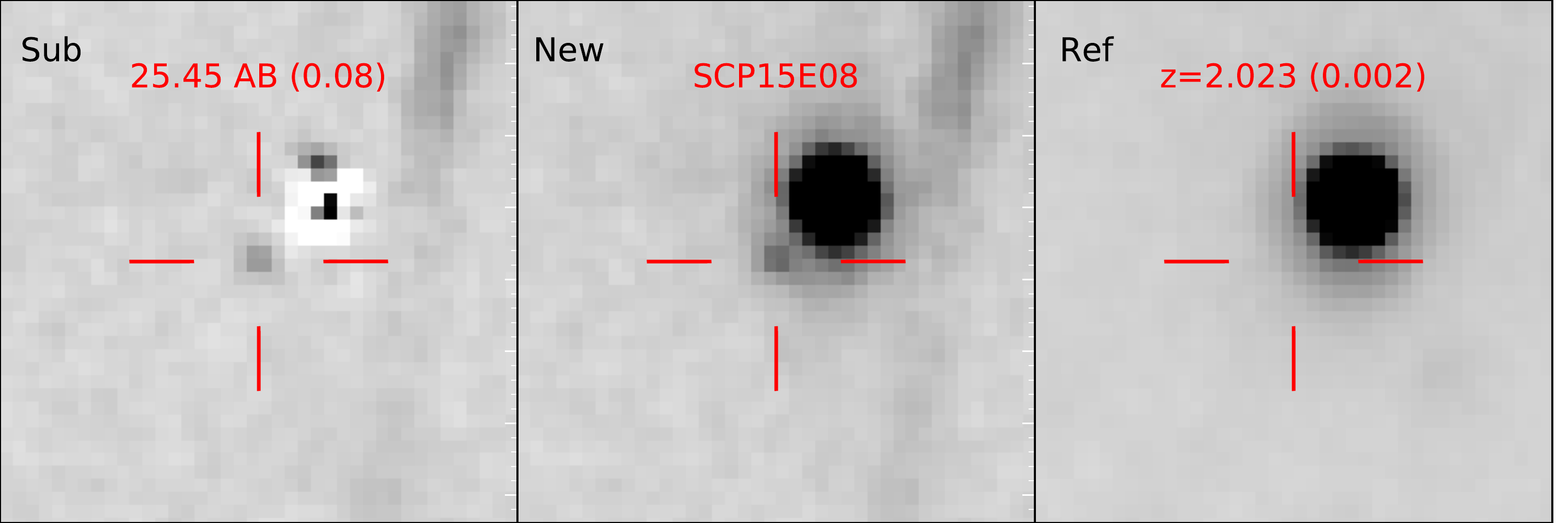}
\includegraphics[width=0.32 \textwidth]{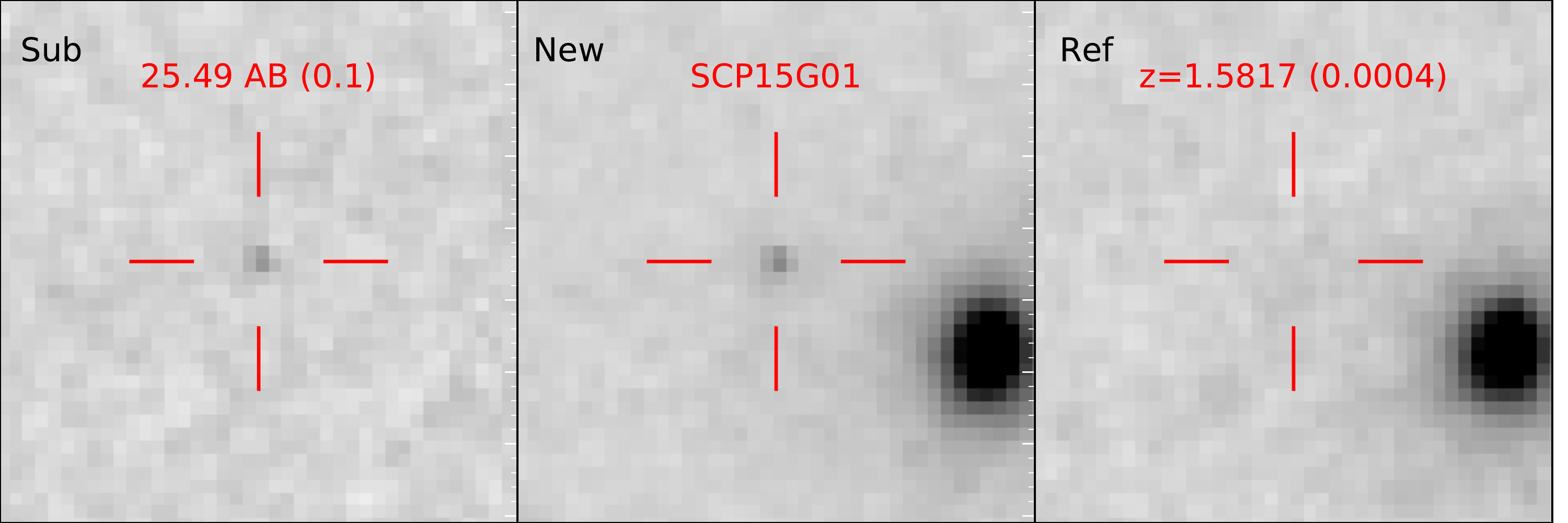}

\includegraphics[width=0.32 \textwidth]{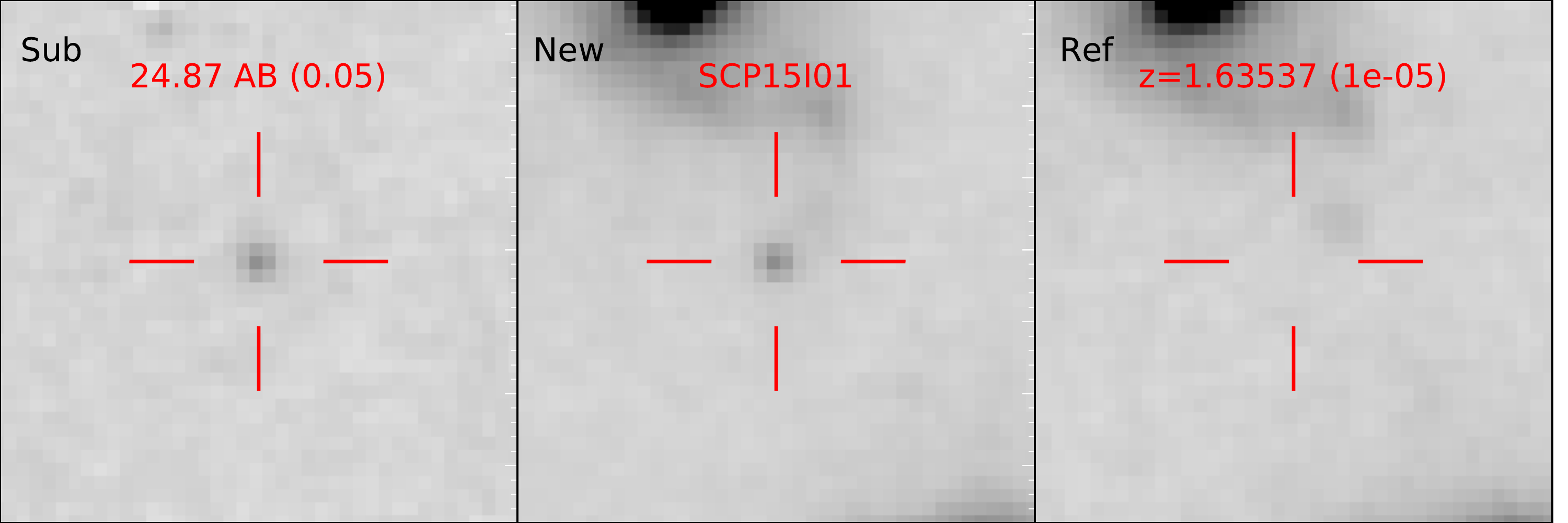}
\includegraphics[width=0.32 \textwidth]{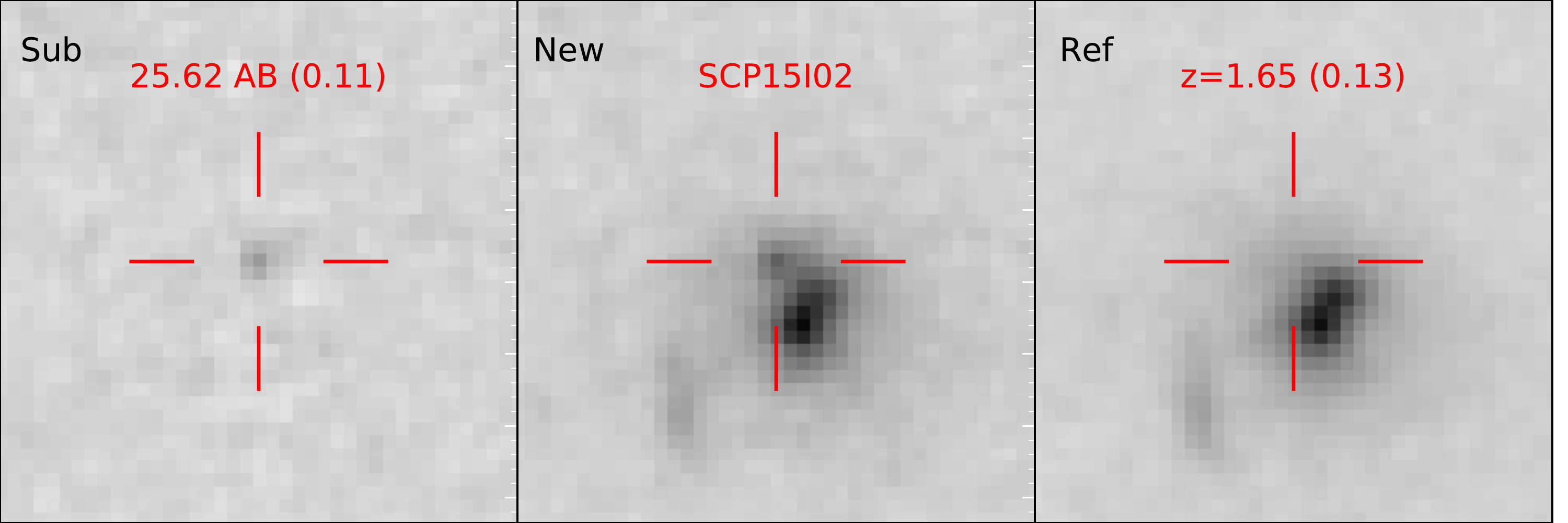}
\includegraphics[width=0.32 \textwidth]{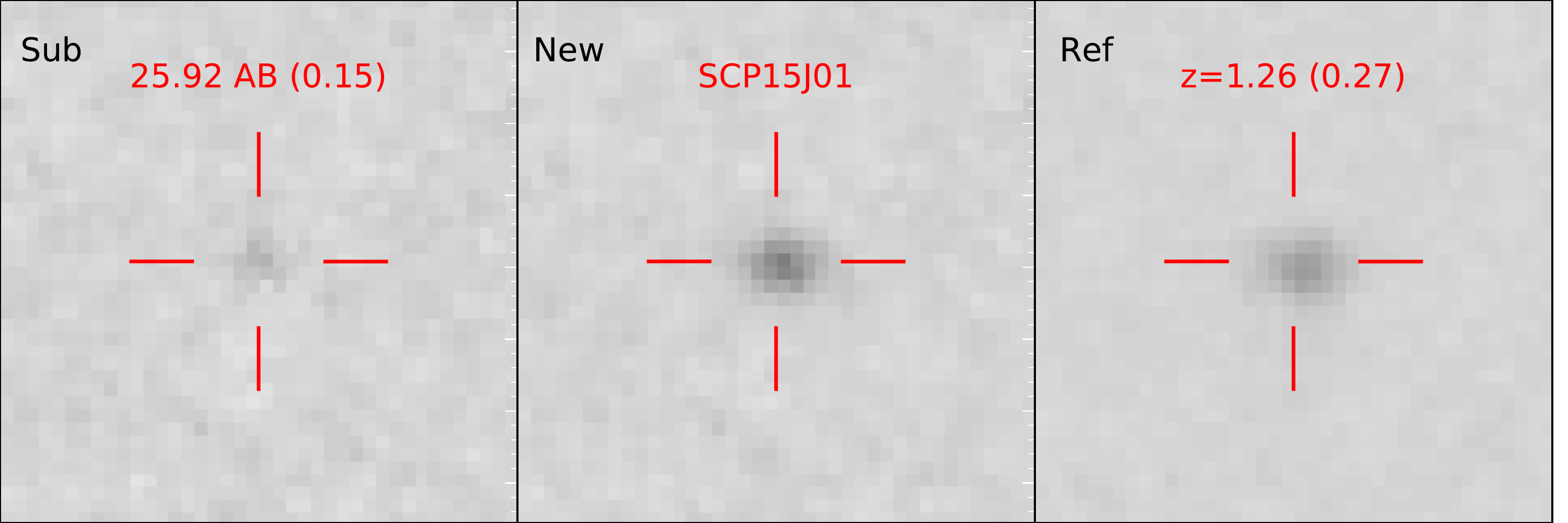}

\includegraphics[width=0.32 \textwidth]{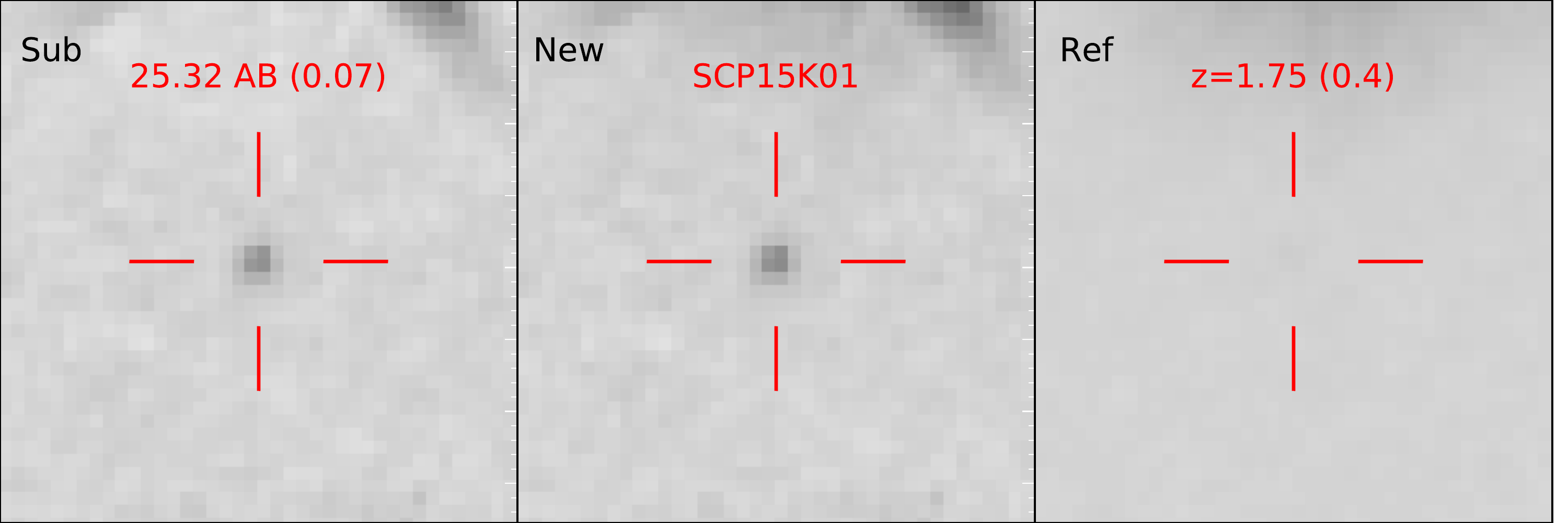}
\includegraphics[width=0.32 \textwidth]{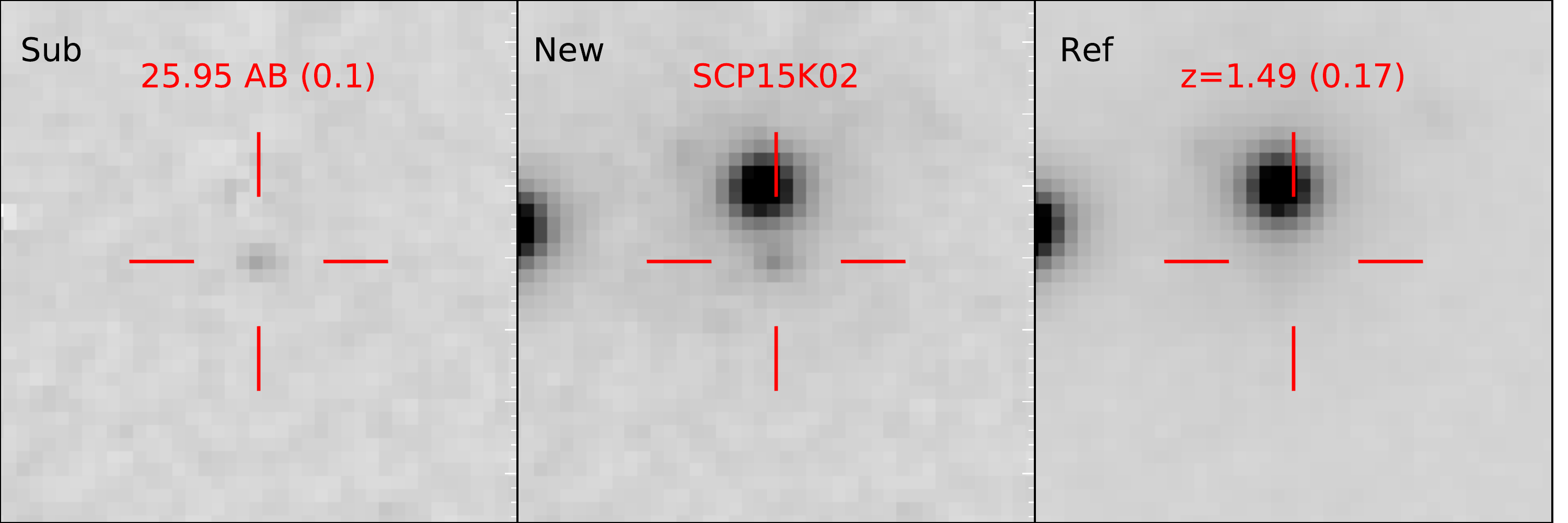}
\includegraphics[width=0.32 \textwidth]{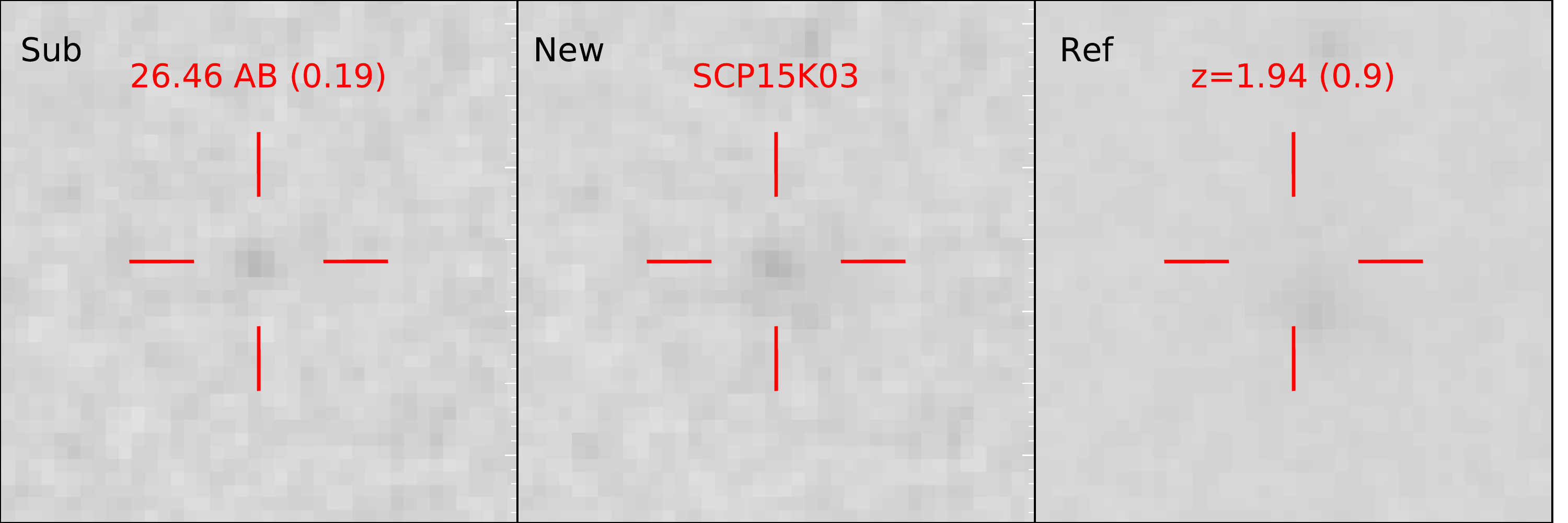}

\end{centering}
\end{figure*}

\begin{figure*}[h]
\begin{centering}

\includegraphics[width=0.32 \textwidth]{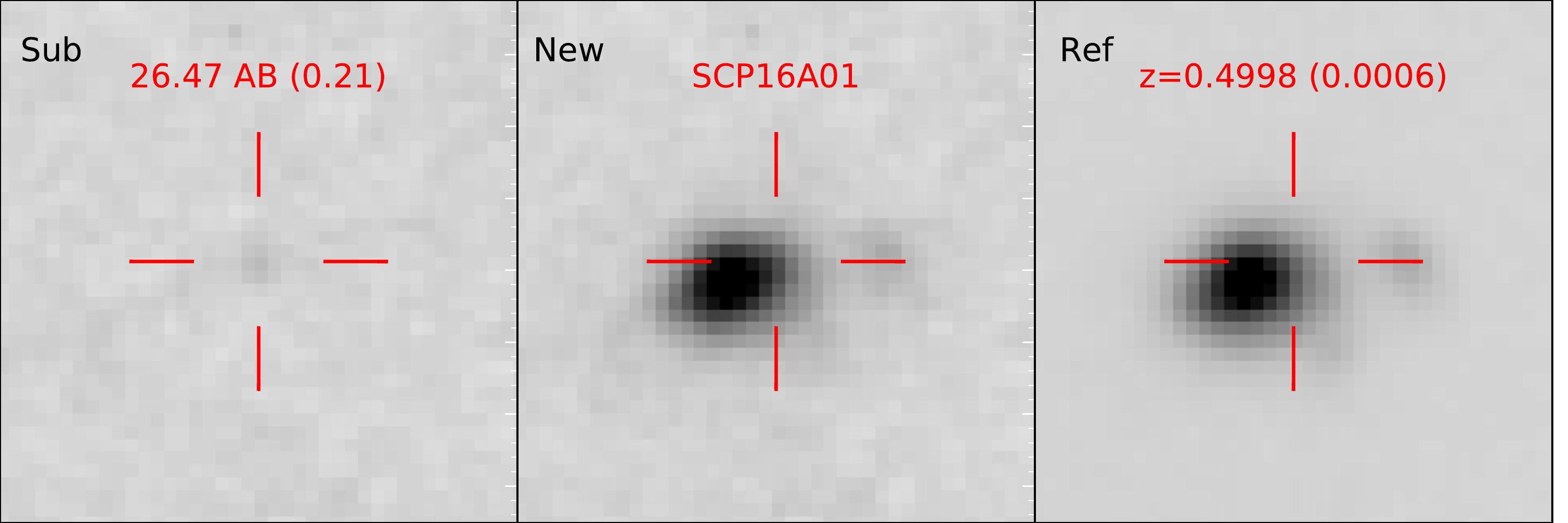}
\includegraphics[width=0.32 \textwidth]{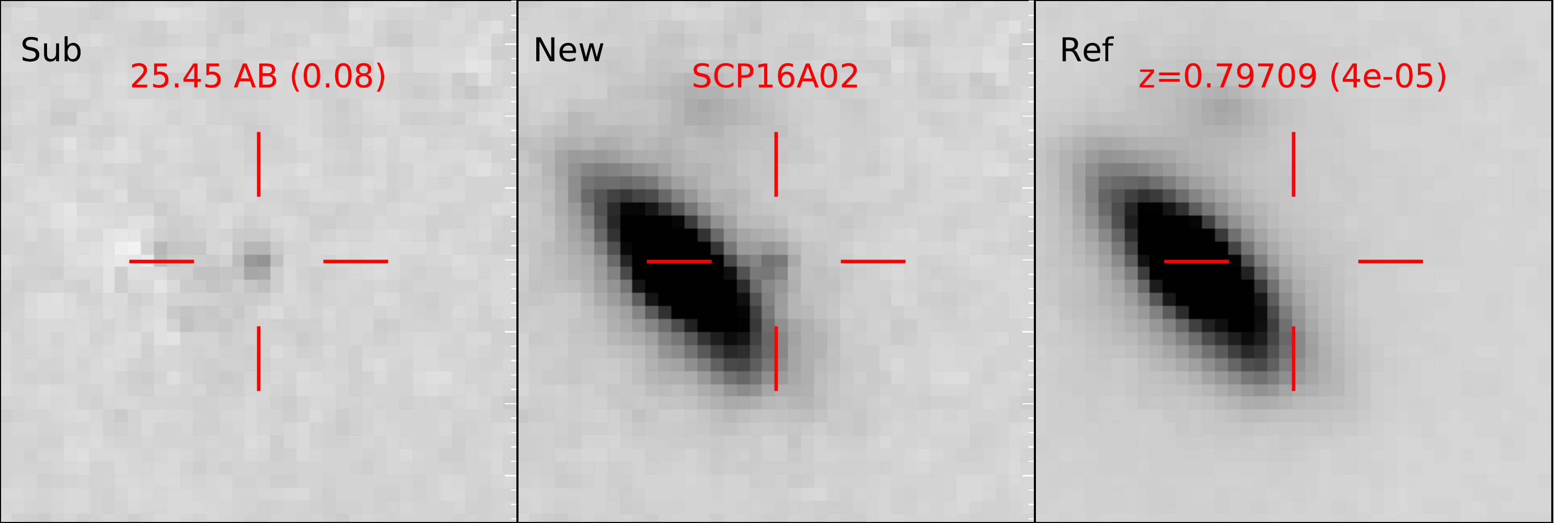}
\includegraphics[width=0.32 \textwidth]{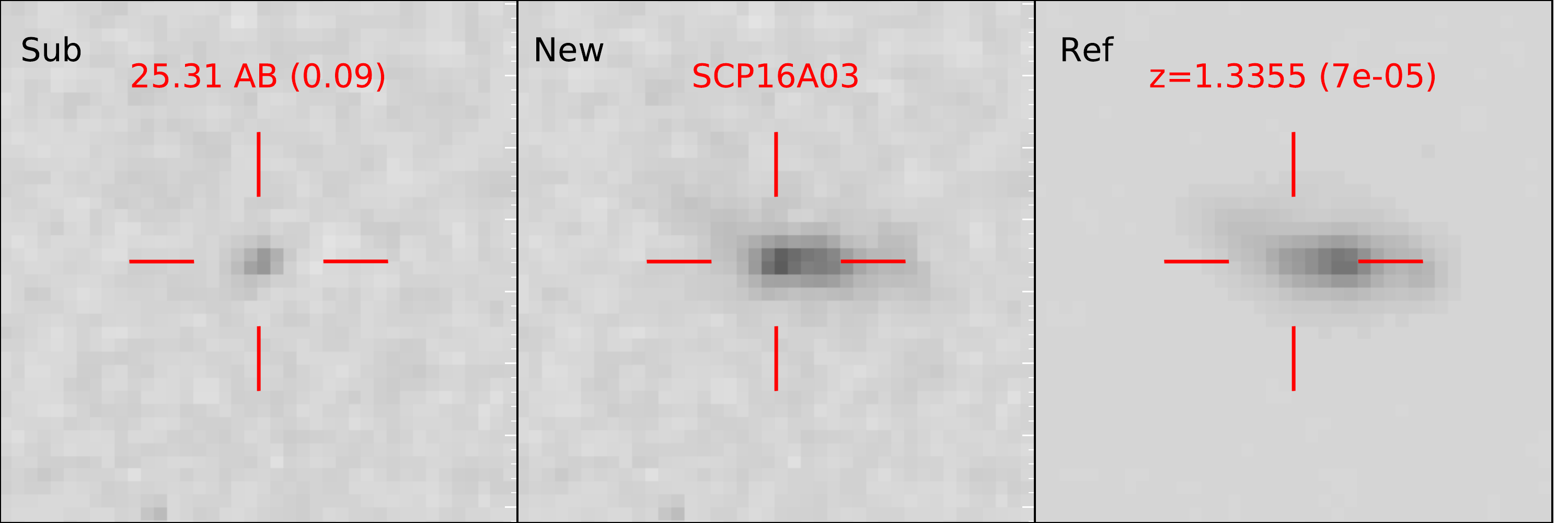}

\includegraphics[width=0.32 \textwidth]{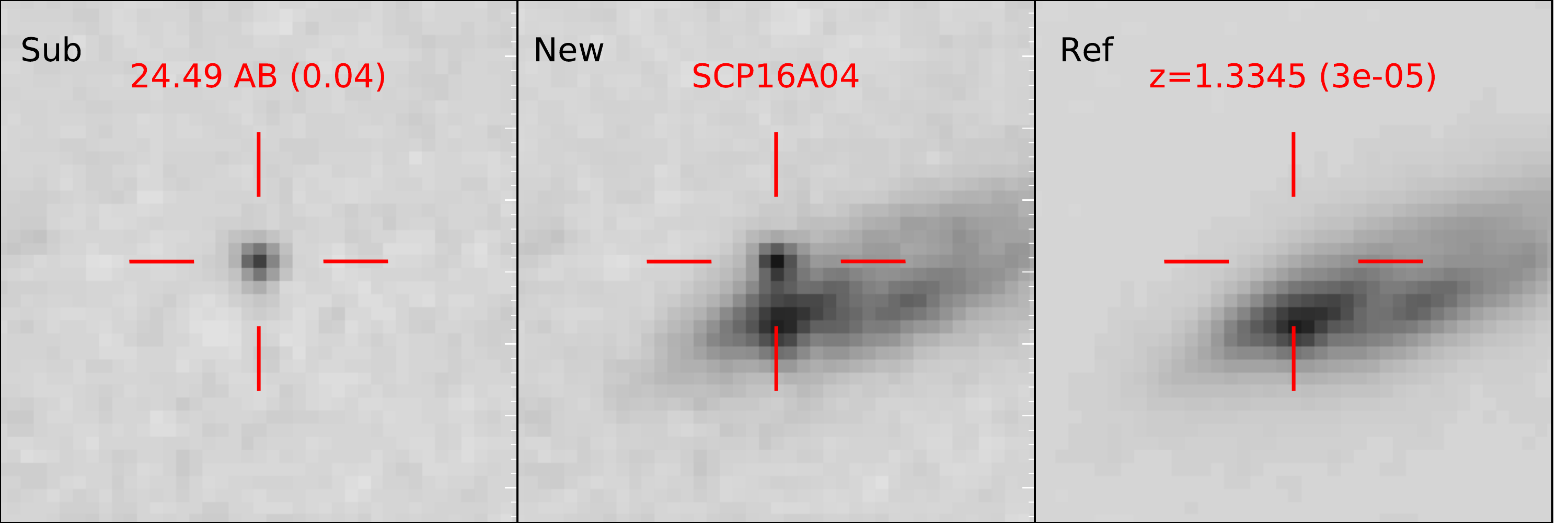}
\includegraphics[width=0.32 \textwidth]{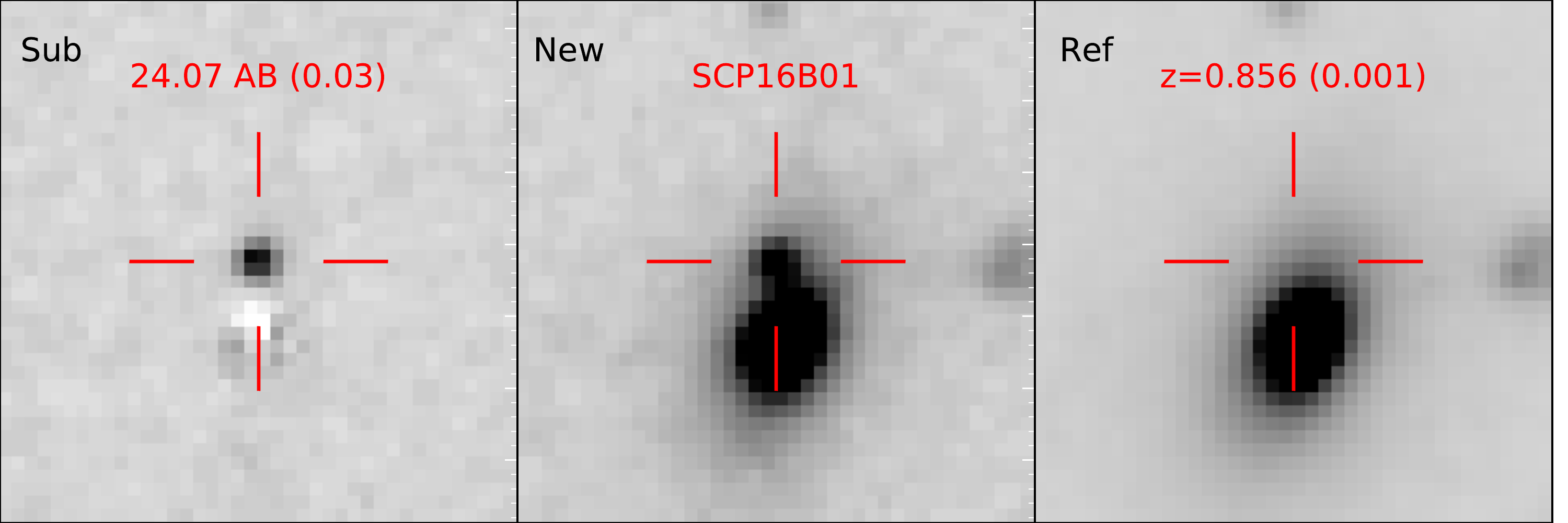}
\includegraphics[width=0.32 \textwidth]{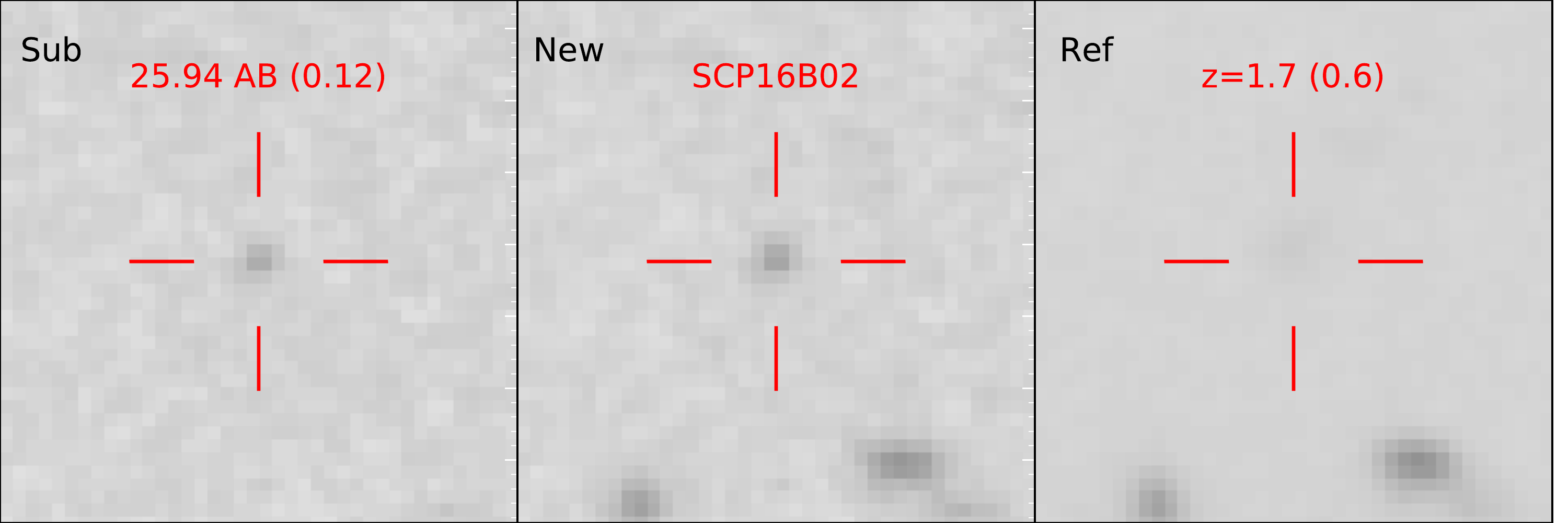}

\includegraphics[width=0.32 \textwidth]{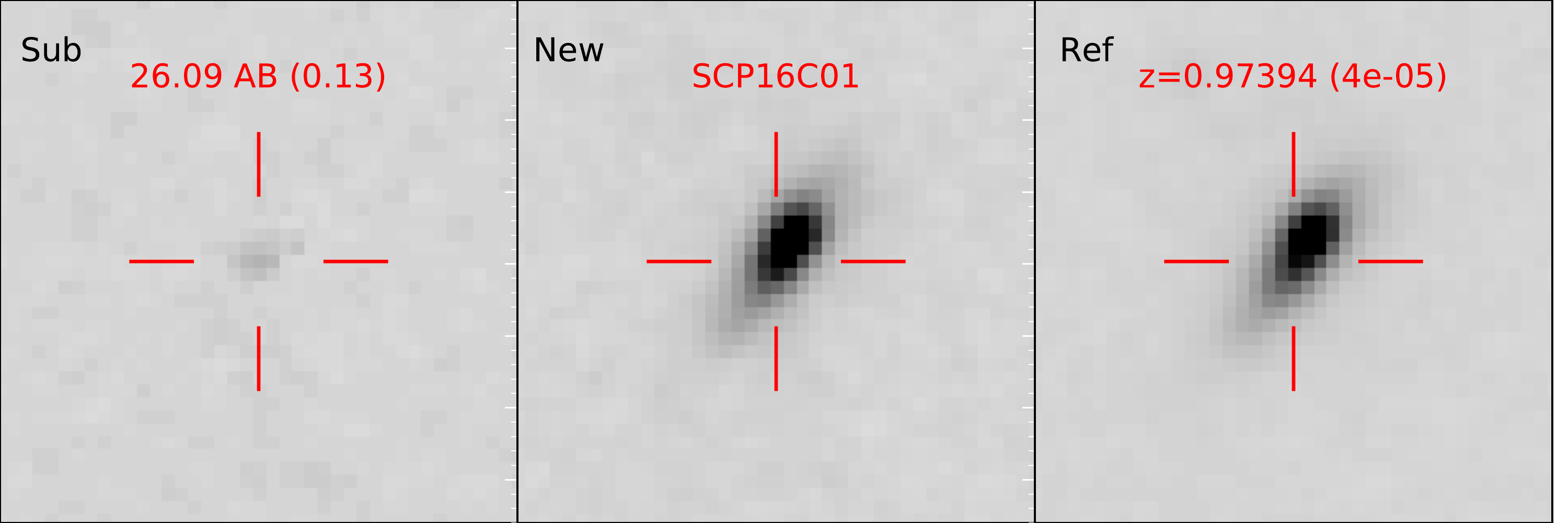}
\includegraphics[width=0.32 \textwidth]{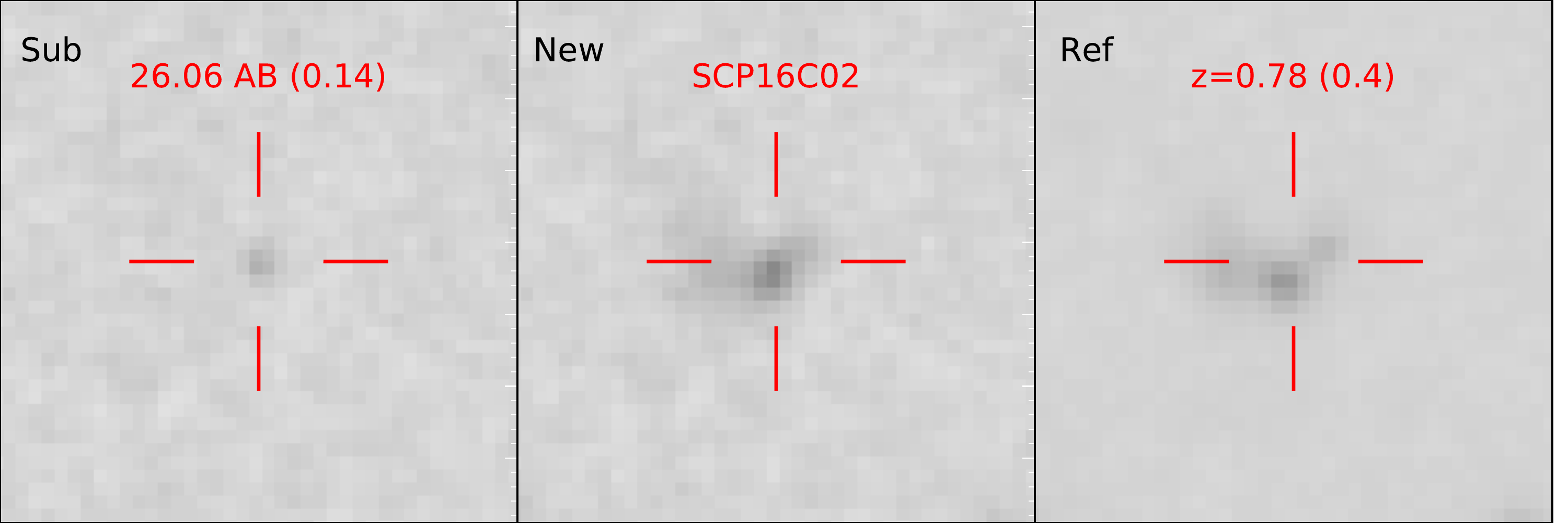}
\includegraphics[width=0.32 \textwidth]{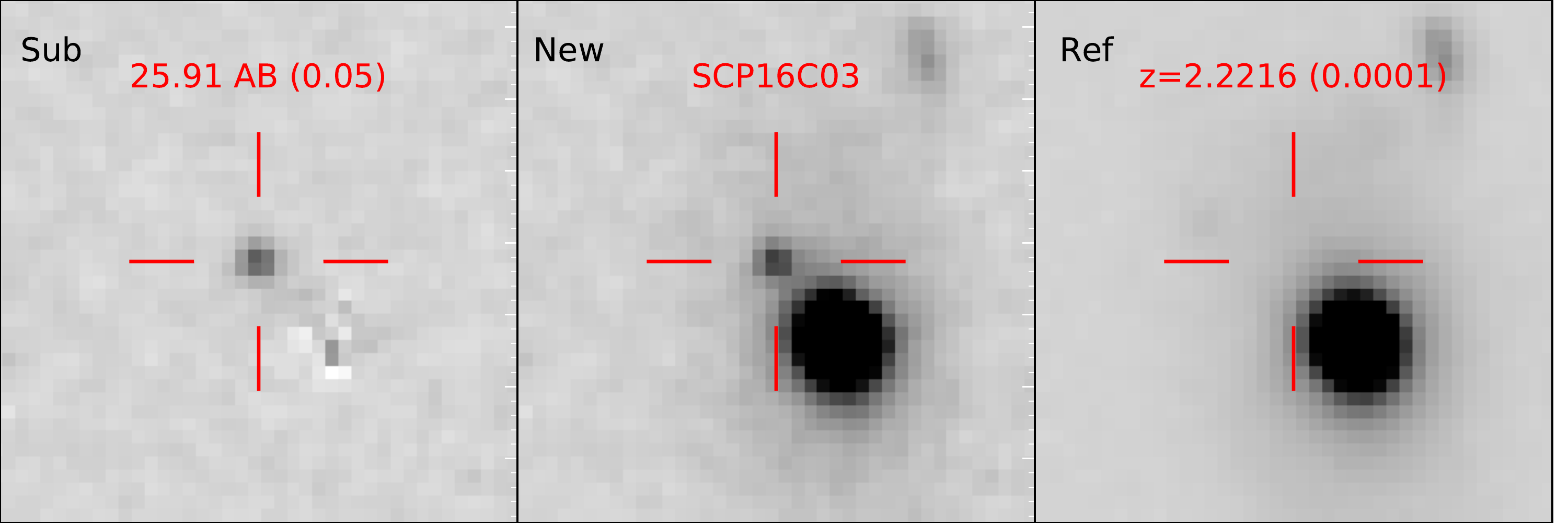}

\includegraphics[width=0.32\textwidth]{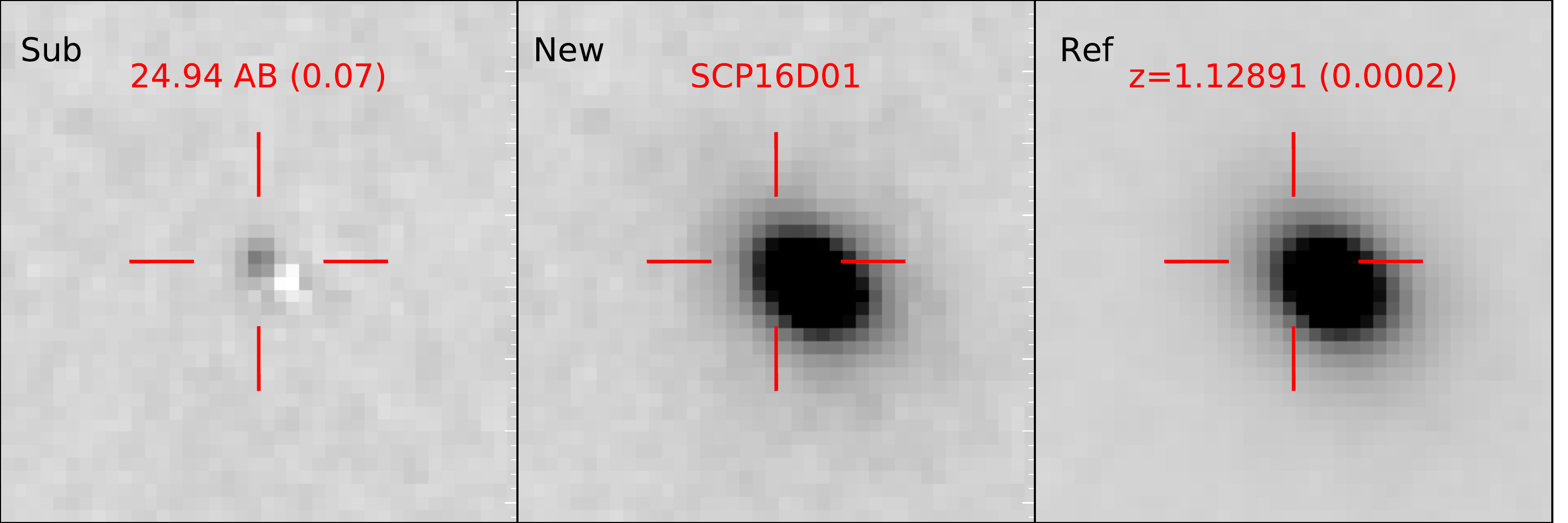}
\includegraphics[width=0.32\textwidth]{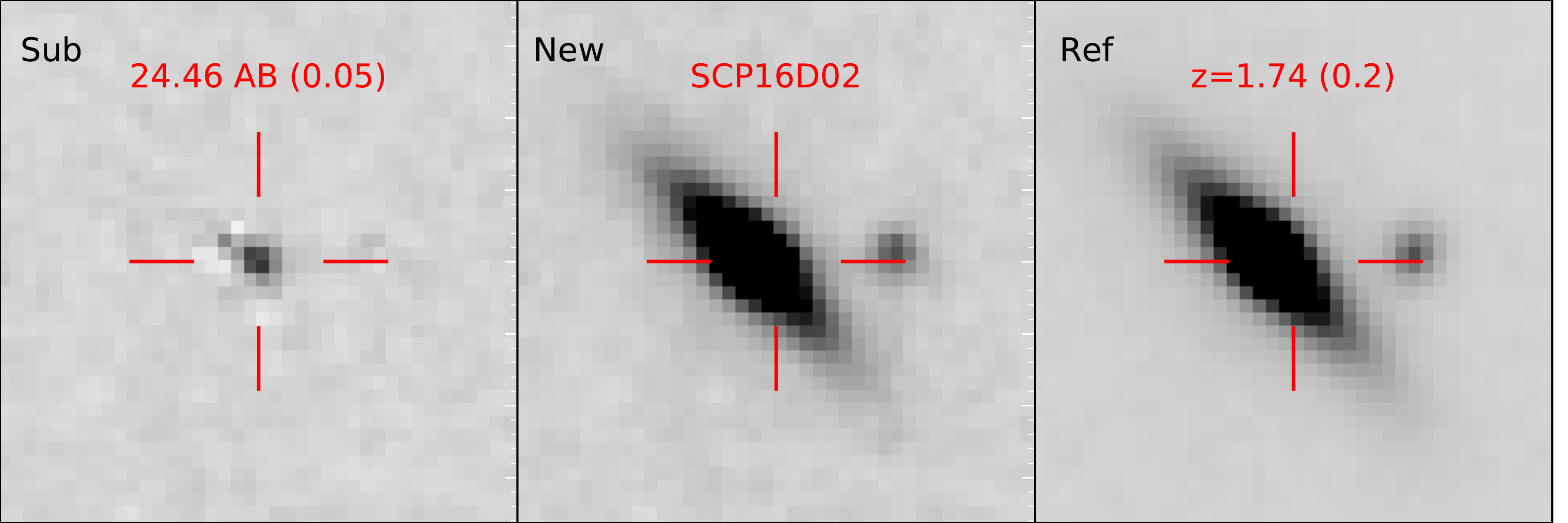}
\includegraphics[width=0.32\textwidth]{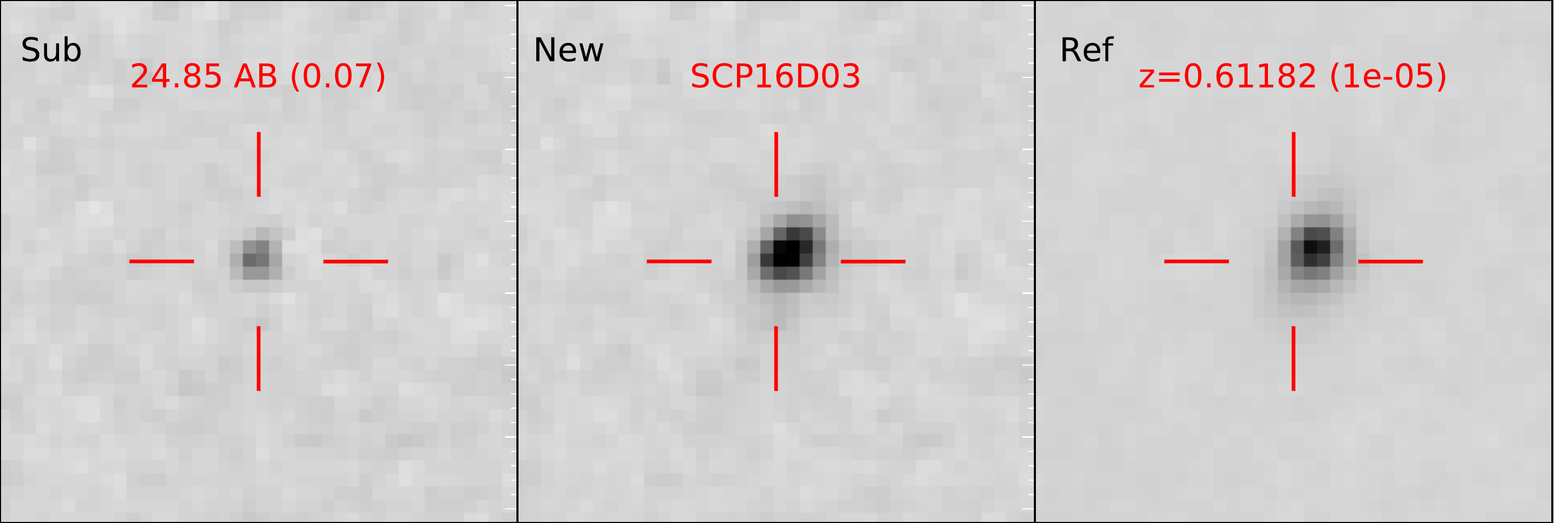}

\includegraphics[width=0.32 \textwidth]{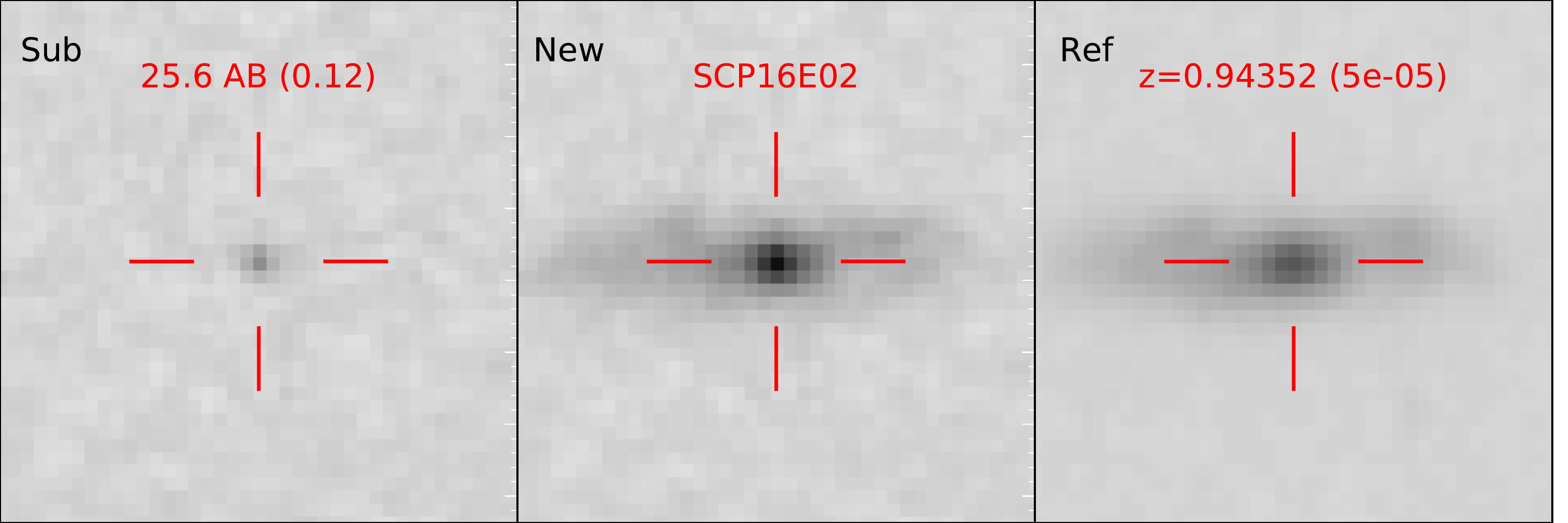}
\includegraphics[width=0.32\textwidth]{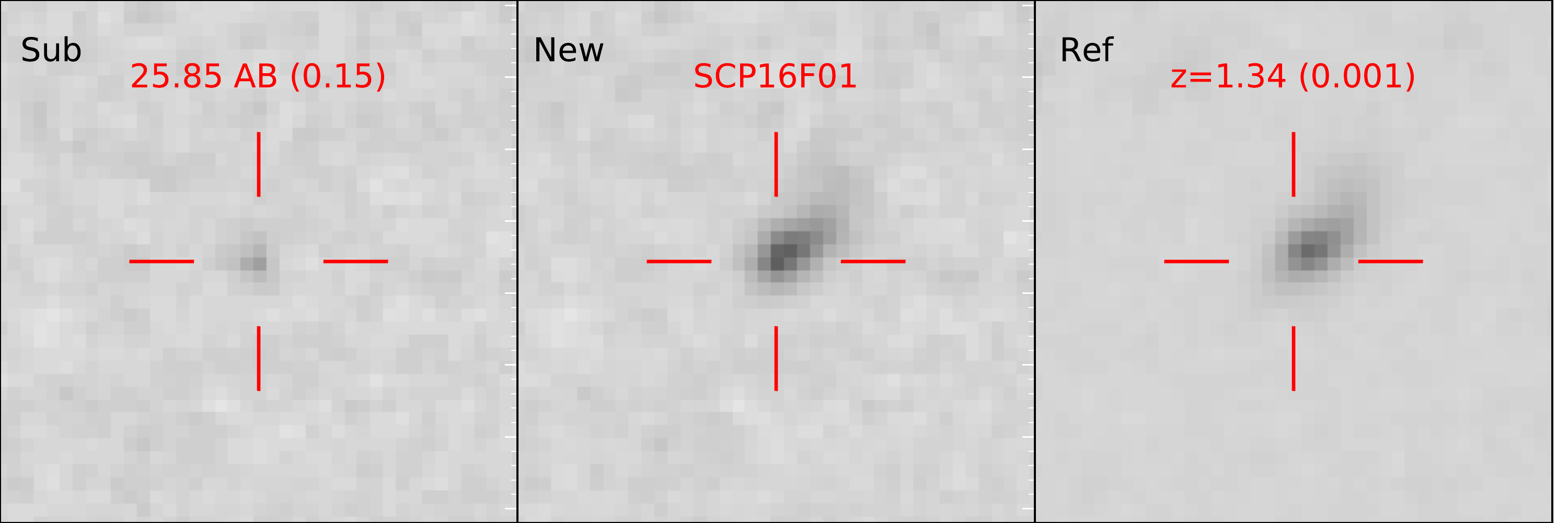}
\includegraphics[width=0.32 \textwidth]{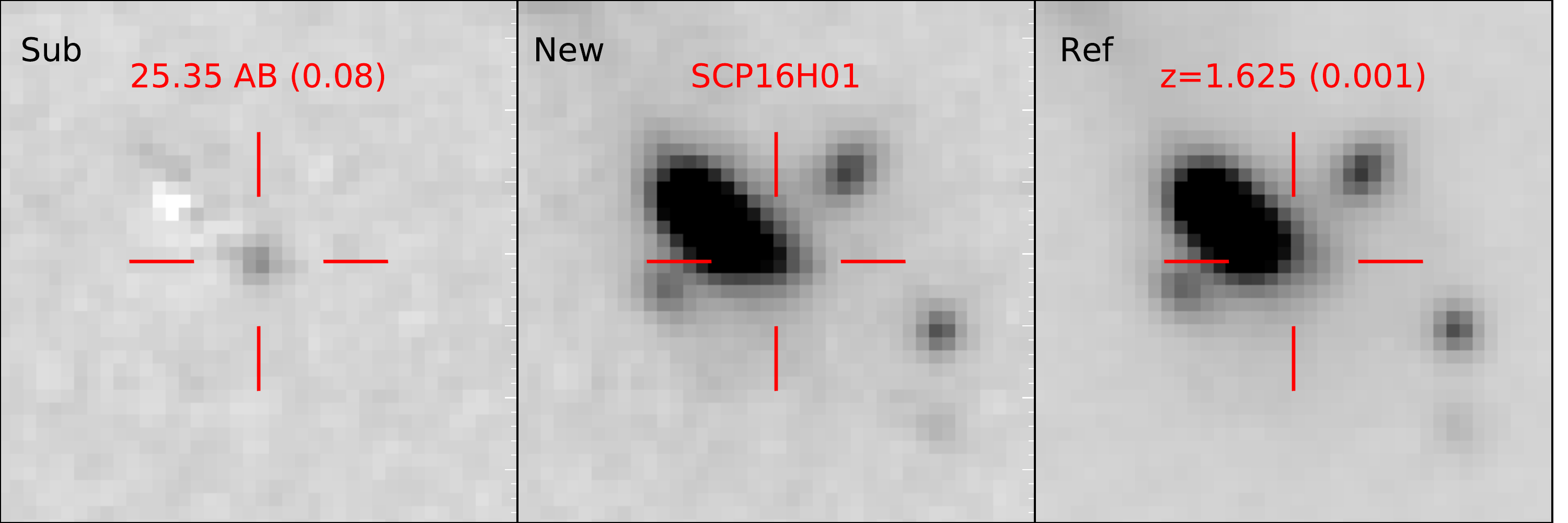}

\includegraphics[width=0.32 \textwidth]{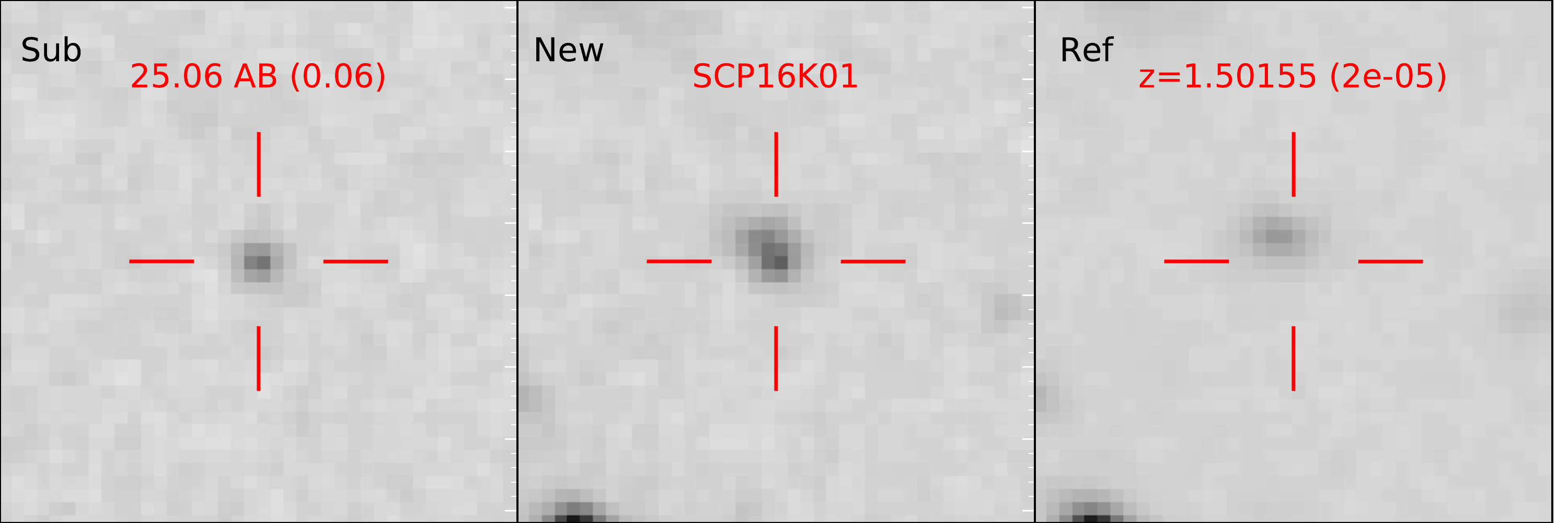}
\includegraphics[width=0.32 \textwidth]{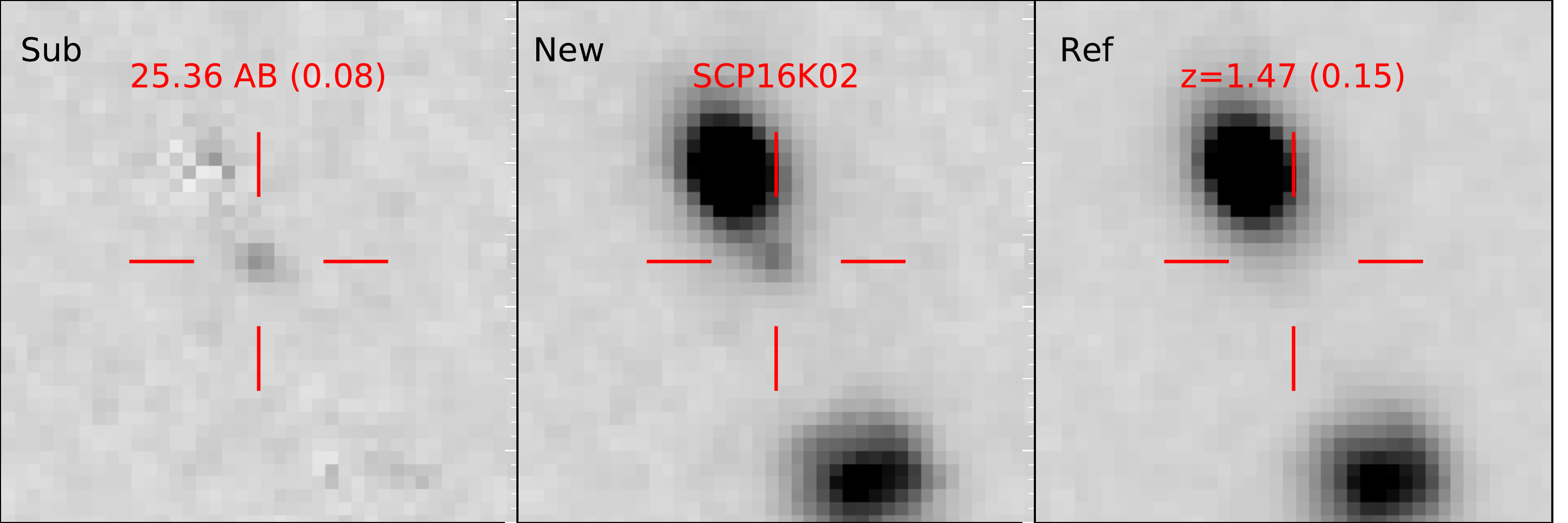}
\includegraphics[width=0.32 \textwidth]{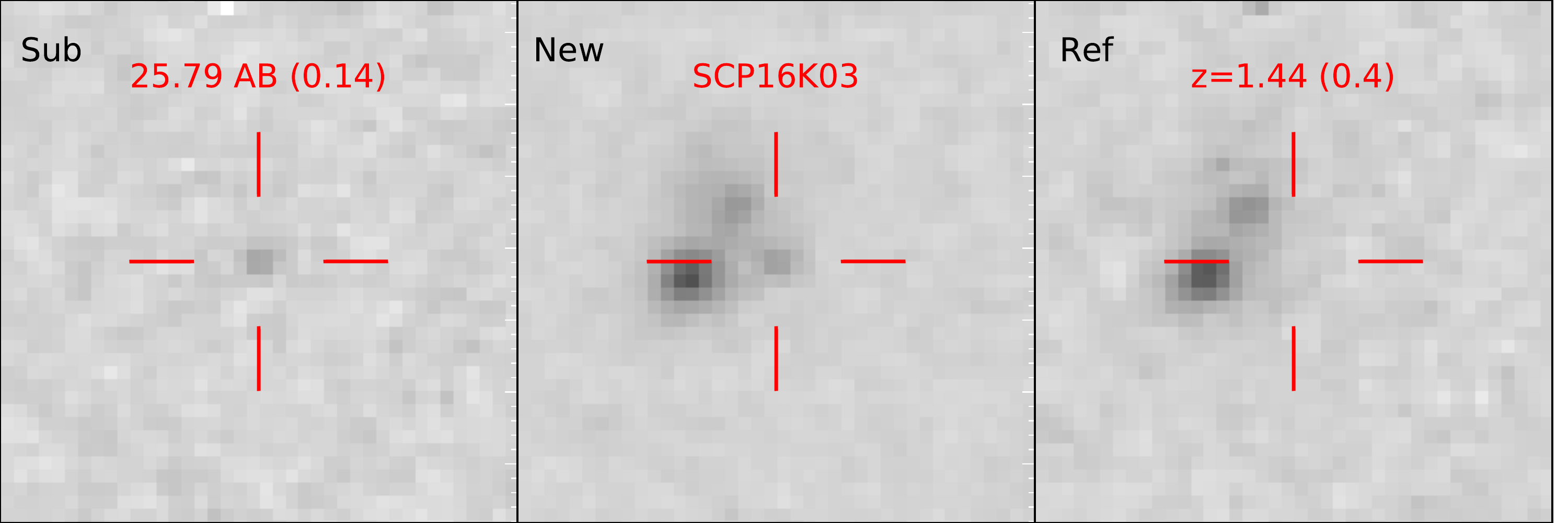}

\includegraphics[width=0.32 \textwidth]{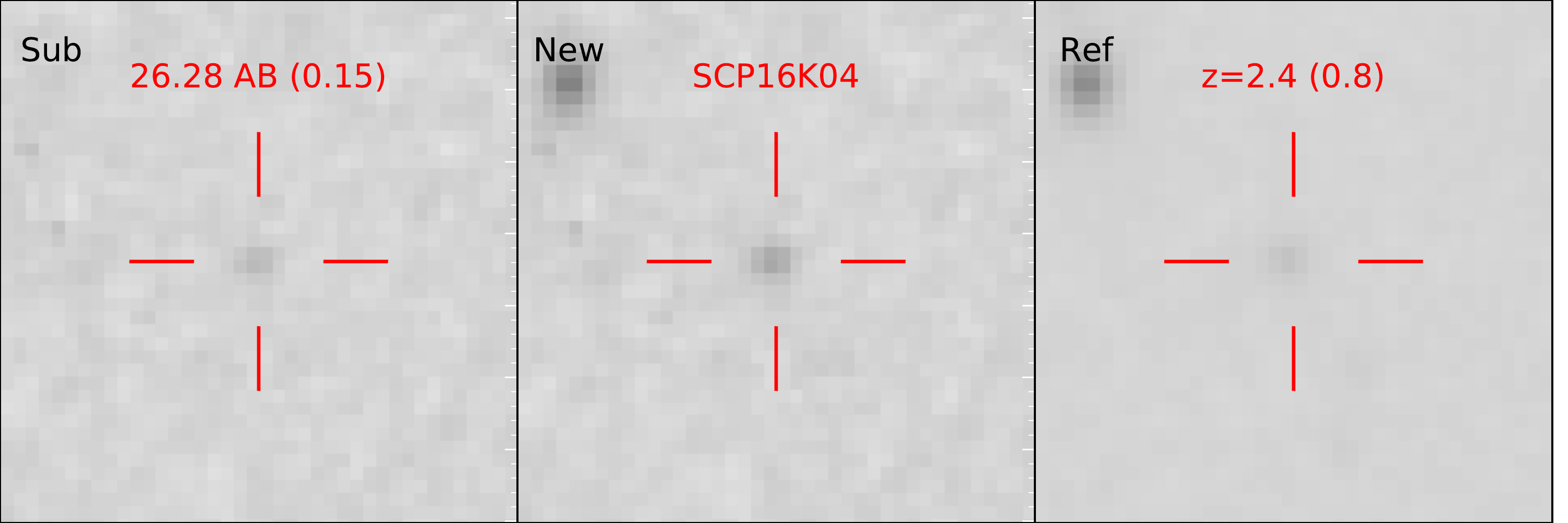}
\includegraphics[width=0.32 \textwidth]{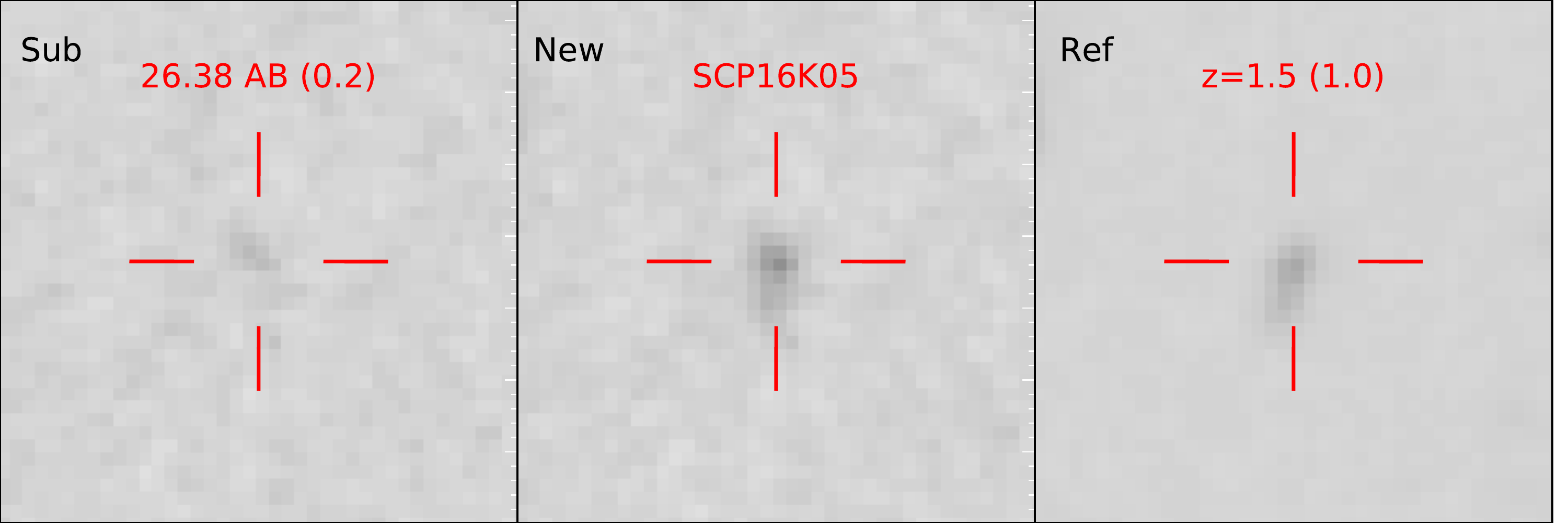}
\includegraphics[width=0.32 \textwidth]{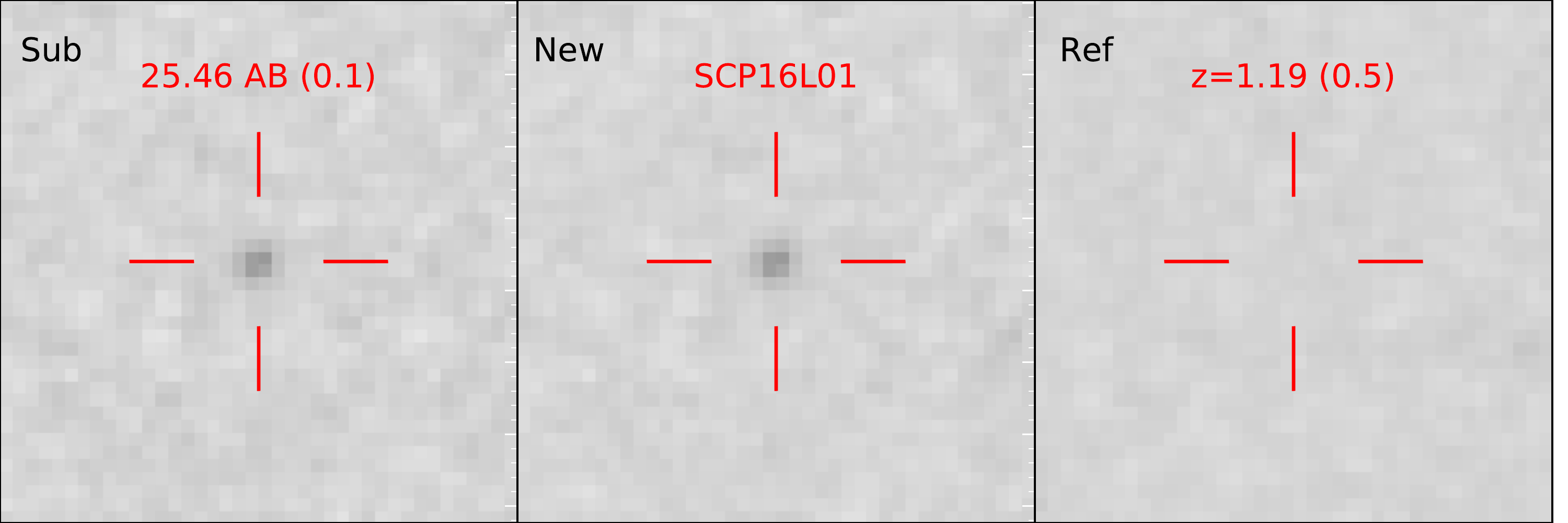}

\caption{Postage stamp ($3\farcs6$ by $3\farcs6$) images for each SN. Each set of three panels shows a subtraction (for the epoch with the maximum SN flux), the new image, and a reference. We also provide the AB magnitude at discovery (with uncertainty in parentheses), and the redshift (again with uncertainty in parentheses). This figure is based on image resampling and subtraction, but we use the images without subtracting the reference to compute SN fluxes (scene-modeling), so the artifacts visible for some galaxies do not affect photometry.}
\label{fig:postagestamps}
\end{centering}
\end{figure*}

\section{The Supernova Sample}
\label{sec:sn_sample}
\begin{figure*}[h]
\begin{centering}
\includegraphics[width=0.98 \textwidth]{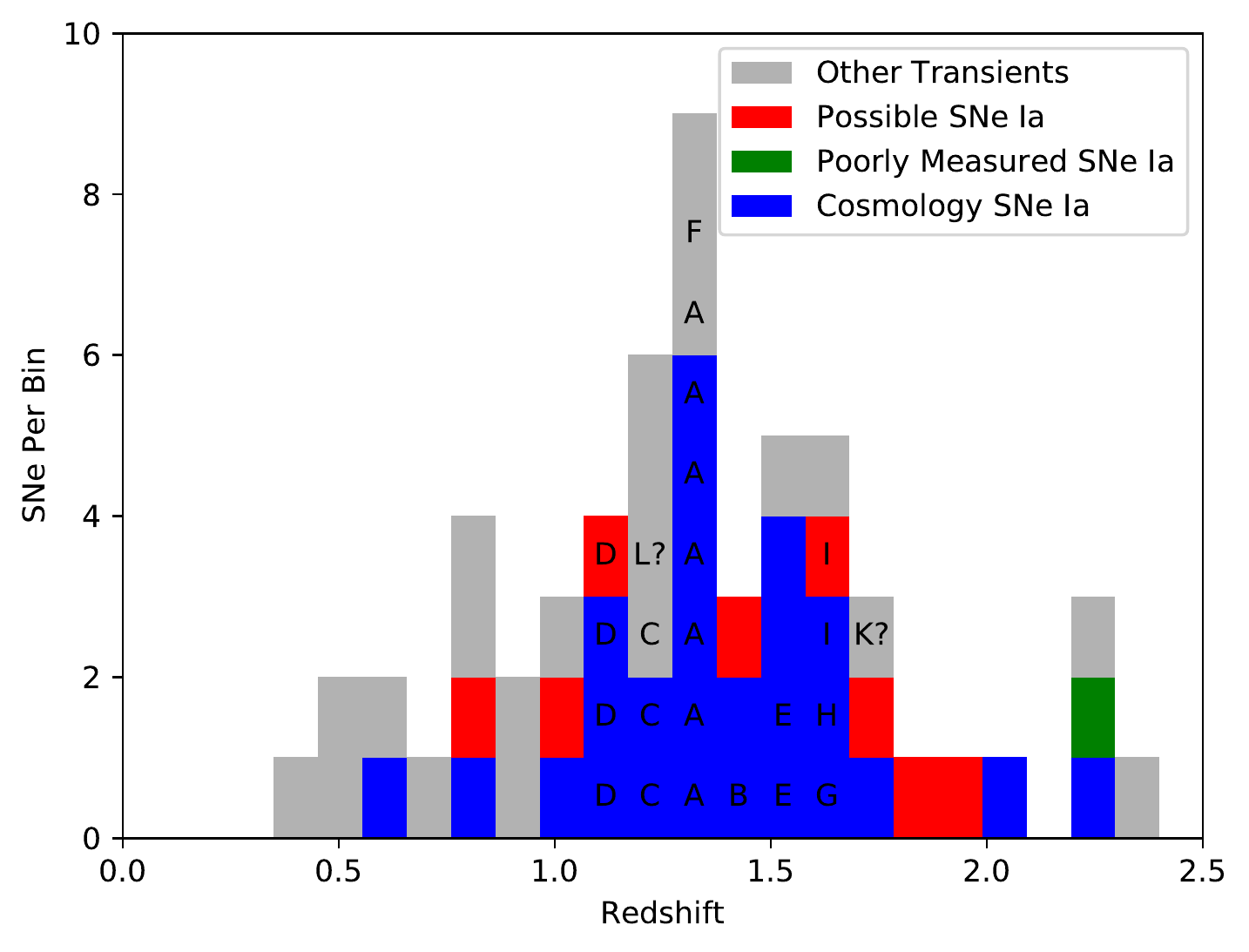}
    \caption{The redshift distribution of transients discovered in See Change. We color according to thresholds defined in Section~\ref{sec:sn_sample}. Blue bin coloring indicates a \gold or \silver in the likely cosmology sample (distance modulus uncertainty $< 0.35$ magnitude, although most cosmology SNe are much better measured than this), green is for \gold or \silver with poorly measured distances, red is for \bronze (with any measurement quality), and gray is for other transients. For SNe whose redshifts match the galaxy cluster, the cluster abbreviation from Table~\ref{tab:clusters} is shown in black text. Two transients with undetected hosts are shown with a question mark next to the cluster. As expected, massive galaxy clusters are an efficient way to discover high-redshift SNe~Ia with the small FoV of \HST, with one of the clusters by itself yielding 6 of the SNe Ia.}
\label{fig:zdist}
\end{centering}
\end{figure*}

In the See Change \HST survey, we identified \ntransients objects that we classified as real transients. SN-live (``new''), reference, and subtraction images for each candidate are shown in Figure~\ref{fig:postagestamps}, and the redshift distribution and cluster membership (of the SNe Ia) are shown in Figure~\ref{fig:zdist}. Details for each transient, including redshifts and a preliminary SN~Ia classification, are presented in Table~\ref{tab:SNe}. The SN type in Table~\ref{tab:SNe} is a mapping of the SN~Ia probability as estimated from the combined probabilities given by the light-curve classification procedure in \citet{rubin18}, an \textit{a posteriori} version of the single-epoch typing (using light curve fit restraints and running on the fluxes near peak magnitude) presented in Section~\ref{sec:single_epoch_typing}, and our best knowledge of the redshift. A ``\gold'' has $\mathrm{P(Ia)} \geq 0.9$, a ``\silver'' has $0.75 \leq \mathrm{P(Ia) <0.9}$, a ``\bronze'' has $0.25 \leq \mathrm{P(Ia)<0.75}$, and a \pyrite has $\mathrm{P(Ia)<0.25}$. Since the cosmology analysis will be performed with the Unified Nonlinear Inference for Type~Ia cosmologY (UNITY) framework \citep{rubin15} which builds a robust outlier model, a sample including the ``\gold'' and ``\silver'' SNe is the starting point for the cosmology sample. Not including the lensed SN of \citet{rubin18}, SCP16C03, this cosmology sample consists of $\ncosmo$ SNe at redshift $0.86<z<2.29$. The best measured SNe have distance modulus uncertainties around $\sigma_{\mu} \sim0.07$~mag, and the average SN for our sample has $\sigma_{\mu} \sim0.2$~mag. Of this sample of \ncosmo, \ncosmospec have host-galaxy-based spectroscopic redshifts, and the remaining \ncosmonospec have reasonably well-measured photo-$z$'s with uncertainties on redshift in the range of 0.1--0.2. The See Change SN Ia sample is therefore the largest homogeneous and cosmologically useful sample at $z>1$ to date. 

Table~\ref{tab:lensing} shows our estimates of lensing amplification $\mu$ (in magnitudes) for the SNe behind our galaxy clusters, computed in the same way as the analysis of SCP16C03 by \citet{rubin18}. In short, we sample over mass and redshift (assuming Gaussian uncertainties), assume a Navarro-Frenk-White \citep{nfw96} surface mass density profile, and compute the amplification for each mass/redshift realization. We quote a value and uncertainties based on the 16th, 50th, and 84th percentile of these simulated amplifications. Most of the significantly lensed SNe are not well measured or not SNe~Ia; the exceptions are SCP16C03 (discussed by \citealt{rubin18}) and (possibly) SCP16D02. Due to the small magnifications over most of the WFC3 FoV, and the compensating effect of lensing decreasing the volume \citep{sullivan00}, it is unlikely that the number of discovered transients was significantly modified due to lensing \citep[e.g., ][]{barbary12}.

\section{Discussion and Conclusions}
\label{sec:conclusion}
We present the ``See Change'' SN survey, using \HST WFC3-IR and UVIS to efficiently discover and characterize \nIa SNe~Ia at $z \gtrsim 1$. This strategy produced \SnePerOrbit SNe Ia per \HST orbit, a factor of $\sim 8$ higher than a field search for SNe Ia (see Figure~\ref{fig:snRates}). \orbitPercent of our orbits contained an active high-redshift SN~Ia within 22 rest-frame days of peak brightness, while the other orbits provide reference images of the host galaxy with sufficient signal-to-noise to subtract without signal degradation. This is a highly efficient use of orbits when the search itself was also performed with \HST. This result establishes the feasibility of the higher-efficiency space-based SN Ia search using this strategy of targeting clusters at very high redshift.

Many of the observation, reduction, and analysis elements that we introduced in this survey are likely to be useful in future space-based supernova searches. We selected clusters for observation based on the most massive clusters available at $z > 1$, accounting for \HST visibility windows that allowed for cadenced observations. We used a unique observing strategy to maximize our ability to discover transients, with dither patterns designed to minimize the impact of the WFC3-IR ``blobs'' and maximize signal-to-noise in subtraction images combined across $F105W$ and $F140W$. We also took special care to place the $F105W$ exposures in each orbit to minimize the impact of Helium emission from the Earth's limb, which can cause elevated background up to $5\times$ the zodiacal levels \citep{brammer14}. We developed a random forest classifier for the undersampled \HST imaging, and present the features in the model that were most important in assessing the likelihood of each detection in the image being a real SN. We found that our classifier was nearly indistinguishable from human searchers down to a false positive rate of $\sim 0.5\%$ with an $\sim 80\%$ true positive rate. This will be important for future undersampled space-based transient surveys such as \textit{Roman}, where human review of every candidate will be unfeasible. We monitored our detection efficiency in real-time by planting simulated SNe in each epoch using the WFC3-IR PSF, finding a 50\% detection efficiency at around 26.6 AB mag. We also estimated the lensing magnification of all SNe that were potentially behind each observed galaxy cluster; when combined with our detection efficiency curve, we found it unlikely that lensing significantly enhanced our number of detected transients.

The SN light curves and cosmology are presented in more detail in the companion cosmology paper \citep{companionpaper}, which will use the updated UNITY framework to integrate the See Change sample into the world sample of SNe~Ia. Based on photometric typing including light curve shape, color, and magnitude, we will add \ncosmo SNe~Ia to the Hubble diagram at $z>1$, significantly reducing the uncertainty on the dark energy density at $z>1$ from SN~Ia measurements.

\acknowledgments

Based on \HST observations under programs \PIDList. The primary search and followup observations are available from the STScI MAST Portal under DOI 10.17909/t9-z53g-zn60. The authors wish to recognize and acknowledge the very significant cultural role and reverence that the summit of Maunakea has always had within the indigenous Hawaiian community.  We are most fortunate to have the opportunity to conduct observations from this mountain. We thank the Keck Observatory technical staff for their support, especially for machining our slit masks on short notice. Some of the data presented herein were obtained at the W. M. Keck Observatory, which is operated as a scientific partnership among the California Institute of Technology, the University of California and the National Aeronautics and Space Administration. The Observatory was made possible by the generous financial support of the W. M. Keck Foundation. Based in part on observations made with the Gran Telescopio Canarias (GTC), installed at the Spanish Observatorio del Roque de los Muchachos of the Instituto de Astrofísica de Canarias, in the island of La Palma. Based in part on observations obtained at the Gemini Observatory, which is operated by the Association of Universities for Research in Astronomy, Inc., under a cooperative agreement with the NSF on behalf of the Gemini partnership: the National Science Foundation (United States), National Research Council (Canada), CONICYT (Chile), Ministerio de Ciencia, Tecnolog\'{i}a e Innovaci\'{o}n Productiva (Argentina), Minist\'{e}rio da Ci\^{e}ncia, Tecnologia e Inova\c{c}\~{a}o (Brazil), and Korea Astronomy and Space Science Institute (Republic of Korea). Based on observations collected at the European Organisation for Astronomical Research in the Southern Hemisphere under ESO programme(s) 294.A-5025(A), 095.A-0830(A, B, C), 096.A-0926(B, C), 097.A-0442(A, B, C) and 0100.A-0851(A). GW acknowledges support from the National Science Foundation through grant AST-1517863, by HST program numbers GO-13677/14327.01 and GO-15294, and by grant number 80NSSC17K0019 issued through the NASA Astrophysics Data Analysis Program (ADAP). GA and RG acknowledge support from HST program NASA HST-GO-14163.002-A, and by grant number NNH16AC25I issued through NASA ADAP. M. J. Jee acknowledges support for the current research from the National Research Foundation (NRF) of Korea under the programs 2017R1A2B2004644 and 2020R1A4A2002885. H. Hildebrandt is supported by a Heisenberg grant of the Deutsche Forschungsgemeinschaft (Hi 1495/5-1) as well as an ERC Consolidator Grant (No. 770935). Support for program numbers GO-13677/14327.01 and GO-15294 was provided by NASA through a grant from the Space Telescope Science Institute, which is operated by the Association of Universities for Research in Astronomy, Incorporated, under NASA contract NAS5-26555. This work was also partially supported by the Office of Science, Office of High Energy Physics, of the U.S. Department of Energy, under contract no. DE-AC02-05CH11231.

\facilities{\Hubble, Keck:I (LRIS), Keck:I (MOSFIRE), VLT, Gemini, GTC, Subaru}

\software{
astropy \citep{Astropy},
DrizzlePac \citep{gonzaga12},
iPython \citep{ipython},
Matplotlib \citep{matplotlib}, 
Numpy \citep{numpy}, 
SEP \citep{sep:paper},
scikit-learn \citep{scikit-learn}, 
SciPy \citep{scipy}
}

\startlongtable
\begin{deluxetable}{cccccccc}
\tablecaption{Transients discovered in the See Change SN Survey.\label{tab:SNe}}
\tablehead{\colhead{SN} & \colhead{RA} & \colhead{Dec} & \colhead{Discovery Epoch (UTC)} & \colhead{Host $z$} & \colhead{$z$source} & \colhead{Type} & \colhead{P$_\mathrm{Ia}$}}
\startdata
SCP15A01 & 02:05:46.346 & $-$58:28:54.943 & 2015-01-18 08:10 & $1.3 \pm 1.0$ & photo & \gold & 100\% \\
SCP15A02 & 02:05:45.672 & $-$58:28:16.949 & 2015-02-22 08:46 & $0.89662 \pm 0.00001$ & spec\tablenotemark{a} & \pyrite & 0\% \\
SCP15A03 & 02:05:44.381 & $-$58:30:01.901 & 2015-03-27 14:48 & $1.3395 \pm 0.0003$ & spec\tablenotemark{a} & \gold & 100\% \\
SCP15A04 & 02:05:51.130 & $-$58:29:27.384 & 2015-06-05 23:50 & $1.33450 \pm 0.00003$ & spec\tablenotemark{a,b} & \gold & 100\% \\
SCP15A05 & 02:05:48.119 & $-$58:29:26.685 & 2015-07-09 18:40 & $1.3227 \pm 0.0002$ & spec\tablenotemark{a,b} & \gold & 99\% \\
SCP15A06 & 02:05:43.092 & $-$58:29:36.728 & 2015-07-09 18:40 & $1.3128 \pm 0.0005$ & spec\tablenotemark{a} & \gold & 100\% \\
SCP15A07 & 02:05:49.295 & $-$58:29:41.019 & 2015-09-16 15:42 & $0.50163 \pm 0.00005$ & spec\tablenotemark{a} & \pyrite & 0\% \\
SCP16A01 & 02:05:44.320 & $-$58:29:06.964 & 2016-05-08 08:41 & $0.4998 \pm 0.0006$ & spec\tablenotemark{a} & \pyrite & 0\% \\
SCP16A02 & 02:05:45.844 & $-$58:27:50.001 & 2016-05-08 08:41 & $0.79709 \pm 0.00004$ & spec\tablenotemark{a} & \bronze & 70\% \\
SCP16A03 & 02:05:52.390 & $-$58:29:40.282 & 2016-06-12 22:19 & $1.3355 \pm 0.0001$ & spec\tablenotemark{a} & \pyrite & 0\% \\
SCP16A04 & 02:05:51.086 & $-$58:29:26.910 & 2016-06-29 00:34 & $1.33450 \pm 0.00003$ & spec\tablenotemark{a,b} & \gold & 100\% \\
SCP15B01 & 14:32:18.103 & +32:52:53.225 & 2015-01-20 21:29 & $1.4 \pm 0.2$ & photo & \gold & 100\% \\
SCP15B02 & 14:32:16.048 & +32:52:48.948 & 2015-07-17 05:55 & $2.29456 \pm 0.00003$ & spec\tablenotemark{c} & \gold & 99\% \\
SCP15B03 & 14:32:16.367 & +32:53:17.440 & 2015-07-17 05:55 & $1.86 \pm 0.50$ & photo & \bronze & 36\% \\
SCP15B04 & 14:32:22.929 & +32:53:01.129 & 2015-07-17 05:55 & $1.68 \pm 1.00$ & photo & \pyrite & 0\% \\
SCP15B05 & 14:32:18.758 & +32:52:48.648 & 2015-12-18 19:01 & $2.238920 \pm 0.000002$ & spec\tablenotemark{c} & \pyrite & 1\% \\
SCP16B01 & 14:32:20.100 & +32:53:52.908 & 2016-05-03 17:58 & $0.856 \pm 0.001$ & spec\tablenotemark{c,d} & \gold & 100\% \\
SCP16B02 & 14:32:18.022 & +32:52:55.876 & 2016-07-11 20:10 & $1.7 \pm 0.6$ & photo & \bronze & 42\% \\
SCP14C01 & 10:14:04.056 & +00:38:11.610 & 2014-12-26 13:13 & SN: $1.23 \pm 0.01$ & spec\tablenotemark{e,a} & \gold & 100\% \\
SCP15C01 & 10:14:05.174 & +00:37:16.387 & 2015-02-27 11:45 & $0.97177 \pm 0.00001$ & spec\tablenotemark{a,g} & \gold & 100\% \\
SCP15C02 & 10:14:07.620 & +00:38:08.443 & 2015-03-30 17:16 & $0.75060 \pm 0.00005$ & spec\tablenotemark{a,d} & \pyrite & 0\% \\
SCP15C03 & 10:14:07.237 & +00:37:29.418 & 2015-10-19 02:01 & $1.23376 \pm 0.00001$ & spec\tablenotemark{c,d} & \pyrite & 18\% \\
SCP15C04 & 10:14:06.530 & +00:38:47.081 & 2015-12-23 19:43 & $1.2305 \pm 0.0003$ & spec\tablenotemark{a} & \gold & 100\% \\
SCP16C01 & 10:14:06.748 & +00:39:03.977 & 2016-01-27 17:46 & $0.97394 \pm 0.00004$ & spec\tablenotemark{c,a} & \pyrite & 0\% \\
SCP16C02 & 10:14:08.349 & +00:38:52.372 & 2016-01-27 17:46 & $0.78 \pm 0.40$ & photo & \pyrite & 0\% \\
SCP16C03 & 10:14:06.375 & +00:38:25.370 & 2016-02-28 19:25 & $2.2216 \pm 0.0001$ & spec\tablenotemark{b} & \gold & 98\% \\
SCP15D01 & 21:06:11.783 & $-$58:45:20.318 & 2015-04-07 19:48 & $0.5682 \pm 0.0001$ & spec\tablenotemark{a} & \gold & 100\% \\
SCP15D02 & 21:06:07.313 & $-$58:44:02.724 & 2015-05-09 17:24 & $1.1493 \pm 0.0002$ & spec\tablenotemark{a,b} & \bronze & 47\% \\
SCP15D03 & 21:06:02.299 & $-$58:44:54.031 & 2015-08-14 13:46 & $1.1130 \pm 0.0003$ & spec\tablenotemark{a} & \gold & 100\% \\
SCP15D04 & 21:06:00.343 & $-$58:45:48.510 & 2015-09-15 07:17 & $1.13 \pm 0.50$ & photo & \gold & 94\% \\
SCP16D01 & 21:06:07.125 & $-$58:45:11.480 & 2016-02-28 16:29 & $1.12891 \pm 0.00020$ & spec\tablenotemark{a} & \gold & 97\% \\
SCP16D02 & 21:06:03.439 & $-$58:44:16.008 & 2016-03-24 05:20 & $1.74 \pm 0.20$ & photo & \silver & 81\% \\
SCP16D03 & 21:05:57.151 & $-$58:45:12.277 & 2016-05-27 22:37 & $0.61182 \pm 0.00001$ & spec\tablenotemark{a} & \pyrite & 0\% \\
SCP15E01 & 20:41:04.766 & $-$44:52:01.272 & 2015-05-09 18:59 & $0.83981 \pm 0.00001$ & spec\tablenotemark{a} & \pyrite & 0\% \\
SCP15E02 & 20:40:58.767 & $-$44:51:02.555 & 2015-06-12 17:35 & $0.35 \pm 0.20$ & photo & \pyrite & 0\% \\
SCP15E03 & 20:41:02.934 & $-$44:51:31.555 & 2015-06-12 17:35 & $1.17 \pm 0.20$ & photo & \pyrite & 0\% \\
SCP15E04 & 20:40:59.697 & $-$44:52:05.029 & 2015-09-26 12:30 & $1.04 \pm 0.40$ & photo & \bronze & 52\% \\
SCP15E05 & 20:41:01.123 & $-$44:51:12.210 & 2015-09-26 12:30 & $1.32 \pm 0.20$ & photo & \pyrite & 3\% \\
SCP15E06 & 20:40:57.739 & $-$44:51:06.524 & 2015-10-14 08:17 & $1.4811 \pm 0.0004$ & spec\tablenotemark{a} & \gold & 100\% \\
SCP15E07 & 20:41:03.306 & $-$44:51:39.474 & 2015-04-12 17:31 & $1.4858 \pm 0.0003$ & spec\tablenotemark{a,b} & \gold & 99\% \\
SCP15E08 & 20:41:03.775 & $-$44:51:28.391 & 2015-10-29 04:27 & $2.023 \pm 0.002$ & spec\tablenotemark{a} & \gold & 92\% \\
SCP16E02 & 20:41:05.729 & $-$44:51:15.900 & 2016-05-19 22:18 & $0.94352 \pm 0.00005$ & spec\tablenotemark{a} & \pyrite & 0\% \\
SCP16F01 & 00:35:47.711 & $-$43:11:57.774 & 2016-06-17 08:32 & $1.340 \pm 0.001$ & spec\tablenotemark{f} & \pyrite & 0\% \\
SCP15G01 & 00:44:02.946 & $-$20:33:49.712 & 2015-07-19 22:13 & $1.5817 \pm 0.0004$ & spec\tablenotemark{c,b} & \gold & 100\% \\
SCP16H01 & 03:30:55.991 & $-$28:42:46.560 & 2016-02-04 11:23 & $1.625 \pm 0.001$ & spec\tablenotemark{b} & \gold & 100\% \\
SCP15I01 & 02:24:26.281 & $-$03:23:32.603 & 2015-08-25 18:48 & $1.63537 \pm 0.00001$ & spec\tablenotemark{c} & \gold & 100\% \\
SCP15I02 & 02:24:27.593 & $-$03:22:57.852 & 2015-11-05 11:50 & $1.65 \pm 0.13$ & photo & \bronze & 64\% \\
SCP15J01 & 10:49:30.024 & +56:40:45.655 & 2015-03-14 13:00 & $1.26 \pm 0.27$ & photo & \pyrite & 21\% \\
SCP15K01 & 14:26:35.329 & +35:07:31.543 & 2015-04-10 17:44 & -- & hostless & \pyrite & 0\% \\
SCP15K02 & 14:26:32.146 & +35:08:48.471 & 2015-11-04 22:24 & $1.49 \pm 0.17$ & photo & \gold & 96\% \\
SCP15K03 & 14:26:28.889 & +35:07:31.842 & 2015-12-09 23:34 & $1.94 \pm 0.90$ & photo & \bronze & 32\% \\
SCP16K01 & 14:26:38.283 & +35:08:26.239 & 2016-01-14 00:57 & $1.50155 \pm 0.00002$ & spec\tablenotemark{c} & \gold & 100\% \\
SCP16K02 & 14:26:33.098 & +35:07:11.808 & 2016-02-18 14:42 & $1.47 \pm 0.15$ & photo & \gold & 92\% \\
SCP16K03 & 14:26:27.706 & +35:07:23.926 & 2016-06-03 19:28 & $1.44 \pm 0.40$ & photo & \bronze & 29\% \\
SCP16K04 & 14:26:34.815 & +35:07:46.311 & 2016-07-09 07:45 & $2.4 \pm 0.8$ & photo & \pyrite & 5\% \\
SCP16K05 & 14:26:36.921 & +35:08:10.360 & 2016-07-09 07:45 & $1.5 \pm 1.0$ & photo & \pyrite & 1\% \\
SCP16L01 & 11:42:44.876 & +15:27:51.725 & 2016-03-10 17:55 & -- & hostless & \pyrite & 0\% \\
\enddata
\tablenotetext{}{SNe with lensing estimates greater than 0 mag were de-lensed for the typing estimates shown here}
\tablenotetext{a}{VLT FORS2; \cite{vltpaper} }
\tablenotetext{b}{VLT X-shooter; \cite{vltpaper} }
\tablenotetext{c}{KECK MOSFIRE}
\tablenotetext{d}{KECK LRIS}
\tablenotetext{e}{GTC OSIRIS}
\tablenotetext{f}{HST GRISM}
\tablenotetext{g}{GEMINI GMOS-S}
\end{deluxetable}

\begin{deluxetable}{ccccccc}
\tablecaption{Possibly Lensed SNe from See Change \label{tab:lensing}}
\tablehead{\colhead{SN} & \colhead{$z_{\mathrm{SN}}$} & \colhead{Cluster} & \colhead{$z_{\mathrm{CL}}$} & \colhead{$M_{200}$} & \colhead{offset(\arcsec)} & \colhead{$\mu$}}
\startdata
SCP15A01 & $1.3 \pm 1.0$ & SPT0205$^{\mathrm{a}}$ & 1.32 & $8.8 \pm 1.4 \times 10^{14} M_{\odot}$ & 11.79\arcsec & $-0.00^{+0.0}_{-0.83}$ \\
SCP15B01 & $1.4 \pm 0.2$ & ISCSJ-1432+3253$^{\mathrm{b}}$ & 1.4 & $1.4 \pm 0.6 \times 10^{14} M_{\odot}$ & 14.77\arcsec & $-0.00^{+0.0}_{-0.02}$ \\
SCP15B02 & $2.29456 \pm 0.00003$ & ISCSJ-1432+3253$^{\mathrm{b}}$ & 1.4 & $1.4 \pm 0.6 \times 10^{14} M_{\odot}$ & 34.12\arcsec & $-0.09^{+0.03}_{-0.03}$ \\
SCP15B03 & $1.86 \pm 0.50$ & ISCSJ-1432+3253$^{\mathrm{b}}$ & 1.4 & $1.4 \pm 0.6 \times 10^{14} M_{\odot}$ & 26.29\arcsec & $-0.08^{+0.05}_{-0.06}$ \\
SCP15B04 & $1.68 \pm 1.00$ & ISCSJ-1432+3253$^{\mathrm{b}}$ & 1.4 & $1.4 \pm 0.6 \times 10^{14} M_{\odot}$ & 58.58\arcsec & $-0.02^{+0.02}_{-0.02}$ \\
SCP15B05 & $2.23892 \pm 0.00002$ & ISCSJ-1432+3253$^{\mathrm{b}}$ & 1.4 & $1.4 \pm 0.6 \times 10^{14} M_{\odot}$ & 19.95\arcsec & $-0.20^{+0.07}_{-0.07}$ \\
SCP16B02 & $1.7 \pm 0.6$ & ISCSJ-1432+3253$^{\mathrm{b}}$ & 1.4 & $1.4 \pm 0.6 \times 10^{14} M_{\odot}$ & 12.43\arcsec & $-0.14^{+0.14}_{-0.15}$ \\
SCP16C03 & $2.2216 \pm 0.0001$ & MOO-1014$^{\mathrm{c}}$ & 1.23 & $5.6 \pm 0.6 \times 10^{14} M_{\odot}$ & 18.61\arcsec & $-0.61^{+0.2}_{-0.16}$ \\
SCP15D04 & $1.13 \pm 0.50$ & SPT-CLJ2106-5844$^{\mathrm{d}}$ & 1.13 & $12.7 \pm 0.21 \times 10^{14} M_{\odot}$ & 75.21\arcsec & $-0.00^{+0.0}_{-0.08}$ \\
SCP16D02 & $1.74 \pm 0.20$ & SPT-CLJ2106-5844$^{\mathrm{d}}$ & 1.13 & $12.7 \pm 0.21 \times 10^{14} M_{\odot}$ & 28.84\arcsec & $-0.49^{+0.14}_{-0.13}$ \\
SCP15E08 & $2.023 \pm 0.002$ & SPT2040$^{\mathrm{e}}$ & 1.48 & $5.8 \pm 1.4 \times 10^{14} M_{\odot}$ & 45.25\arcsec & $-0.10^{+0.02}_{-0.02}$ \\
SCP15I02 & $1.65 \pm 0.13$ & SPARCSJ0224$^{\mathrm{i}}$ & 1.63 & $4.9 \pm 4.0 \times 10^{14} M_{\odot}$ & 36.27\arcsec & $-0.00^{+0.0}_{-0.03}$ \\
SCP15K01 & $1.75 \pm 0.40$ & IDCSJ1426$^{\mathrm{k}}$ & 1.75 & $4.3 \pm 1.1 \times 10^{14} M_{\odot}$ & 60.48\arcsec & $-0.00^{+0.0}_{-0.02}$ \\
SCP15K03 & $1.94 \pm 0.90$ & IDCSJ1426$^{\mathrm{k}}$ & 1.75 & $4.3 \pm 1.1 \times 10^{14} M_{\odot}$ & 71.43\arcsec & $-0.05^{+0.03}_{-0.04}$ \\
SCP16K03 & $1.44 \pm 0.40$ & IDCSJ1426$^{\mathrm{k}}$ & 1.75 & $4.3 \pm 1.1 \times 10^{14} M_{\odot}$ & 87.29\arcsec & $-0.01^{+0.01}_{-0.02}$ \\
SCP16K04 & $2.4 \pm 0.8$ & IDCSJ1426$^{\mathrm{k}}$ & 1.75 & $4.3 \pm 1.1 \times 10^{14} M_{\odot}$ & 44.58\arcsec & $-0.13^{+0.07}_{-0.08}$ \\
SCP16K05 & $1.5 \pm 1.0$ & IDCSJ1426$^{\mathrm{k}}$ & 1.75 & $4.3 \pm 1.1 \times 10^{14} M_{\odot}$ & 51.52\arcsec & $-0.00^{+0.0}_{-0.06}$ \\
SCP16L01 & $1.19 \pm 0.50$ & MOO-J1142$^{\mathrm{l}}$ & 1.19 & $11.0 \pm 0.2 \times 10^{14} M_{\odot}$ & 40.02\arcsec & $-0.00^{+0.0}_{-0.21}$ \\
\enddata
References for Cluster masses -- \textit{a}: \citealt{stalder13}, \textit{b}: \citealt{brodwin13}, \textit{c}: \citealt{brodwin15}, \textit{d}: \citealt{foley11}, \textit{e}: \citealt{bayliss14}, \textit{i}: \citealt{muzzin13}, \textit{k}: \citealt{brodwin12}, \textit{l}: \citealt{gonzalez15}
\end{deluxetable}

\end{document}